\documentclass[prb,reprint,aps,superscriptaddress,longbibliography]{revtex4-2}
\usepackage[colorlinks=true,linkcolor=blue,citecolor=blue,urlcolor=blue]{hyperref}

\usepackage{graphicx}
\usepackage{bm}
\usepackage{braket}
\usepackage{amsmath}
\usepackage{amssymb}
\usepackage{graphics}
\usepackage{enumerate}
\usepackage{siunitx}
\usepackage[dvipsnames]{xcolor}

\begin{document}

\title{Tunneling spectroscopy of two-dimensional superconductors with the quantum twisting microscope}

\author{Nemin Wei}
\affiliation{\mbox{Department of Physics and Yale Quantum Institute, Yale University, New Haven, Connecticut 06520, USA}}

\author{Felix von Oppen}
\affiliation{\mbox{Dahlem Center for Complex Quantum Systems and Fachbereich Physik, Freie Universit\"at Berlin, 14195 Berlin, Germany}}

\author{Leonid I.\ Glazman}
\affiliation{\mbox{Department of Physics and Yale Quantum Institute, Yale University, New Haven, Connecticut 06520, USA}}

\date{\today}

\begin{abstract}
    The ongoing discoveries of graphene-based superconductors underscore the quest to understand the structure of new superconducting orders. We develop a theory that 
    facilitates the use of the quantum twisting microscope (QTM) for that purpose. This work investigates momentum-conserving tunneling across a planar junction formed by a normal monolayer graphene tip and a superconducting graphene sample within the QTM setting. We show that the bias dependence of the zero-temperature tunneling conductance exhibits singularities that provide momentum-resolved information about the Bogoliubov quasiparticle spectra, including the superconducting gap. Using a model of superconducting twisted bilayer graphene (TBG), we illustrate that simultaneously tuning the tip doping level and the tip-sample twist angle allows for measuring the momentum-resolved superconducting gap in TBG. Our results indicate that momentum-conserving tunneling spectroscopy with the QTM is a  promising method for exploring superconductivity in two-dimensional van der Waals materials.
\end{abstract}

\maketitle

\section{Introduction}

Tunneling spectroscopy is a powerful experimental tool for characterizing the superconducting state \cite{wolf2011principles}. It has been widely applied to determine the energy gap of superconductors \cite{giaever1960energy,giaever1960electron}, to probe their pairing symmetry \cite{fischer2007scanning,sukhachov2023andreev,biswas2025andreev}, and to confirm phonon-mediated pairing in conventional superconductors by precisely reconstructing the Eliashberg function from tunneling data \cite{mcmillan1965lead,parks2018superconductivity}. 

Determining the structure of the superconducting order parameter is a pivotal problem in the study of graphene-based superconductivity \cite{cao2018unconventional,park2021tunable,hao2021electric,zhang2022promotion,park2022robust,burg2022emergence,su2023superconductivity,uri2023superconductivity,zhou2021superconductivity,zhou2022isospin,han2025chiral}. For superconducting magic-angle twisted bilayer and trilayer graphene, the observation of a `V'-shaped tunneling gap provides compelling evidence for their unconventional, non $s-$wave nature \cite{oh2021evidence,kim2022evidence,park2025simultaneous,kim2025resolving}. However, significant uncertainties remain, including whether the superconducting gap is strongly anisotropic but fully open \cite{tanaka2025superfluid}, or intrinsically nodal \cite{banerjee2025superfluid}, and whether the order parameter exhibits nematicity \cite{cao2021nematicity}. Definitive progress will require measurements of quasiparticle excitation spectra across a wide range of momenta in the moir\'e Brillouin zone (mBZ).

The development of the quantum twisting microscope (QTM) \cite{inbar2023quantum} enables momentum-resolved tunneling spectroscopy of two-dimensional van der Waals materials by forming a twistable planar tunnel junction between van der Waals tip and sample. Coherent tunneling in a QTM conserves energy and in-plane momentum, allowing momentum-resolved measurements of energy dispersions by probing the tunneling conductance as a function of bias voltage and twist angle \cite{birkbeck2024measuring,lee2025revealing,pichler2024probing,xiao2024theory,wei2025dirac}.
Recent cryogenic QTM measurements with a monolayer graphene (MLG) tip have imaged the electron spectral function of magic-angle twisted bilayer graphene (MATBG), unveiling strong renormalization effects driven by interactions  \cite{xiao2025interacting}. This breakthrough suggests that a low-temperature QTM holds considerable promise for probing superconductivity in MATBG. 

QTM tunneling 
reveals the dispersion of electronic excitations in the sample
most directly if the tip has tiny Fermi pockets (centered around the Dirac momenta in the case of a MLG tip). Under this condition, the (zero-temperature) differential tunneling conductance $dI/dV$ measures the single-particle spectral function of the sample at the momenta of the pointlike Fermi pockets
\cite{wei2025theory, jang2017full}. As a function of twist angle between tip and sample, these pockets probe the Brillouin zone (BZ) of the sample along an arc in reciprocal space. 

\begin{figure}
    \centering
    \includegraphics[width=1\linewidth]{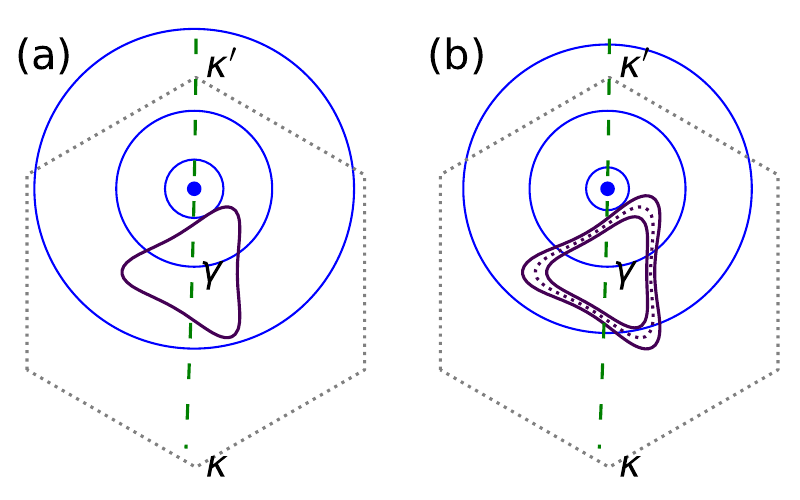}
    \caption{Schematic momentum-space diagrams showing the mBZ (gray dotted) and equal-energy contours of a moir\'e graphene sample (black) as well as the Fermi circles of the MLG tip at different carrier densities (blue). The blue dot represents a Dirac point of the MLG tip, moving along the dashed green arc as the tip is twisted relative to the sample. (a) Zero-bias energy- and momentum-conserving tunneling between two normal states occurs at intersection points of the Fermi lines of tip and sample. 
    As the doping level of the tip increases, the onset of tunneling is marked by the smallest blue circle, signaled by a singularity in the zero-bias tunneling conductance $dI(0)/dV$ \cite{eisenstein1991probing}, whereas $dI(0)/dV$ vanishes for a sufficiently large tip Fermi circle, such as the largest blue circle. (b) A superconducting gap suppresses the zero-bias tunneling. When the tip Fermi circle intersects the Fermi line of the sample (black dotted) at momenta $\bm p^*$, the tunneling conductance sharply increases when the bias reaches the superconducting gap, $-eV=\pm|\Delta_{\bm p^*}|$. Adjusting the size of the tip Fermi circle can scan the gap along the Fermi line of the sample. At larger biases $V^*$, the equal-energy contours of Bogoliubov quasiparticles splits into two concentric lines (black solid), and if either of them is in tangency with the tip Fermi circle (e.g., the smallest and biggest blue circles), $dI(V^*)/dV$ becomes singular, c.f. Eq.~\eqref{eq:dIdV_v*}.  }
    \label{fig:bz_schematic}
\end{figure}

A dual-gated design equips the QTM with independent control over the doping levels of tip and sample \cite{inbar2023quantum}. This allows one to continuously ``inflate'' the Fermi pockets of the tip, thereby extending the accessible momentum range, see Fig.~\ref{fig:bz_schematic} for illustration. Given the small size of the sample mBZ, varying the Fermi wave vector of the tip can scan the momentum states over a substantial portion of mBZ. At the same time, a finite extent of the tip's Fermi pocket introduces ambiguity in identifying the momenta of the measured quasiparticle states \cite{wei2025dirac}.
This makes it desirable to develop protocols capable of directly acquiring momentum-resolved quasiparticle spectra
from tunneling data, thereby minimizing the need for fitting to numerical simulations based on model Hamiltonians.

Motivated by these opportunities and challenges, we theoretically investigate elastic momentum-conserving tunneling across 
tunnel junctions between a normal-state MLG and superconducting MATBG in the QTM geometry.
We analyze the bias-dependent singularities in the differential conductance $dI/dV$ and their evolution with tip doping level. We propose that the momentum-resolved superconducting gap along the Fermi line of the sample can be extracted by tracing the singularities in $dI/dV$ as a function of the twist angle and tip doping.
Our results show that QTM tunneling spectroscopy can, in principle, determine whether a moir\'e superconductor is nodal and if so, locate the node(s) within the mBZ.
We focus on normal state-superconductor tunneling, which is suitable for spectroscopic analysis. Another approach to probe the order parameter symmetry was suggested in an earlier work on Josephson coupling in twistable superconductor-superconductor junctions \cite{xiao2023probing}.

The paper is organized as follows. In Sec.~\ref{sec:formalism}, we present the formalism for elastic momentum-conserving tunneling in a normal metal-superconductor junction. Section \ref{sec:fs_singularity} analyzes the singularity in $dI/dV$ and its connection to the momentum-resolved energy dispersion of superconducting quasiparticles, followed by a discussion of the effects of quasiparticle broadening in Sec.~\ref{sec:broadening}. We demonstrate the formalism first on a toy model (Sec.~\ref{sec:toymodel}) and then on a continuum model of twisted bilayer graphene (TBG) sufficient for capturing superconductivity at the BCS level (Sec.~\ref{sec:tbg}).
In that section, we identify signatures of various superconducting gap structures. Finally, limitations of our theory are discussed in Sec.~\ref{sec:conclusion}.

\section{Formalism of elastic momentum-conserving tunneling}\label{sec:formalism}

In this section, we study elastic momentum-conserving tunneling between a normal-metal tip and a superconducting sample. 
We describe elastic interlayer tunneling using a tunneling Hamiltonian $H_{\text{tun}}$ that preserves spin and in-plane momentum. 
The tunneling current can be derived using linear response theory \cite{mahan2013many},
%
\begin{align}\label{eq:i_basic}
    I=&\frac{2\pi e}{\hbar}\sum_{\sigma}\sum_{\bm k\lambda} \sum_{\bm k^{\prime}\lambda^{\prime}} |\langle \bm k^{\prime}\lambda^{\prime}\sigma| H_{\text{tun}}|  \bm k\lambda\sigma \rangle|^2\notag\\
    &\times \int d \omega\ (f_{\omega}-f_{\omega+e V}) A_{\lambda^{\prime}\sigma}^{T}\left(\bm k^{\prime}, \omega+e V\right)A_{\lambda\sigma}^{S} \left(\bm k, \omega\right),
\end{align}
%
%
Here, $|\bm k\lambda\sigma\rangle$ represents a Bloch state in the sample with quasimomentum $\bm k$, band index $\lambda$, and spin $\sigma$. Primed variables refer to the corresponding quantities in the tip. $f_{\omega}$ denotes the Fermi-Dirac distribution, and $A^{T/S}$ denote the spectral functions of the tip and sample, respectively. The spectral function of the superconductor $A_{\lambda\sigma}^{S}(\bm k, \omega) = -\text{Im}G_{\lambda\sigma}^{R}(\bm k, \omega)/\pi$ can be determined from the retarded Green's function 
\begin{equation}\label{eq:GR}
    G_{\lambda\sigma}^{R}(\bm k, \omega) = -i\int_{0}^{\infty} dt\ \langle\{c_{\bm k\lambda\sigma}(t), c_{\bm k\lambda\sigma}^{\dagger}(0)\}\rangle e^{i\omega t},
\end{equation}
where $c_{\bm k\lambda\sigma}$ annihilates the sample's Bloch state $|\bm k\lambda \sigma\rangle$. 

The spin and orbital structure of Cooper pairs in graphene-based superconductors remain controversial. We restrict our analysis to BCS pairing between time-reversed states in a single Bloch band. 
The band index $\lambda$ is suppressed henceforth to shorten the notation. 
The mean-field Hamiltonian of the superconductor reads
\begin{align}\label{eq:h_sc}
\hat{H}_{\lambda}&=\sum_{\bm k}\left(c_{\bm k\uparrow}^{\dagger}, c_{-\bm k\downarrow}\right)\left(\begin{array}{cc}
\xi_{\bm k\uparrow} & \Delta_{\bm k} \\
\Delta_{\bm k}^* & -\xi_{-\bm k\downarrow}
\end{array}\right)\left(\begin{array}{c}
c_{\bm k\uparrow} \\
c_{-\bm k\downarrow}^{\dagger}
\end{array}\right),
\end{align}
where $\xi_{\bm k\sigma}$ denotes normal-state energy dispersion relative to the Fermi level of the sample, and $\Delta_{\bm k}$ represents the gap function. 
The Green's function obeys
%
%
\begin{align}\label{eq:GR_infinite}
    G_{\uparrow}^{R}(\bm k, \omega)^{-1} & = \omega+i\eta-\xi_{\bm{k}\uparrow}-\frac{|\Delta_{\bm k}|^2}{\omega+i\eta  + \xi_{-\bm{k}\downarrow}}.
\end{align}
The spectral function 
\begin{align}\label{eq:spectral_+-}
    A_{\uparrow}^{S}(\bm k, \omega) = |u_{\bm k}|^2\delta\left(\omega-\epsilon^{S}_{\bm k\uparrow}\right) + |v_{\bm k}|^2 \delta\left(\omega+\epsilon^{S}_{-\bm k\downarrow}\right),
\end{align}
consists of two Bogoliubov quasiparticle peaks at energies $\epsilon^{S}_{\bm k\uparrow}$ and $-\epsilon^S_{-\bm k\downarrow}$ with weights given by the coherence factors $|u_{\bm k}|^2$ and $|v_{\bm k}|^2=1-|u_{\bm k}|^2$, respectively. Due to the time-reversal symmetry of the normal state, $\xi_{\bm k\uparrow}=\xi_{-\bm k\downarrow}$, we have
\begin{gather}
    \epsilon^{S}_{\bm k\uparrow}= \epsilon^{S}_{-\bm k\downarrow} =\sqrt{\xi_{\bm k\uparrow}^2+|\Delta_{\bm k}|^2},\label{eq:dispersion}\\
    |u_{\bm k}|^2 = \frac{1}{2}\Big(1+\frac{\xi_{\bm k\uparrow}}{\sqrt{\xi_{\bm k\uparrow}^2+|\Delta_{\bm k}|^2}}\Big).
\end{gather}
The corresponding calculation for the spin-down component yields $A_{\downarrow}^{S}(-\bm k,\omega) = A_{\uparrow}^{S}(\bm k,\omega)$.
For superconductors with a time-reversal symmetric normal state and intraband pairing only, we thus find that the two time-reversed flavors ($\uparrow$, $\downarrow$) contribute equally to the tunneling current. This holds regardless of whether the superconductor spontaneously breaks time-reversal symmetry. We can then focus on a single spin component and omit the spin indices in the following.

In evaluating the differential conductance $dI/dV$, we make the simplifying assumption of negligible screening in the tip electrode. 
Then, the bias voltage $V$ controls the Fermi energy $\mu_T$ of the tip electrons, while the electrostatic band energy offset $\phi$ is independent of $V$. 
Consequently, we only need to  differentiate the Fermi-Dirac function in Eq.~\eqref{eq:i_basic} with respect to $V$ to obtain the differential conductance.
Pulling together all of these considerations, we obtain the zero-temperature tunneling conductance
\begin{align}\label{eq:didv_basic}
    \frac{dI}{dV}=\frac{2\pi e^2 N_f}{\hbar}\sum_{\bm k} \sum_{\bm k^{\prime}\lambda^{\prime}} &|\langle \bm k^{\prime}\lambda^{\prime}| H_{\text{tun}}|  \bm k\lambda \rangle|^2 \notag\\
    &\times A_{\lambda^{\prime}}^{T}\left(\bm k^{\prime}, 0\right)A^{S} \left(\bm k, -eV\right),
\end{align}
where $N_{f}=2$ denotes the flavor degeneracy. A few comments are in order: 
(i) In graphene-based superconductors with well-defined valley quantum numbers, the two time-reversed Bloch states occupy opposite valleys. If each valley has twofold spin degeneracy such as in intervalley spin-singlet superconductors, Eq.~\eqref{eq:didv_basic} can be further simplified by restricting $\bm k$ to a single valley and using $N_f=4$. (ii) While the above analysis considers pairing between opposite spins, Eq.~\eqref{eq:didv_basic} also holds for equal-spin pairing between two Bloch states related by orbital time reversal, if we identify $\sigma$ as the valley pseudospin and restrict $\bm k$ to a single valley in the derivations. This type of pairing has been conjectured for certain graphene-based superconductors \cite{zhou2022isospin,huang2022pseudospin,qin2021inplane,dong2024superconductivity,dong2023transformer}. (iii) The simplification of the electrostatics made in our derivation is not essential for the $dI/dV$ singularity discussed in later sections. For instance, Ref.~\cite{wei2025dirac} found the same type of singularity in a different limit, where $\phi$ varies with $V$ but $\mu_T$ is fixed.

\section{Fermi edge singularities}\label{sec:fs_singularity}

\begin{figure}
    \centering
    \includegraphics[width=1\linewidth]{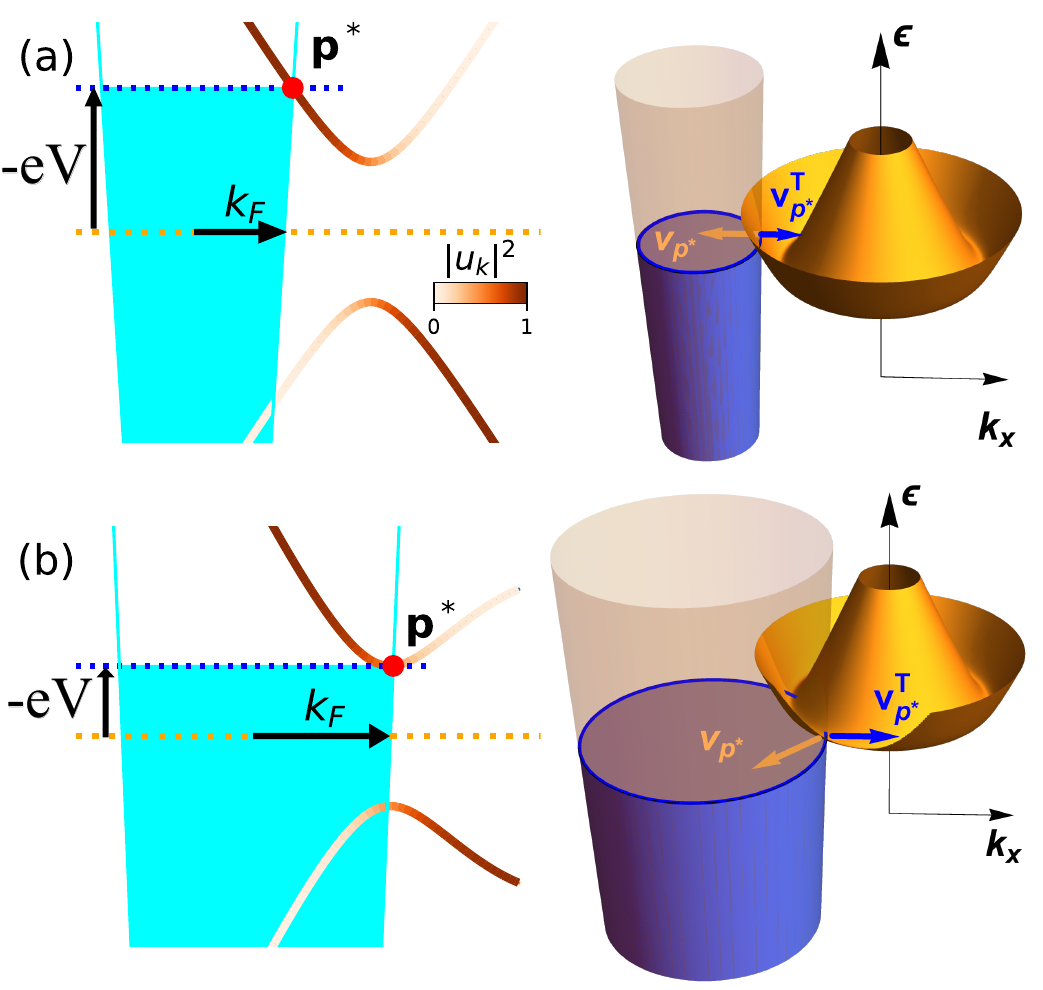}
    \caption{Schematics of the energy band alignment of the normal metallic tip (blue) and superconducting sample (orange). On the left, the dashed blue and orange lines represent the Fermi levels of the tip and sample, respectively. The bias $-eV$ controls the tip's Fermi wave vector, while $k_F$ denotes the Fermi wave vector at zero-bias. The intersections of tip's Fermi line and the Bogoliubov bands marked by the red dots generates a finite differential conductance due to the elastic momentum-conserving tunneling. The color scale of Bogoliubov quasiparticles represents the coherence factor $|u_{\bm k}(\epsilon)|^2$, see Eq.~\eqref{eq:uk}. On the right, the 3D plots illustrate the tangency between the tip energy bands and the upper Bogoliubov band at $\bm p^*$. An isotropic superconducting gap $\Delta$ is assumed. (a) For $-eV>\Delta$, the Fermi velocity $\bm v_{\bm p^*}$ (orange arrow) of the superconductor is parallel to $\bm v_{\bm p^*}^T$ (blue arrow).  (b) For $-eV=\Delta$, the $\bm p^*$ lies on the Fermi line of the superconductor's normal state, and the Fermi velocities $\bm v_{\bm p^*}$ and $\bm v_{\bm p^*}^{T}$ are not necessarily in parallel.}
    \label{fig:schematic}
\end{figure}

In this section, we study the singularity in tunneling conductance arising from the nonanalytic bias dependence of the phase space for tunneling and the discontinuity of the Fermi-Dirac function across the Fermi level. 

Figure~\ref{fig:schematic}(a) shows a representative band alignment configuration that yields a tunneling-conductance singularity at bias $-eV$ above the isotropic superconducting gap $\Delta$.
In this plot, the tip's Fermi line and the quasiparticle dispersion touch at the wave vector $\bm p^*$, marked by a red dot. 
Adjusting the tip's Fermi wave vector can cause the point of tangency $\bm p^*$ to scan the quasiparticle dispersion (orange curve), revealing it through the bias at which the tunneling conductance singularity occurs. Compared to the normal state, the superconducting order opens a gap along the Fermi line of the sample, yielding a singularity at $-eV=\Delta$ if the Fermi lines of the tip and sample cross; see Fig.~\ref{fig:schematic}(b). This singularity is particularly useful for measuring the superconducting gap along the Fermi line of the sample.
Figure~\ref{fig:schematic}(a) and (b) can both be interpreted as instances of tangency between the tip Fermi line and the quasiparticle dispersion surface in the three-dimensional energy-momentum space, so the resulting singularities of the tunneling conductance can be analyzed in a unified framework, which is detailed below.

We model the tip as a free-electron system. Assuming that the Fermi level resides within a single band $\lambda'$ with energy dispersion $\xi_{\bm k'}^{T}$ relative to the Fermi level, the spectral function takes the form $A_{\lambda'}^{T}(\bm k',\omega) = \delta(\omega-\xi_{\bm k'}^{T})$. The spectral function of the superconductor is given in Eq.~\eqref{eq:spectral_+-}  with $\epsilon_{\bm k\uparrow}^S=\epsilon_{-\bm k\downarrow}^S \equiv\epsilon_{\bm k}^S$. Neglecting Umklapp processes,
the tunneling matrix element can be written as $\langle \bm k^{\prime}\lambda^{\prime}| H_{\text{tun}}|  \bm k\lambda \rangle =T(\bm k)\delta_{\bm k',\bm k}$ and the differential conductance Eq.~\eqref{eq:didv_basic} reduces to
\begin{align}
    \frac{dI}{dV} = \frac{2\pi e^2 N_f\Omega}{\hbar} \sum_{s=\pm}&\int \frac{d^2k}{(2\pi)^2} |T(\bm k)|^2 |u_{\bm k}(s\epsilon_{\bm k}^{S})|^2\notag\\
    &\ \times \delta( \xi_{\bm k}^T)\delta (\xi_{\bm k}^T - s\epsilon_{\bm k}^{S} - eV).\label{eq:dIdV_sc_fs1}
\end{align}
%
Here, $\Omega$ stands for the junction area, and the notation
\begin{equation}\label{eq:uk}
|u_{\bm k}(\epsilon)|^2 = 
\begin{cases}
|u_{\bm k}|^2,\ \epsilon\geq 0,  \\
|v_{\bm k}|^2,\ \epsilon<0. \\
\end{cases}
\end{equation}
is introduced for compactness. The tunneling conductance originates from intersections of the tip's Fermi line $\xi_{\bm k}^T=0$ and the contour lines $\pm\epsilon_{\bm k}^S=-eV$ of the sample's quasiparticle dispersion. At characteristic bias voltage(s) $V^*$, intersections
coalesce into a point of tangency. Notice that this happens in both Fig.\ \ref{fig:schematic}(a) and Fig.\ \ref{fig:schematic}(b).
At the corresponding wave vector $\bm p^*$, the quasiparticles of tip and sample have parallel group velocities, $\bm v_{\bm p^*}^T\parallel \bm v_{\bm p^*}^S =\partial_{\bm p^*}\epsilon_{\bm p^*}^S/\hbar$. This condition includes the case $\bm v_{\bm p^*}^S=0$ at the gap edge of an isotropic superconductor, as shown in Fig.~\ref{fig:schematic}(b).

Assuming that the singular bias dependence of $dI/dV$ at bias $V^*$ is completely contributed by wave vectors $\bm k = \bm p^*+\bm q$ in the vicinity of a single $\bm p^*$, we adopt local coordinates $\bm q = q_{\parallel}\hat{\bm v}^T + q_{\perp}\hat{z}\times\hat{\bm v}^T$ defined by the tip dispersion. Here, the unit vector $\hat{\bm v}^T$ points along the direction of $\bm v_{\bm p^*}^T$.  Near $\bm p^*$, the normal-state dispersions of tip and sample read
\begin{align}
    \xi_{\bm p^*+\bm q}^{T} &\approx \xi_{\bm p^*}^{T}  + \hbar  v_{\bm p^*}^T q_{\parallel} + \frac{\hbar^2q_{\perp}^2}{2m_{\bm p^*}^T}, \label{eq:dispersion_tip}\\
    \xi_{\bm p^*+\bm q} &\approx \xi_{\bm p^*} + \hbar v_{\bm p^*,\parallel}q_{\parallel} + \hbar v_{\bm p^*,\perp}q_{\perp} + \frac{\hbar^2q_{\perp}^2}{2m_{\bm p^*}}. \label{eq:dispersion_sample_normal}
%
\end{align}
Here, $m_{\bm p^*}^{T}$ denotes the effective mass of the tip's electron in the direction transverse to its Fermi velocity, while $m_{\bm p^*}$ denotes the effective mass in the normal state of the sample along the same direction. We eliminate the first Dirac-delta in Eq.~\eqref{eq:dIdV_sc_fs1} by integrating over $q_{\parallel}$. This restricts the tunneling electrons to the tip's Fermi line $\bm p^* + \bm q (q_{\perp})$, with 
\begin{equation}
\bm q (q_{\perp}) \approx -\frac{\hbar^2 q_{\perp}^2}{2m_{\bm p^*}^Tv_{\bm p^*}^T} \hat{\bm v}^T + q_{\perp} \hat{z}\times \hat{\bm v}^T.
\end{equation}
We can then expand the quasiparticle dispersion of the sample along $\bm p^*+\bm q(q_{\perp})$ to quadratic order in $q_{\perp}$,
\begin{equation}\label{eq:beta_def}
    |\epsilon_{\bm p^*+\bm q(q_{\perp})}^S|^2 = |\epsilon_{\bm p^*}^S|^2 + \beta_{\bm p^*}\hbar^2 v_{\bm p^*}^2q_{\perp}^2 + \mathcal{O}(q_{\perp}^3). 
\end{equation}
Note that unlike in Eq.~\eqref{eq:dispersion_sample_normal}, the first-order term in $q_{\perp}$ vanishes due to the tangency condition $\bm v_{\bm p^*}^T\parallel \bm v_{\bm p^*}^S$.
Inserting Eq.~\eqref{eq:beta_def} into the second Dirac-delta in Eq.~\eqref{eq:dIdV_sc_fs1}, we arrive at the asymptotic expression 
\begin{align}
    \frac{dI}{dV} 
    &\xrightarrow{V\rightarrow V^*} \frac{e^2}{h}\frac{ N_f\Omega|T(\bm p^*)|^2}{\hbar^2 |v_{\bm p^*}^T v_{\bm p^*}|} |u_{\bm p^*}(-eV^*)|^2\notag\\
    &\qquad\quad\times\sqrt{\frac{2V^*}{\beta_{\bm p^*}(V -V^*)}}\Theta\left(\frac{\beta_{\bm p^*}(V -V^*)}{V^*}\right) \label{eq:dIdV_v*}
\end{align}
for the differential conductance.
%
We conclude that the tangency between the tip's Fermi line and the contour line of the quasiparticle dispersion  generically leads to an inverse-square-root singularity in $dI/dV$. We emphasize that this result applies to the situations depicted in Fig.\ \ref{fig:schematic}(a) as well as (b).

The sign and magnitude of $\beta_{\bm p^*}$ are important in characterizing the $dI/dV$ singularity. We express $\beta_{\bm p^*}$ in terms of the band structure parameters by inserting Eqs.~\eqref{eq:dispersion} and~\eqref{eq:dispersion_sample_normal} into Eq.~\eqref{eq:beta_def},
%
which yields the relation
\begin{equation}\label{eq:beta}
    \beta_{\bm p^*} =  \frac{\alpha_{\bm p^*}\xi_{\bm p^*}}{m_{\bm p^*}v_{\bm p^*}^2} +\frac{v_{\bm p^*, \perp}^2}{v_{\bm p^*}^2} - \frac{\partial_{{\parallel}}|\Delta_{\bm p^*}|^2}{2m_{\bm p^*}^Tv_{\bm p^*}^Tv_{\bm p^*}^2} + \frac{\partial_{{\perp}}^2|\Delta_{\bm p^*}|^2}{2v_{\bm p^*}^2}.
\end{equation}
Here we use the shorthand notations $\alpha_{\bm p^*} = 1-m_{\bm p^*}v_{\bm p^*,\parallel}/m_{\bm p^*}^Tv_{\bm p^*}^T$ and $\partial_{i}f_{\bm p^*}\equiv \partial_{\hbar q_{i}}f_{\bm p^*+\bm q}|_{\bm q=0}$ ($i=\parallel,\perp$).
%
The first two terms in Eq.~\eqref{eq:beta} are not entirely independent as the tangency condition $\bm v_{\bm p^*}^T\parallel \bm v_{\bm p^*}^S$ requires 
\begin{equation}\label{eq:tangency_constraint}
    \xi_{\bm p^*}v_{\bm p^*,\perp} = -\partial_{{\perp}}|\Delta_{\bm p^*}|^2.
\end{equation}
Equations~\eqref{eq:dIdV_v*}-\eqref{eq:tangency_constraint} are the main results of this section. In the following, we apply them to analyze the $dI/dV$ singularities associated with Fig.~\ref{fig:schematic} in more detail. 


In Fig.~\ref{fig:schematic}(a), the point of tangency $\bm p^*$ is away from the Fermi line of the sample, $\xi_{\bm p^*}\neq 0$. Equations~\eqref{eq:beta} and~\eqref{eq:tangency_constraint} imply that $\beta_{\bm p^*}=\alpha_{\bm p^*} \xi_{\bm p^*}/m_{\bm p^*}v_{\bm p^*}$ because of the constant gap and $v_{\bm p^*,\perp}=0$.
%
Interestingly, for $\bm p^*$ on opposite sides of the $\xi_{\bm k}=0$ line, the opposite signs of $\xi_{\bm p^*}$ and $\beta_{\bm p^*}$ lead to a qualitative difference in $dI/dV$. If $\beta_{\bm p^*}>0$, the differential conductance $dI/dV$ shows an inverse-square-root divergence as $|V|$ approaches $|V^*|$ from above. If $\beta_{\bm p^*}<0$, the divergence occurs as $|V|$ approaches $|V^*|$ from below [see Eq.~\eqref{eq:dIdV_v*}].

In Fig.~\ref{fig:schematic}(b), $\bm p^*$ lies on the intersection between the Fermi lines of the tip and sample, $\xi_{\bm p^*} = 0$ \footnote{Figure~\ref{fig:schematic}(b) contains two intersections between the Fermi lines of tip and sample. We focus on one of them.}. The right plot of panel (b) shows that the Fermi velocities of the tip and sample at $\bm p^*$ can be non-parallel, $v_{\bm p^*,\perp}\neq 0$.
As a result, $\beta_{\bm p^*}= v_{\bm p^*,\perp}^2/v_{\bm p^*}^2>0$. 
Because $|u_{\bm p^*}|^2= 1/2$, the differential conductance $dI/dV$ exhibits two `particle-hole' symmetric peaks at biases $-eV^*=\pm \Delta$, associated with the upper and lower Bogoliubov quasiparticle dispersion, respectively.


These two types of $dI/dV$ singularities cross in the $(k_F, V)$ plane when the Fermi lines of the tip and sample become tangent, $\xi_{\bm p^*}=0$ and $v_{\bm p^*,\perp}=0$, at certain Fermi wave vectors $k_F^0$ and biases $-eV^*=\pm\Delta$. This leads to a vanishing $\beta_{\bm p^*}$ and a stronger divergence of tunneling conductance. Using Eq.~\eqref{eq:dIdV_sc_fs1} and the higher-order expansion of the energy dispersion, $|\epsilon_{\bm p^*+\bm q(q_{\perp})}^S|^2 = (\alpha_{\bm p^*} \hbar^2q_{\perp}^2/2m_{\bm p^*})^2 + \Delta^2$, we find that as $V\rightarrow V^*$, 
\begin{equation}\label{eq:didv_3/4}
    \frac{dI(k_F^{0},V)}{dV} \propto |V-V^*|^{-\frac{3}{4}}\Theta(|V|-|V^*|).
\end{equation}

As shown further in Sec.\ \ref{sec:toymodel}, the above consideration aided by Fig.~\ref{fig:schematic} directly carries over to superconductors with anisotropic gaps, provided the momentum dependence of the superconducting gap is weak,
%
\begin{equation}\label{eq:assumption}
    \left|\frac{\partial_{\bm p^*}|\Delta_{\bm p^*}|}{\hbar v_{\bm p^*}}\right|
    \ll 1.
\end{equation}
In this case, Eqs.~\eqref{eq:tangency_constraint} and~\eqref{eq:assumption} imply that the tangency of the tip's Fermi line and Bogoliubov quasiparticle dispersion takes the form of either Fig.~\ref{fig:schematic}(a) $(v_{\bm p^*,\perp}/v_{\bm p^*}\ll 1)$ or Fig.~\ref{fig:schematic}(b) $(\xi_{\bm p^*}\approx 0)$.
The associated $dI/dV$ singularities occupy different regions in the $(k_F, V)$ plane. 
The former appears over a large range of biases ($|eV|\gtrsim |\Delta_{\bm p^*}|$), 
while the latter occurs at
%
\begin{equation}\label{eq:v*_largevp}
    -eV^*(k_{F})\approx\pm|\Delta_{\bm p^*(k_{F})}|.
\end{equation}
According to Eq.~\eqref{eq:assumption}, it varies within an energy range much narrower than the sample Fermi energy as $k_{F}$ changes. The crossover region, where $\xi_{\bm p^*}\approx 0$ and $v_{\bm p^*, \perp}\approx 0$, can feature an enhanced strength of the $dI/dV$ singularity due to the smallness of $\beta_{\bm p^*}$ (see Eq.~\eqref{eq:beta}). 

\section{Effects of quasiparticle broadening}\label{sec:broadening}

Finite quasiparticle lifetime in either the tip or the sample broadens the $dI/dV$ singularities in elastic momentum-conserving tunneling \cite{zheng1993tunneling}. However, a strong mismatch between the quasiparticle velocities in the tip and in the narrow-band superconductor being probed can result in markedly different broadening effects due to quasiparticle relaxation in the two systems. Heuristically, an energy broadening $\gamma_{T}$ in the tip corresponds to a momentum uncertainty $\gamma_{T}/|\bm v_{\bm p^*}^T|$. Due to in-plane momentum conservation, this momentum uncertainty maps onto an energy resolution of order $\gamma_{T} |\bm v_{\bm p^*}^S|/|\bm v_{\bm p^*}^T|\ll \gamma_{T}$ when measuring the sample's energy spectrum \cite{wei2025dirac}.

The suppression of tip-induced broadening is particularly pronounced when $\bm p^*$ lies on the Fermi line of the sample ($\xi_{\bm p^*}=0$), where $|\bm v_{\bm p^*}^{S}|\sim |\partial_{\bm p^*} |\Delta_{\bm p^*}||/\hbar \ll |\bm v_{\bm p^*}^T|$. 
In this scenario, 
we can derive a simple formula for the tunneling conductance that accounts for a finite relaxation rate $\eta$ of the Bogoliubov quasiparticles in the sample. 


To this end, we define the set $\mathcal{S}(k_F,\theta)$ as the collection of wave vectors $\bm p^*$ of the intersections between the Fermi lines of the tip and sample (at zero bias \footnote{It does not matter whether the intersections between the Fermi lines of the tip and sample are chosen at zero bias, or at $-eV=|\Delta_{\bm p^*}|$ or $-|\Delta_{\bm p^*}|$. Using the other two definitions changes $\bm p^*$ by $|\delta\bm p^*|\sim |\Delta_{\bm p^*}|/\hbar|\bm v_{\bm p}^T|$ and changes $|\Delta_{\bm p^*}|$ by $\delta\Delta_{\bm p^*}\sim |\Delta_{\bm p^*}|\partial_{\bm p^*}|\Delta_{\bm p^*}|/\hbar v_{\bm p^*}^{T}\ll |\Delta_{\bm p^*}|$.} and at tip-sample twist angle $\theta$). The differential conductance is given by the sum of contributions from wave vectors around these intersections. We linearize the normal-state dispersion as $\xi_{\bm p^{*}+\bm q}^{(T)} =\hbar \bm v_{\bm p^*}^{(T)}\cdot\bm q$. Note that $A^T(\bm p^*+\bm q, 0)$ changes only along $\hat{\bm v}^T$ and concentrates in a narrow range of $q_{\parallel}$ of width $\sim\gamma_T/\hbar|\bm v_{\bm p}^T|$. Within this narrow $q_{\parallel}$ window, we treat $A^{S}(\bm p^*+\bm q,\omega)$ as a function of $q_{\perp}$ only. The momentum integrals in Eq.~\eqref{eq:i_basic} can then be carried out along these two orthogonal directions separately,
\begin{align}
    \frac{dI}{dV} 
    &\approx \sum_{\bm p^*\in \mathcal{S}(k_F,\theta)} \frac{e^2N_f\Omega}{2\pi\hbar^3}|T(\bm p^*)|^2 \int dq_{\parallel} A^{T}(\bm p^* + q_{\parallel}\hat{\bm v}^T, 0)\notag\\
    &\qquad\qquad \times \int d q_{\perp}\ A^{S}(\bm p^*+ q_{\perp}\hat{z}\times\hat{\bm v}^{T}, -eV). \label{eq:dIdV_ii_broadening}
\end{align}
It is clear that the bias dependence of the tunneling conductance arises from the second integral, and is unaffected by the spectral function of the electrons in the tip. We model the tip spectral function as a Lorentzian, $A^T(\bm k, \omega) = \gamma_T/\pi[(\omega - \xi_{\bm k}^T)^2+\gamma_T^2]$, and introduce 
\begin{equation}\label{eq:G_p*}
    G_{\bm p^*} \equiv \frac{e^2}{h}\frac{N_f\Omega|T(\bm p^*)|^2}{\hbar^2|\bm v_{\bm p^*}^T\times \bm v_{\bm p^*}|}, 
\end{equation}
which represents the zero-bias tunneling conductance of the normal state in the sample contributed by $\bm p^*$. Equation~\eqref{eq:dIdV_ii_broadening} then reduces to
\begin{equation}
    \frac{dI}{dV}\approx \sum_{\bm p^*\in \mathcal{S}(k_F,\theta)}G_{\bm p^*} \text{Re}\Big[\frac{-eV + i\eta}{\sqrt{(-eV+i\eta)^2 - |\Delta_{\bm p^*}|^2}} \Big] ,\label{eq:dIdV_ii_broadening2}
\end{equation}
where 
the branch cut of the square root is chosen to be $[0,+\infty)$.
Equation~\eqref{eq:dIdV_ii_broadening2} reproduces the same singularity as the asymptotic formula, Eq.~\eqref{eq:dIdV_v*}, without broadening $(\eta = 0^+)$. 
In the presence of finite broadening, it resembles the Dynes formula for the tunneling density of states in superconductors \cite{dynes1978direct}, but crucially retains the momentum-resolved information about the gap function. 


%
%

%

\section{A toy model}\label{sec:toymodel}
\begin{figure}
    \centering
    \includegraphics[width=1\linewidth]{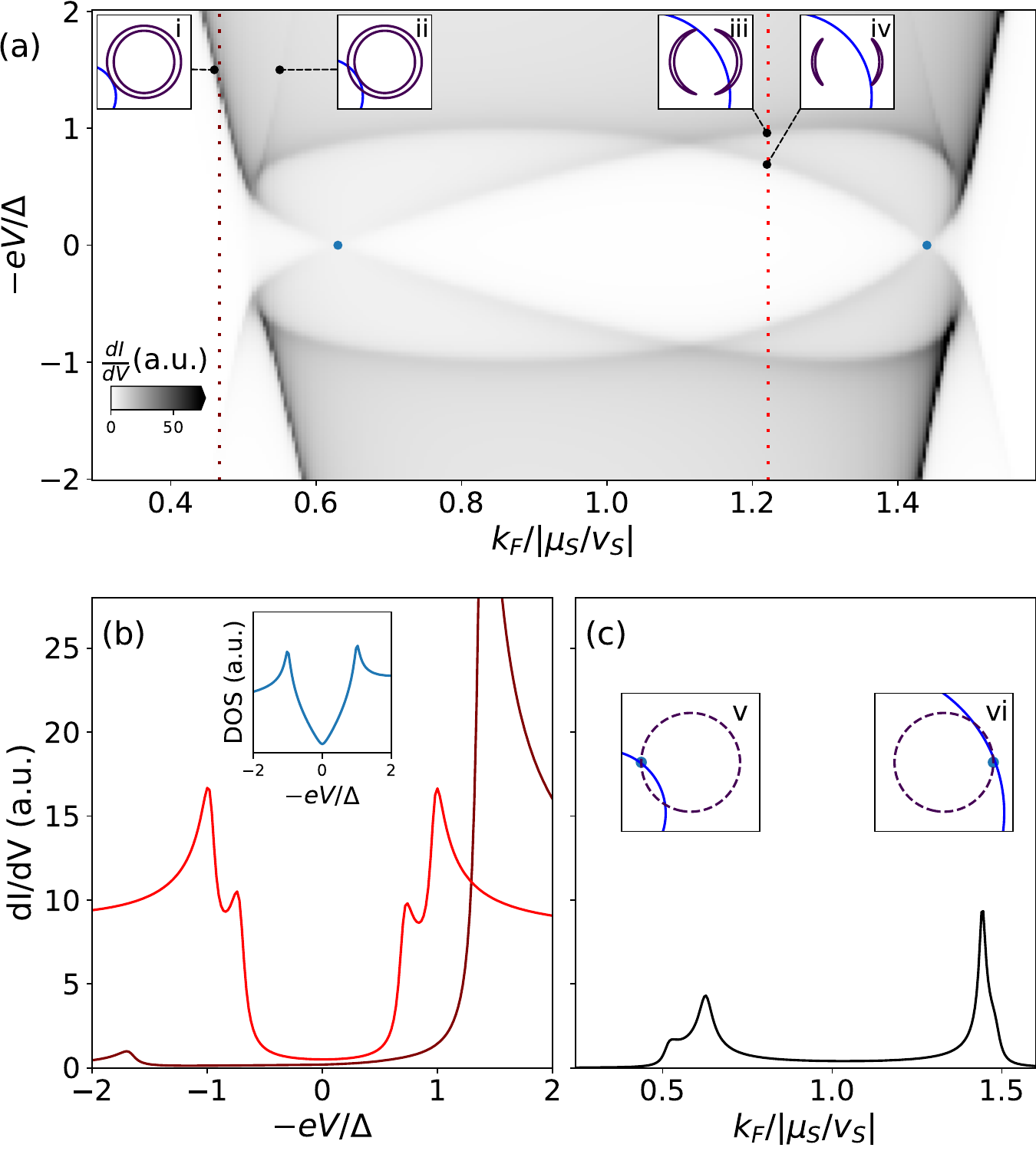}
    \caption{(a) Simulated $dI/dV$ map for a tunnel junction between a nodal superconductor and a monolayer graphene tip with varying bias voltage $V$ and Fermi wave vector $k_{F}$ of the tip at zero bias. Close to the Fermi energy $\mu_S$ of the sample, the gap function $\Delta_{\bm k} = \Delta\sin\theta_{\bm k}$ with $\mu_S/\Delta = 15$. The insets illustrate that $dI/dV$ exhibit peaks as the Fermi line of the tip (blue circle) touches the equal-energy contours of the Bogoliubov bands (black curves). (b) Vertical line cuts of panel (a) for different $k_{F}$ of the tip. The inset shows the lineshape of the tunneling density of states $\propto \sum_{\bm k}A^S(\bm k,-eV)$. (c) A horizontal line cut of panel (a) at zero bias. In the inset, the solid blue and dashed black lines represent the Fermi line of the tip and sample, respectively. They intersect at the nodes of the gap function marked by the dots, leading two peaks in zero-bias differential conductance. For these plots, 
    we use broadening parameters $\gamma_T=2\eta=0.1\Delta$.}
    \label{fig:didv_py}
\end{figure}

To illustrate our analysis in Sec.~\ref{sec:fs_singularity}, we consider the quasiparticle features for superconductors with anisotropic gap functions
within a toy model. We assume constant tunneling matrix elements, $T(\bm k)=w$, and isotropic normal-state dispersions in both sample and tip, $\xi_{\bm k}=\hbar v_S |\bm k| -\mu_{S}$ and $\xi_{\bm k}^{T}=\hbar v_F(|\bm k-\Delta \bm K| - k_F) + eV$, respectively. We account for the twist angle by displacing the center of the tip's Fermi circle 
relative to that of the sample by $\Delta \bm K$ in the direction of $(-\sqrt{3}/2,-1/2)^{\text{t}}$. For the sample, we choose the Fermi velocity $v_S=0.1v_F$, Fermi wave vector $\mu_S/v_S = |\Delta \bm K|/2$, and a $p-$wave nodal gap function $\Delta_{\bm k}=\Delta\sin{\theta_{\bm k}}$ near the Fermi level. We numerically compute $dI/dV$ in Eq.~\eqref{eq:didv_basic} as a function of the bias $-eV$ and the tip's Fermi wave vector $k_F$ at zero bias. The results are presented in Fig.~\ref{fig:didv_py}. 

Figure~\ref{fig:didv_py}(a) shows a gray region depicting a nonzero tunneling conductance facilitated by the intersections of  
the Fermi line of the tip with the Bogoliubov quasiparticle dispersion of the sample. 
We observe that the tunneling spectrum of the superconductor exhibits multiple lines of $dI/dV$ peaks. While the peak positions are approximately symmetric about zero bias, the intensities are in general asymmetric. The $dI/dV$ peaks arise from the tangency of the tip's Fermi line and the equal-energy contours of the superconductor, 
as shown in the insets. The lines of $dI/dV$ peaks in the $(k_F, V)$ plane map out the Bogoliubov dispersion along the trajectories of different points of tangency, while their intensities contain information on the coherence factors of the Bogliubov quasiparticles.

At large bias, $-eV\gtrsim\Delta$, we observe two branches of $dI/dV$ peaks on either side of the grey region, due to the tangency between the tip's Fermi line and two concentric equal-energy contours of the superconductor, see inset (i) and (ii). The $dI/dV$ peak associated with inset (ii) diminishes rapidly as the bias increases, as a result of the decreasing coherence factor. The $dI/dV$ singularities at large bias also exhibit markedly asymmetric intensities about zero bias because of different coherence factors of the upper and lower Bogoliubov quasiparticle dispersion. This is exemplified by the brown curve in Fig.~\ref{fig:didv_py}(b). 
The weaker peak is induced by superconductivity, while the stronger one persists even in the normal state.  

At small bias, $|eV|\leq \Delta$, the equal-energy contours of the $p$-wave superconductor split into two crescent-shaped pockets, as shown in the inset (iii) and (iv).  
The tip's Fermi line may become tangent to the top or bottom of the crescent-shaped equal-energy contours at wave vectors $\bm p_{1,2}^*$ at biases $-eV\approx \pm |\Delta_{\bm p_{1,2}^*}|$ (cf. Eq.~\eqref{eq:v*_largevp}). Thus, provided that the normal-state tunneling conductance is finite, anisotropic gaps generally give rise to multiple coherence peaks in $dI(V)/dV$. 
The $dI/dV$ spectrum along the red line cut is plotted as the red curve in Fig.~\ref{fig:didv_py}(b). 
It shows that the coherence peaks are approximately symmetric about zero bias, and are separated by a hard gap, 
consistent with Eq.~\eqref{eq:dIdV_ii_broadening2} but in contrast to the `$V$-shaped' tunneling density of states in the inset of Fig.~\ref{fig:didv_py}(b). Varying $k_F$ shifts $\bm p_{1,2}^*$ and yields the four branches of $dI/dV$ peaks in the center of the grey region in Fig.~\ref{fig:didv_py}(a), which scans the magnitude of gap function $|\Delta_{\bm p_{1,2}^*}|$ along the Fermi line of the sample.

In this way, the two nodes of the $p$-wave superconductor can be identified from the tunneling spectrum. As plotted in Fig.~\ref{fig:didv_py}(c), the zero-bias $dI/dV$ exhibits two peaks when the tip's Fermi line crosses the two nodes (see inset (v) and (vi)). 
According to Eqs.~\eqref{eq:G_p*} and~\eqref{eq:dIdV_ii_broadening2}, without quasiparticle relaxation, the zero-bias differential conductance is zero unless the tip’s Fermi line intersects a node, in which case its magnitude becomes finite and depends on the angle of intersection between the Fermi lines of the tip and sample. 

\section{Superconducting TBG}\label{sec:tbg}

In this section, we study the elastic momentum-conserving tunneling between MLG and the superconducting states in TBG. Due to significant uncertainties regarding the electronic structures of both the superconducting and the parent states in TBG, we are not aiming at quantitatively predicting its tunneling spectra. Instead, we demonstrate that
Eq.~\eqref{eq:dIdV_ii_broadening2} remains useful for extracting the momentum-resolved superconducting gap along the Fermi line of the sample, even if the sample has more complicated wave functions and electron spectra than the toy model of Sec.~\ref{sec:toymodel}.
Furthermore, we discuss how to identify the location of superconducting nodes in the BZ, without detailed knowledge of the fermiology in TBG.

For simplicity, we use the non-interacting Bistritzer-MacDonald model of $1.05^{\circ}-$TBG with Dirac velocity $v_F=10^6$~\si{\meter\per\second} in both graphene layers, and the intra-and inter-sublattice tunneling parameters $w_0=77$\si{meV} and $w_1=110$\si{meV}, respectively. 
The momentum separation between the Dirac points of the top and bottom layers in TBG is denoted as $k_M$.
We focus on filling factor $\nu=-3$, 
where TBG has a single $\bm\gamma-$centered Fermi pocket. 

Suppose that the tip is twisted clockwise relative to the top layer of TBG by a small angle $\theta$. The $K$-valley Dirac point of the tip is twisted to $\bm {K}_{\theta} = O(\theta) \bm {K_t}$, where $\bm {K_{t}}$ is the top-layer Dirac point of TBG and $O(\theta)$ denotes the rotation matrix. We define three reciprocal lattice vectors $\bm G_{1,2,3}$ of the top-layer graphene in TBG such that $\bm {K_t} + \bm G_{n}= O(2 \pi(n-1)/3)\bm K_{\bm t}$ correspond to three $K$-valley corners of the first BZ. They are rotated to $\bm G_n'= O(\theta)\bm G_{n}$ in the tip. 
In the band basis, the matrix elements of momentum-conserving tunneling between the single-particle states at wave vector $\bm k$ in the sample and $\bm k'$ in the tip 
read \cite{wei2025dirac}
\begin{align}\label{eq:h_tun}
    \langle \bm k' \lambda'\text{T} | \hat{H}_{\text{tun}} |\bm k\lambda\text{S}\rangle = \sum_{n=1}^{3} &\sum_{\bm g}T_{\lambda'\lambda}(\bm k+\bm g + \bm G_{n})\notag\\
    &\times\delta_{\bm k'+\bm G_n', \bm k+\bm g + \bm G_n },
\end{align}
with
\begin{align}\label{eq:matrix_elements}
        T_{\pm,\lambda}(\bm p+\bm G_n) = \frac{w}{\sqrt{2}}& \left[1\pm e^{\frac{2\pi i(n-1)}{3}-i\varphi_{\bm p+\bm G_n-\bm G_n'}} \right]\notag\\
        &\times \left[\psi_{\bm p tA}^{\lambda}+e^{-\frac{2\pi i(n-1)}{3}}\psi_{\bm p tB}^{\lambda}\right].
\end{align}
Here $\bm k$ and $\bm k'$ stay in the $K$-valley, are referenced with respect to the $\Gamma$ point of graphene, and $\bm k$ is restricted to one mBZ.
Moreover, $w$ represents the tunneling strength, $\varphi_{\bm k'}$ stands for the angle between $\bm k'-\bm K_{\theta}$ and $\bm K_{\theta}$, the location of the tip Dirac point in the reference frame of the TBG layer;
$\psi_{\bm k+\bm g l\sigma}^{\lambda}$ denotes the normalized reciprocal-space Bloch wave functions, where $\bm g$ sums over the moir\'e reciprocal lattice vectors of TBG, 
$l=t,b$ indicate the top and bottom layers of TBG, respectively, and $\sigma$ labels the sublattice.

The three Bragg scatterings within the first Brillouin zone of graphene ($n=1,2,3$) contribute equally to the tunneling current in a $C_{3z}$-invariant system. Spontaneous breaking of $C_{3z}$-symmetry can, in principle, triple the number of singularities in the tunneling spectrum \cite{wei2025dirac}. This effect should be considered in a QTM study of  graphene-based nodal superconductors, given that many theories predict \cite{wu2018theory,wu2019topological,kozii2019nematic,lake2022pairing,lothman2022nematic,yu2023euler,liu2024electron,wang2024molecular} coexistence of nodal pairing and nematicity in these superconductors. For instance, when the orbital part of the superconducting order parameter involves a single irreducible representation of the $D_6$ point group of MATBG, 
the Ginzburg-Landau theory \cite{kozii2019nematic,lake2022pairing} requires nematic gap functions to be nodal.

Due to the distinct matrix elements associated with the conduction and valence bands of the tip, probing with either band may highlight different features in the tunneling spectra. 
In the following, we assume the Fermi level of the tip is parked in the conduction band ($\lambda'=+$) at zero bias. 

\subsection{Normal state of TBG}
\begin{figure}[h!]
    \centering
    \includegraphics[width=1\linewidth]{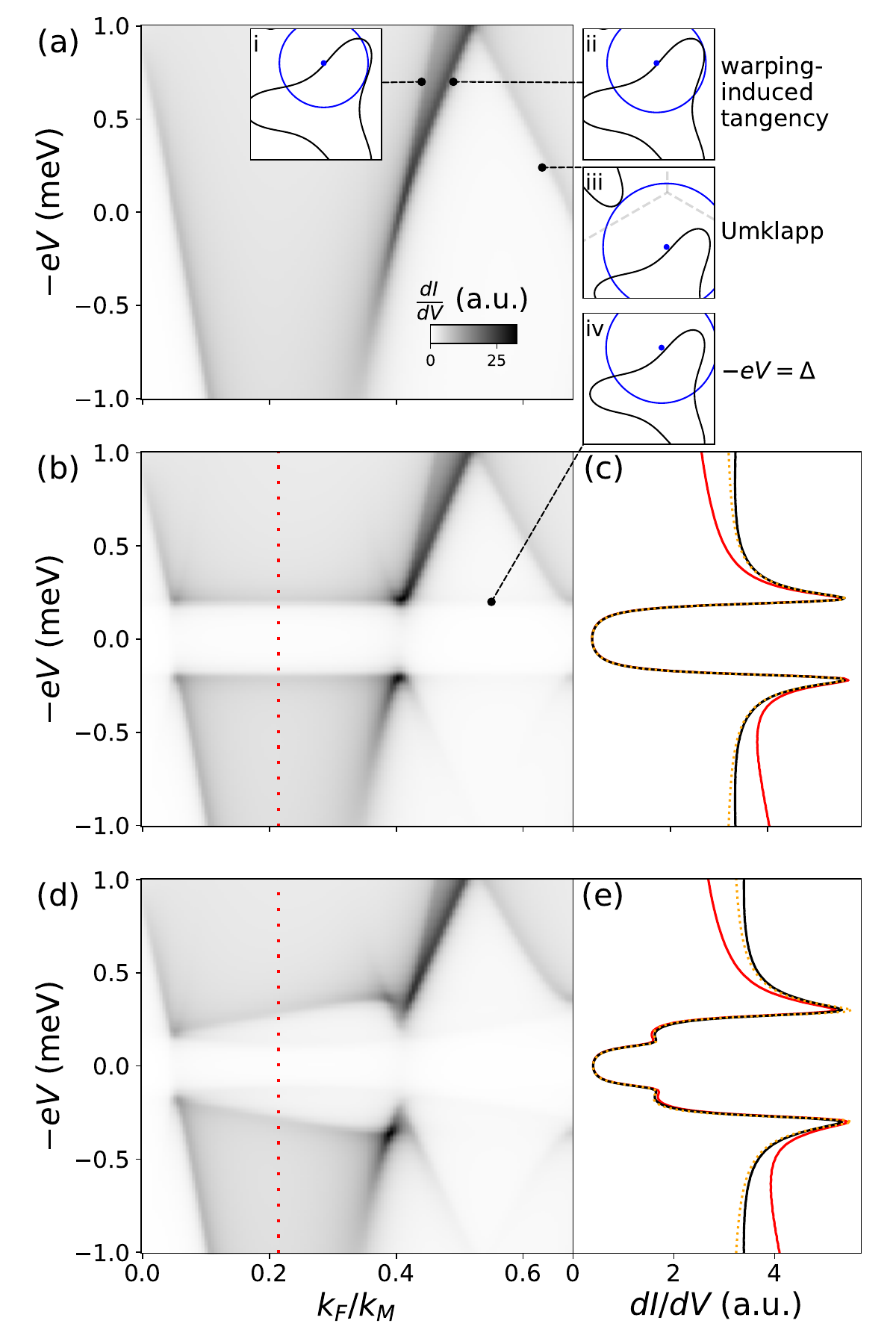}
    \caption{Simulated tunneling conductance between a MLG tip and a $1.05^{\circ}-$TBG in three different phases: normal state (a), an $s-$wave superconductor (b,c) with a constant gap $\Delta=0.2$\si{meV}, and a chiral superconductor (d,e) whose gap function is given by Eq.~\eqref{eq:delta_chiral} with $\Delta=0.2$~\si{meV}. The filling factor of TBG flat bands $\nu=-3$. The insets illustrate the tangency of the (black) equal-energy contours of TBG and the (blue) Fermi lines of the tip at several points in the $(k_F, -eV)$ space, see the black dots in panel (a) and (b). We plot grey dashed lines in one inset to indicate the boundary of mBZ. (c) The red line is a vertical line cut marked by the red dotted line in panel (b). The black line is calculated from Eq.~\eqref{eq:dIdV_ii_broadening2}. It fits well with the symmetrized data represented by the orange line. (e) The same as (c) except using the line cut from panel (d). For these plot, we use a non-interacting band structure of the TBG and a fixed position of the tip Dirac point at twist angle $\theta=0.57^{\circ}$ relative to the top layer of TBG, see the blue dots in the insets. The broadening parameters $\gamma_{T}=5\Delta$ for the tip and $\eta=\Delta/8$ for the sample.}
    \label{fig:didv_tbg_s}
\end{figure}

Most qualitative differences between the tunneling spectra of TBG and the toy model in Sec.~\ref{sec:toymodel} are not tied to superconductivity. Since these will not be the main focus of our discussion, we explain these features separately using the normal-state tunneling spectrum depicted in Fig.~\ref{fig:didv_tbg_s}(a). We fix the Dirac point of the tip at $\bm K_{\theta=0.57^{\circ}}$, whose position in the mBZ is marked by the blue dot in the insets. This position 
is close to but not directly
on the Fermi line of the sample. Therefore, for sufficiently small $|V|$ and $k_{F}$, the Fermi lines of the tip and sample do not intersect, and the differential conductance vanishes. Increasing $|V|$ or $k_{F}$ leads to an onset feature at a threshold bias voltage $V=V^*(k_F)$ when the Fermi line of the tip comes into contact with the equal-energy contours of the sample at energy $-eV^*$. As $k_{F}$ increases further, Fig.~\ref{fig:didv_tbg_s}(a) exhibits several novel features that are absent in the toy model in Sec.~\ref{sec:toymodel}.

First, at $k_{F}/k_M\gtrsim 0.4$ (this value is specific for the example we are considering), we find a narrow wedge-like region with an enhanced $dI/dV$. This region is confined from two sides by peaks in $dI/dV$ originating from the
tangency of the tip Fermi line and the trigonally warped energy bands of TBG, see the insets (i) and (ii) in Fig.~\ref{fig:didv_tbg_s}(a).  
As the energy decreases, the equal-energy contours of TBG shrink to the $\bm\gamma$ point and become more isotropic. Therefore, the wedge-like region terminates at a sufficiently negative bias in this figure.  

Second, the tunneling conductance in the area to the right of the wedge-like region is noticeably lower than in the area to the left, although the tip’s Fermi line continuously intersects the TBG bands. This contrast can be attributed to variations in the tunneling matrix elements as the intersection points shift along the Fermi lines.

Additionally, $dI/dV$ shows an enhancement in the top right corner of Fig.~\ref{fig:didv_tbg_s}(a). This arises from the onset of Umklapp scattering in TBG. As the tip Fermi line increases in size, it starts to intersect the energy contours in neighboring mBZs of the TBG, see the inset (iii).

\subsection{Nodeless gap function}\label{sec:tbg_nodeless}

We now consider tunneling into superconducting TBG assuming an isotropic gap $|\Delta_{\bm k}|=\Delta$. Figure~\ref{fig:didv_tbg_s}(b) shows both types of Fermi edge singularities discussed in Sec.~\ref{sec:toymodel}. For $|eV|\gtrsim\Delta$, $dI/dV$ exhibits an onset feature on the left of the plot, resembling the tunneling spectrum of the normal state. The characteristic bias $-eV^*(k_{F})(\gtrsim \Delta)$ for this $dI/dV$ singularity traces the momentum-resolved Bogoliubov quasiparticle dispersion along a line that crosses the Fermi line of the sample. In addition, $dI/dV$ shows a distinct threshold behavior at a constant bias $-eV=\pm\Delta$. It occurs when the tip's Fermi line intersects the edge of the Bogoliubov quasiparticle dispersions, see the inset (iv). Figure~\ref{fig:didv_tbg_s}(c) plots a line cut of $dI/dV$ along the red dashed line in panel (b). The differential conductance $dI/dV$ exhibits a noticeable particle-hole asymmetry due to the narrow band width and strong energy dependence of the density of states of the TBG flat bands. Nevertheless, the orange line obtained by symmetrizing the data about $V=0$ shows good agreement with the black dashed line calculated using  Eq.~\eqref{eq:dIdV_ii_broadening2}.
 
We have also studied a superconducting state with a $C_3$-invariant anisotropic gap function
\begin{align}
    &\Delta_{\bm k} = \Delta\sum_{j=1}^{3}e^{i\frac{2\pi j}{3}}\cos(\bm k\cdot \bm \delta_j), \label{eq:delta_chiral}
\end{align}
where $\bm\delta_1=-a_M\hat{y}=O(\frac{2\pi}{3})\bm\delta_2=O(\frac{4\pi}{3})\bm\delta_3$. This corresponds to a chiral superconductor with Cooper-pair angular momentum $l=1$. The resulting $dI/dV$ map plotted in Fig.~\ref{fig:didv_tbg_s}(d) exhibits a clear gap at zero bias since $\Delta_{\bm k}$ is fully gapped. The line cut in Fig.~\ref{fig:didv_tbg_s}(e) shows a two-gap features due to two intersection points $\bm p_{1,2}^{*}$ between the Fermi lines of the tip and sample, as discussed for the toy model in Sec.~\ref{sec:toymodel}. After symmetrization (orange curve), $dI/dV$ can again be well approximated by Eq.~\eqref{eq:dIdV_ii_broadening2} (black dotted line). 

\subsection{Nematic nodal gap function}\label{sec:nodal}
\begin{figure*}
    \centering
    \includegraphics[width=1\linewidth]{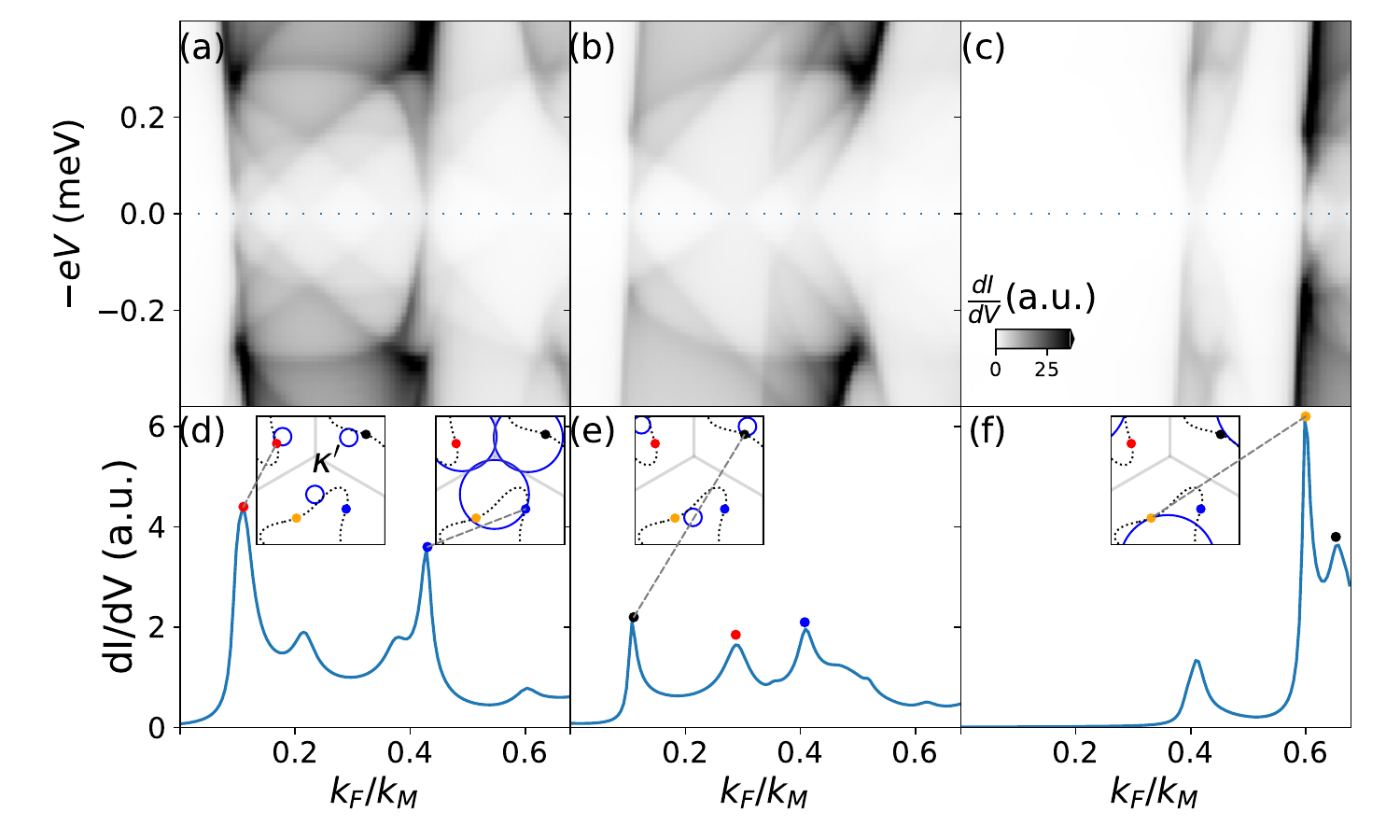}
    \caption{Simulated $dI/dV$ spectra of a nematic and nodal $d-$wave superconducting state in a $1.05^{\circ}-$TBG, 
    whose gap function is given by Eq.~\eqref{eq:delta_nodal} with $\Delta=0.3$~\si{meV}.
    $k_F$ denotes the Fermi wave vector of the tip at zero bias. The tip is twisted clockwise relative to the top layer of TBG by $\theta=0.5^{\circ}$(a), $0.8^{\circ}$(b), and $1.4^{\circ}$(c).  
    The tunneling gap closes at multiple values of $k_F$, which typically leads to nearly symmetric cone-shaped features around zero bias, with a few exceptions, e.g., $k_F\approx 0.42 k_M$ in (a) and $k_F\approx 0.6 k_M$ in (c), see Eq.~\eqref{eq:dIdV_v*_nodal} and the discussions below.
    (d-f) The zero-bias line cuts of (a-c). Peaks in the zero-bias $dI/dV$ are induced by momentum-conserving tunneling between the superconducting nodes and the Fermi circle of the MLG tip. The three Bragg scattering processes in the tunneling Hamiltonian couple the tip's Fermi circle to the wave vectors on the three circles in the mBZs of TBG, shown as the blue circles in the insets. These insets illustrate cases where the blue circles intersect one of the four inequivalent superconducting nodes, marked by dots in four distinct colors on the (dotted) Fermi line of TBG. The same colored dots are also used to label those pronounced $dI/dV$ peaks in panel (d-f), indicating their association with specific superconducting nodes. 
    In the inset, the grey solid lines mark the boundaries of the mBZs that cross at the top-layer Dirac point of TBG ($\bm\kappa'$). We used broadening parameter $\gamma_T=1$~\si{meV} and $\eta=0.015$~\si{meV}.}
    \label{fig:didv_tbg_nodal}
\end{figure*}
In this section, we discuss how to extract information about the superconducting nodes from the tunneling spectrum by considering the gap function
\begin{align}
    \Delta_{\bm k} = \Delta\sum_{j=1}^{3}\cos{\frac{2\pi j}{3}}\cos(\bm k\cdot \bm \delta_j). \label{eq:delta_nodal}
\end{align}
This gap function is the real part of the one in Eq.~\eqref{eq:delta_chiral} and generates a nematic nodal superconductor.

Figure~\ref{fig:didv_tbg_nodal}(a-c) plots $dI/dV$ as a function of the tip's Fermi wave vector $k_F$ at zero bias and as a function of the bias voltage at three different twist angles between tip and sample. The tunneling spectra become more involved than those in Sec.~\ref{sec:tbg_nodeless} because of the splitting of $dI/dV$ singularities induced by the $C_{3z}$ symmetry breaking. Although the $dI/dV$ singularity generally vanishes as the bias approaches zero (see Eq.~\eqref{eq:dIdV_v*}), we can still recognize the closing of the tunneling gap at multiple values of $k_F$, which is a signature of the nodal superconductor.

Figure~\ref{fig:didv_tbg_nodal}(d-f) plots the horizontal line cuts of panels (a-c) at zero bias. The insets illustrate that the $dI/dV$ peaks occur when the (blue) Fermi circle of the tip couples to one of the four superconducting nodes on the (dashed) Fermi line of the sample. Notice that wave vectors of the tip generally couple to three wave vectors in the sample via the three Bragg scatterings in the interlayer tunneling Hamiltonian \cite{wei2025dirac}, and these three wave vectors are related to each other by threefold rotation with respect to $\bm \kappa'$. Therefore, in each inset we drew three blue circles to indicate the states in the mBZ that couple to the tip Fermi circle.

More quantitative information on the superconducting nodes can be obtained from the $dI/dV$ spectra at low bias.
We neglect broadening and denote the characteristic bias voltages of the $dI/dV$ singularity as $V^*=V^*(k_F)$. 
Suppose that the tip's Fermi circle of radius $k_{F}=k_{F}^n$ passes through a superconducting node $\bm p_n$ on the Fermi line of TBG, i.e., $V^*(k_F^n)=0$ and $\bm p^*(k_F^n) =\bm p_n$. 
For a $k_F$ slightly deviating from $k_F^n$, the characteristic bias can be expressed in terms of the linearized normal-state dispersion $\xi_{\bm p^*}\approx \hbar\bm v_{\bm p_n}\cdot(\bm p^*-\bm p_n)$ and gap function $\Delta_{\bm p^*}\approx \hbar\bm v_{\Delta}\cdot(\bm p^*-\bm p_n)$,
\begin{equation}
    -eV^*(k_F) = \pm \hbar\sqrt{[\bm v_{\bm p_n}\cdot(\bm p^*-\bm p_n)]^2 + [\bm v_{\Delta}\cdot(\bm p^*-\bm p_n)]^2}.\notag
\end{equation}
By inserting the $k_F$-dependence of $\bm p^*$ as determined by the tangency condition Eq.~\eqref{eq:tangency_constraint} and the relation $k_{F} \approx k_F^n + \hat{\bm v}^T\cdot (\bm p^*-\bm p_n)$ ($\hat{\bm v}^T = \bm v_{\bm p_n}^T/|\bm v_{\bm p_n}^T|$), we find that the bias voltages $V^*$ vary linearly with $k_F$ for $k_F\approx k_F^n$ and the slopes obey
\begin{equation}
     \left|e\frac{d V^*(k_F)}{d k_F}\right|=\frac{\hbar|\bm v_{\bm p_n}\times \bm v_{\Delta}|}{\sqrt{|\hat{\bm v}^T \times \bm v_{\bm p_n}|^2 + |\hat{\bm v}^T \times \bm v_{\Delta}|^2}}. \label{eq:slope}
\end{equation}
This equation provides a quantitative description of the cone-shaped features in the $dI/dV$ map near zero bias, see Fig.~\ref{fig:didv_tbg_nodal}(a-c). 

The intensity of the $dI/dV$ singularities can be analyzed based on Eq.~\eqref{eq:dIdV_v*}. In the limit of $V/V^*\rightarrow 1^{+}$,
%
%
%
\begin{equation}
    \frac{dI}{dV}
    \propto \frac{|u_{\bm p^*}(V^*)|^2}
    {\sqrt{|\hat{\bm v}^T \times \bm v_{\bm p_n}|^2 + |\hat{\bm v}^T \times \bm v_{\Delta}|^2}}\sqrt{\frac{2V^*}{V -V^*}}. \label{eq:dIdV_v*_nodal}
\end{equation}
In the first term on the right-hand side, the denominator is proportional to $\beta_{\bm q^*}$ of Eq.~\eqref{eq:beta}, simplified by using the linearized dispersion and gap function. 
The numerator involves the coherence factor
\begin{equation}\label{eq:coherence}
   |u_{\bm p^*}(V^*)|^2 = \frac{1}{2} + \frac{s}{2}\frac{|\hat{\bm v}^{T}\times\bm v_{\Delta}|}{\sqrt{|\hat{\bm v}^T \times \bm v_{\bm p_n}|^2 + |\hat{\bm v}^T \times \bm v_{\Delta}|^2}}.
\end{equation}
Here, $s=\text{sgn}(\xi_{\bm p^*})\text{sgn}(-eV^*)\propto \text{sgn}(k_F-k_F^n)\text{sgn}(V^*)$. Typically, $|\bm v_{\bm p_n}|\gg |\bm v_{\Delta}|$, so that the second term in Eq.~\eqref{eq:coherence} is negligible and the intensity of the $dI/dV$ peaks should be approximately symmetric about $k_{F}^n$. 

Notice that some $dI/dV$ peaks near zero bias do not exhibit clear cone-shaped features in Fig.~\ref{fig:didv_tbg_nodal}(a)-(c) (e.g., around $k_{F}^n\approx0.42k_M$). This occurs because the Fermi lines of tip and sample become approximately tangent at the node [e.g., the blue dot in the right inset of panel (d)]. When $\hat{\bm v}^{T}\parallel \bm v_{\bm p_n}$, the coherence factor in Eq.~\eqref{eq:coherence} jumps between $0$ and $1$ as $k_{F}$ changes across $k_{F}^n$, leading to a pronounced asymmetry in $dI/dV$ on the two sides of $k_{F}^n$. At the same time, the slope of the characteristic bias voltage $V^*(k_F)$ and the intensity of $dI/dV$ peaks both increase prominently, according to Eqs.~\eqref{eq:slope} and~\eqref{eq:dIdV_v*_nodal}. For comparison, the peak at $\tilde{k}_F^n\approx 0.41k_M$ in panel (e) is associated with the same superconducting node, but the Fermi lines of the tip and sample are not in tangency. This leads to a weaker, flatter, and more symmetric cone-shaped feature in $dI/dV$ near zero bias in Fig.~\ref{fig:didv_tbg_nodal}(b).

\begin{figure}
    \centering
    \includegraphics[width=1\linewidth]{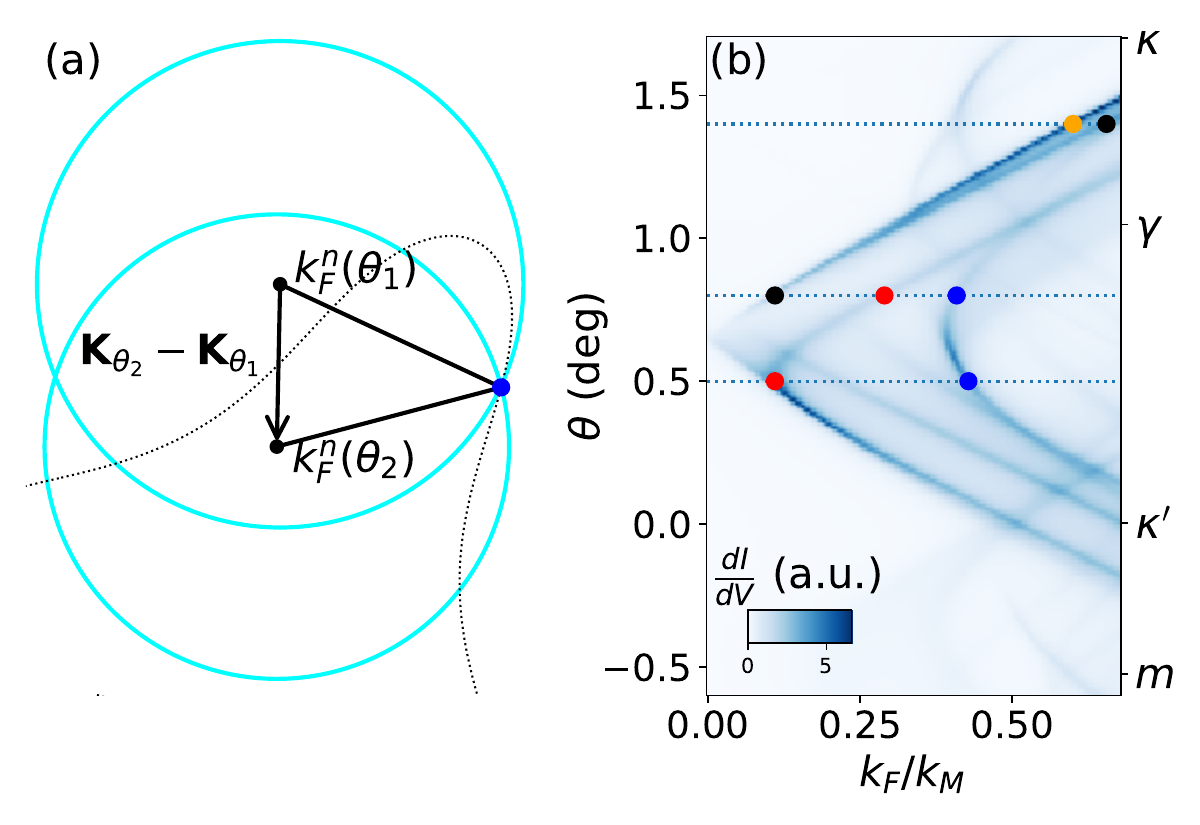}
    \caption{Locating the superconducting nodes in the mBZ using the tunneling spectrum. (a) The black dashed line delineates the Fermi line of the TBG and the blue dots mark the node of the gap function Eq.~\eqref{eq:delta_nodal}. The black dots represent the Dirac point $\bm K_{\theta}$ of the tip at different twist angles $\theta=\theta_{1,2}$. They are connected by an arrow indicating the direction of increasing $\theta$. The blue circles are tip Fermi circles of radius $k_F^n(\theta_{1,2})$, respectively. (b) $dI/dV$ vs. the Fermi wave vector $k_F$ of the tip and the twist angle between the tip and the top layer graphene of the TBG. The three horizontal line cuts correspond to Figure~\ref{fig:didv_tbg_nodal}(d-f). Each superconducting node may generate one or multiple lines of $dI/dV$ peaks. The $k_{F}^n(\theta_{1,2})$ in panel (a) denote the $x$-coordinates of the two blue dots in this map. The intersections of the two circles in panel (a), inferred from these two data points, provides an estimate of the location of the associated superconducting node. }
    \label{fig:nodes}
\end{figure}

In the following, we discuss how to locate the superconducting nodes in the mBZ from the tunneling spectrum, 
even without \textit{a priori} knowledge of the sample's fermiology.
The idea is sketched in Fig.~\ref{fig:nodes}(a). Suppose that a variation of the QTM twist angle
from $\theta_1$ to $\theta_2$ shifts the zero-bias $dI/dV$ peak from $k_F=k_F^{n}(\theta_1)$ to $k_F^{n}(\theta_2)$ in the $dI/dV$ map. A superconducting node $\bm p_n$ should then lie at the intersections between the tip's Fermi circle of radius $k_F^n(\theta_1)$ before the twist and that of radius $k_F^n(\theta_2)$ after the twist. To confirm this expectation, we use the two blue dots in Fig.~\ref{fig:nodes}(b) as the data points and draw the corresponding Fermi circles in panel (a). Apparently, the node marked by the blue dot in panel (a) coincides with one intersection point of two circles. Because the center $\bm K_{\theta}$ of the tip's Fermi circle is precisely determined from the twist angle, the superconducting node $\bm p_n$ can be expressed in terms of measurable parameters by analyzing the triangle in Fig.~\ref{fig:nodes}(a), 
\begin{equation}\label{eq:pn}
    \bm p_n = \bm K_{\theta_1} + k_F^n(\theta_1)(\cos\alpha\ \hat{n} \pm \sin\alpha\ \hat{z}\times\hat{n}),
\end{equation}
with $\hat{n}=(\bm K_{\theta_2}-\bm K_{\theta_1})/|\bm K_{\theta_2}-\bm K_{\theta_1}|$ and
\begin{equation}\label{eq:alpha}
    \alpha=\arccos\frac{|\bm K_{\theta_2}-\bm K_{\theta_1}|^2+k_F^n(\theta_1)^2-k_F^n(\theta_2)^2}{2k_F^n(\theta_1)|\bm K_{\theta_2}-\bm K_{\theta_1}|}.
\end{equation}
The `$\pm$' signs in Eq.~\eqref{eq:pn} reflect a limitation of this ``triangulation'' method. The fact that two circles may intersect at two points introduces an inherent ambiguity in associating a given intersection with the true superconducting node. It is possible to eliminate this ambiguity by exploiting the differences in tunneling matrix elements at these two points and the resulting effects on the intensity of the $dI/dV$ peaks (see Appendix~\ref{sec:extract_gap}). Another way to resolve the ambiguity is by using a tip with stronger trigonal warping at the Fermi level, e.g., bilayer graphene. An additional limitation of the QTM arises if the system lacks $C_{3z}$ symmetry. In these cases, rotating the sample’s band structure by $2\pi/3$ yields an identical tunneling spectrum, meaning the node locations can only be determined up to a three-fold rotational ambiguity. Taken together, using only $dI/dV$ as a function of the twist angle $\theta$ and the Fermi wave vector $k_F$, the location of a superconducting node can be constrained to at most $2\times 3=6$ possible points within the mBZ.

Notice that there are multiple branches of $dI/dV$ peaks in Fig.~\ref{fig:nodes}(b), each associated with a superconducting node. By tracing these lines of peaks, one can constrain the locations of all nodes in the mBZ.

Finally, the ``triangulation" method also applies to the $dI(\theta, k_F)/dV$ spectrum at finite bias $-eV = |\Delta_{\bm p^*}|$ for a generic point $\bm p^*$ on the Fermi line of the sample $(\xi_{\bm k}=0)$, provided that the gap function varies slowly with momentum (cf.\ Eq.~\eqref{eq:assumption}), see Appendix~\ref{sec:extract_gap} for details. In this case, it enables us to locate $\bm p^*$ in the BZ and thus map out the magnitude of the gap function along the $\xi_{\bm k}=0$ contour.




\section{Conclusion and discussion}\label{sec:conclusion}
In this work, we analyzed elastic momentum-conserving tunneling across a normal-superconductor junction in the quantum twisting microscope (QTM) as a function of bias voltage, Fermi wave vector of the tip, and twist angle between tip and sample. We find that $dI(V)/dV$ generally exhibits inverse square-root singularities at bias voltages $-eV^*$ at which the tip's Fermi line becomes tangent to the probed quasiparticle dispersions. In the cases that the Fermi lines of the tip and sample cross, this bias directly measures the momentum-resolved superconducting gap(s) at the crossing point(s), see Fig.~\ref{fig:schematic}(b). Using a non-interacting continuum model of twisted bilayer graphene (TBG) as an example, we show that the simultaneous tunability of the tip's doping level and the twist angle enables extraction of the momentum-resolved superconducting gap along the Fermi line of the sample ($\xi_{\bm k}=0$). In particular, we propose a ``triangulation" method to locate the nodes in the Brillouin zone of a nodal superconductor. Our study shows that the QTM can probe the electronic structures of the sample not only along the line traversed by the tip's Dirac point as a function of twist angle as demonstrated by earlier experiments, but also over an extended area of the Brillouin zone.

To probe the gap along the entire Fermi line of the sample, the tip's Fermi wave vector $k_F$ should roughly exceed the sample's. However, in practice, the tunability of $k_F$ via gating is constrained by electrostatics. When both top and back gates are used as independent controls of the filling of the probed band and the tip, the out-of-plane displacement field $D$ acting on the sample can no longer be adjusted freely. Specifically, changing the tip’s chemical potential by $\delta\mu_T$, at fixed bias and electron density in the sample, modifies $D$ by $\delta D = \epsilon_{\perp}\delta\mu_T /e d$. For a tunneling barrier consisting of three layers of hexagonal boron nitride, with thickness $d=1.2$~\si{nm} \cite{sengottaiyan2024large} and dielectric constant $\epsilon_{\perp}\approx 3.3$, it requires $\delta D\sim 0.4$ \si{\volt\per nm} to induce a change in the tip's Fermi wave vector of $\delta k_F=\delta\mu_T/\hbar v_F\sim 0.22$~\si{\per\nano\meter} ($0.7k_M$ for a $1.05^{\circ}-$TBG). In TBG, the superconducting phase is robust against the displacement field \cite{yankowitz2019tuning}, so that the displacement field is not included in our simulations. However, probing superconducting phases in rhombohedral graphene which are confined to a narrower range of displacement field \cite{zhou2021superconductivity,zhou2022isospin,zhang2023enhanced,zhang2025twist,yang2025impact,han2025chiral} would require a proper device modification to independently control the doping level of the tip and the displacement field.

We caution that for TBG at the magic angle, the nearly vanishing width of the flat-band manifolds might challenge our assumption of intraband pairing and weak momentum dependence of the gap function compared to the band dispersion (Eq.~\eqref{eq:assumption}). Indeed, band-off-diagonal pairing has been proposed in TBG \cite{christos2023nodal,Putzer2025eliashberg,liu2024electron,wang2024molecular,chou2024topological,yu2023euler}. Such superconductors can have exotic energy spectra, such as nodes away from the Fermi line ($\xi_{\bm k}=0$). For them, Eq.~\eqref{eq:dIdV_ii_broadening2} no longer applies, and the bias for the $dI/dV$ singularities might not equal the superconducting gap, even in the low bias regime. However, our method to locate the superconducting nodes remains valid. Furthermore, in Appendix~\ref{sec:twist}, we discuss a generalization of the `triangulation' method that enables band-structure tomography in mBZ without any assumptions on the gap functions.


Our analysis relies on BCS mean-field theory. Strong-coupling superconductors can have frequency-dependent gap functions. The tunneling conductance formula, Eq.~\eqref{eq:dIdV_ii_broadening2}, can be generalized accordingly, see Appendix~\ref{sec:strong-coupling}. Within Eliashberg theory, 
there are no qualitative changes in probing the gap structures. However, the tunneling spectra now contain additional features \cite{schrieffer1963effective,scalapino1966strong} arising from the inelastic scattering of Cooper pairs with the pairing glue such as phonons. The strength of these inelastic features increases with the coupling to the pairing glue. Measuring these inelastic features could therefore shed light on the pairing mechanisms.

In conclusion, this work establishes a theoretical framework for analyzing the tunneling spectra between two-dimensional normal metals and superconductors, with potential implications to future QTM studies of superconductivity in graphene-based heterostructures and twisted transition metal dichalcogenides \cite{xia2025superconductivity,guo2025superconductivity}. 
An interesting topic for future study could be momentum-resolved tunneling spectroscopy of finite-momentum pairing superconductors, as proposed for many flat-band systems \cite{yang2024topological,qin2024chiral,christos2025finite,gaggioli2025spontaneous,gil2025charge,sedov2025probing,chen2025finite}.
\\

\textit{Note added} - We became aware of a related work in advanced stages of preparation \footnote{Y. Waschitz, A. Stern, and Y. Oreg, in preparation.}.

\begin{acknowledgments}
    We thank John Birkbeck, Dmitri Efetov, Shahal Ilani, and Yuval Oreg for helpful discussions. N.W.\ acknowledges support through the Yale Prize Postdoctoral Fellowship in Condensed Matter Theory. Research at Yale was supported by NSF Grant No.\ DMR-2410182 and by the Office of Naval Research (ONR) under Award No.\ N00014-22-1-2764. Research at Freie Universit\"at Berlin and Yale was supported by Deutsche Forschungsgemeinschaft through CRC 183 (project C02 and a Mercator Fellowship).   Research at Freie Universit\"{a}t Berlin was further supported by Deutsche Forschungsgemeinschaft through a joint ANR-DFG project (TWISTGRAPH).
\end{acknowledgments}

\newpage
\appendix


\section{Extracting momentum-resolved gap function on the Fermi line of the sample}\label{sec:extract_gap}
\begin{figure*}
    \centering
    \includegraphics[width=1\linewidth]{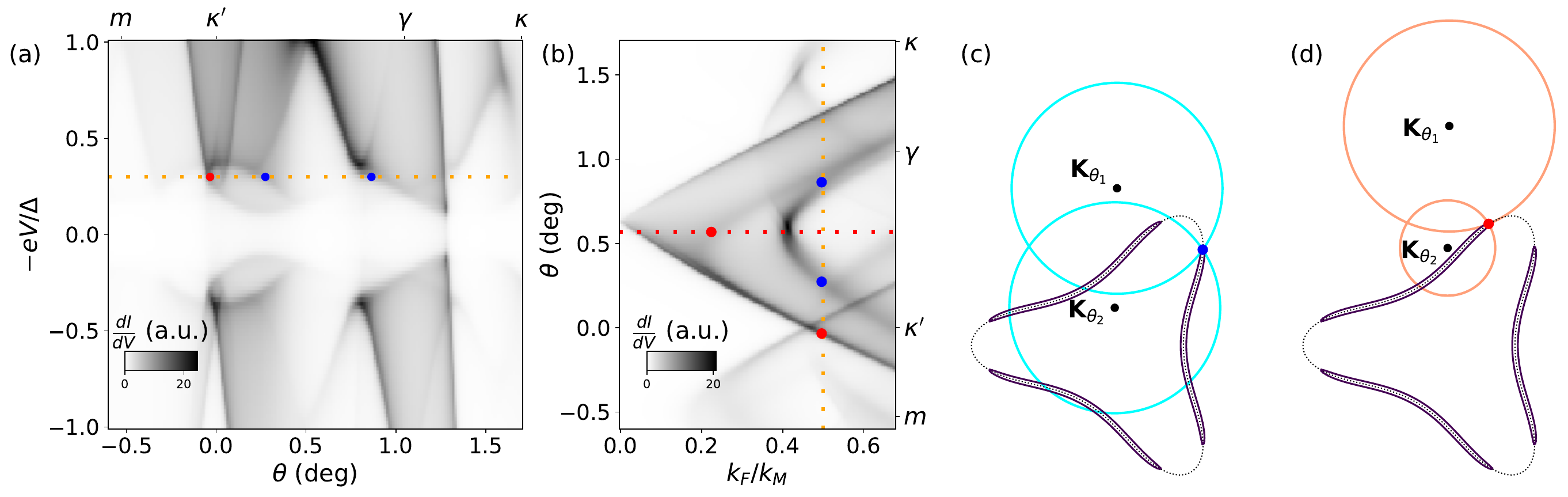}
    \caption{Simulated $dI/dV$ spectra for a nodeless superconducting state in a $1.05^{\circ}-$TBG. The gap function and model parameters are the same as those used in Fig.~\ref{fig:didv_tbg_s}(d). (a) $dI/dV$ vs. the twist angle $\theta$ and bias $-eV$ between the tip and sample. The Fermi level of the tip is parked in the conduction band of MLG with $k_F=0.5k_M$. (b) $dI/dV$ vs. the Fermi wave vector $k_F$ of the tip (at zero bias) and the twist angle $\theta$, with bias fixed at $-eV=0.3$~\si{meV}. The horizontal red line indicates a line cut at $\theta=0.57^{\circ}$. The vertical orange line represents a line cut at $k_F=0.5k_M$, which corresponds to the horizontal line cut in (a). (c) The two blue dots connected by the line $(k_F(\theta), \theta)$ of $dI/dV$ peaks in (b) are used to draw the corresponding Fermi circles (blue) centered at $\bm K_{\theta}$ with radius $k_{F}(\theta)$. Both Fermi circles are tangent to the equal-energy contours $\epsilon_{\bm k}^S=0.3$~\si{meV} (black solid) of the superconductor, and approximately share a common point of tangency $\bm p^*$, marked by the blue point.
    (d) Same as (c), but using the red dots in (b).
    }
    \label{fig:chiral_sc}
\end{figure*}

In this section, we demonstrate that by applying the ``triangulation" method introduced in Sec.~\ref{sec:tbg_nodeless} to the $dI(k_F, \theta)/dV$ map at finite bias voltage $V$, one can locate the wave vector(s) $\bm p^*$ on the Fermi line of the sample that satisfy $|\Delta_{\bm p^*}|=|eV|$. Thus, the magnitude of the momentum-resolved gap function can be measured along the Fermi line of the sample. We restrict the following analysis to a $C_{3z}-$invariant superconductor with an anisotropic gap function to avoid complications in the QTM tunneling spectra caused by $C_{3z}$ symmetry breaking.

First, we recall that when probing a superconducting state, there are two distinct configurations of tangency between the tip Fermi line and Bogoliubov quasiparticle dispersion that are shown in Fig.~\ref{fig:schematic}(a) and (b), respectively. To measure the superconducting gap along the Fermi line of the sample, we should focus on the $dI/dV$ singularities associated with the latter.
As discussed in Sec.~\ref{sec:fs_singularity}, these two types of configurations can be distinguished from tunneling spectra $dI(k_F, V)/dV$. Here, we further point out that they can also be differentiated in the $dI(\theta, V)/dV$ map:
Fig.~\ref{fig:chiral_sc}(a) depicts the $dI/dV$ map as a function of the bias and twist angle $\theta$ between a MLG tip and a gapped superconductor, whose gap function is given by Eq.~\eqref{eq:delta_chiral} with $\Delta=0.2$~\si{meV}. 
We identify two types of $dI/dV$ peaks at finite biases $-eV^*(\theta)$, associated with the tangency configurations in Fig.~\ref{fig:schematic}(a) and (b), respectively: 
\begin{enumerate}
\setlength{\leftmargini}{0pt}
\renewcommand{\labelenumi}{(\Roman{enumi})}
    \item $|eV^*(\theta)|$ changes rapidly with $\theta$ and extends to values much larger than the superconducting gap;  
    \item $|eV^*(\theta)|\lesssim\Delta$ and varies slowly with respect to $\theta$.
\end{enumerate}
The red dot is located in a crossover region between the type-(I) and type-(II) singularities, where the intensity of $dI/dV$ is enhanced (see Sec.~\ref{sec:fs_singularity}).

To decode the momentum-resolved gap structure information from the tunneling conductance, we turn to $dI(k_F, \theta)/dV$ at a fixed bias, such as Fig.~\ref{fig:chiral_sc}(b) for $-eV=0.3$~\si{meV}. In this plot, the $dI/dV$ peaks trace multiple lines. The one that connects the two blue dots is denoted as $k_{F}= k_F(\theta)$. It arises from the tangency between the tip's Fermi circle, $|\bm k- \bm K_{\theta}|=k_F(\theta)$, and the equal-energy contour $\epsilon_{\bm k}^S=-eV$ of the sample. We find that the point of tangency $\bm p^*$ is almost constant within a range of $\theta$ and lies approximately on the Fermi lines of the sample ($\xi_{\bm p^*}=0$), see Fig.~\ref{fig:chiral_sc}(c-d) for illustration. Therefore, we can use the ``triangulation" method to locate $\bm p^*$ and obtain the superconducting gap at $\bm p^*$ ($|\Delta_{\bm p^*}|=|eV|$). The discussion of constraining the node positions in the main text represents the special case of $V=0$.

One concern with the ``triangulation" method is its two-fold ambiguity: the true point of tangency $\bm p^*$ cannot be uniquely identified from the two intersections of the Fermi circles in Fig.~\ref{fig:chiral_sc}(c) and (d). Nevertheless, one could use the tunneling matrix elements to distinguish between these two intersections. First, we note that the lines of $dI/dV$ peaks in Fig.~\ref{fig:chiral_sc}(b) are approximately hyperbolic at small twist angles. At the vertex of each hyperbola, $k_F(\theta)$ reaches a minimum and $\bm p^*-\bm K_\theta$ aligns either parallel or antiparallel to $\bm K_{\theta}$. The antiparallel alignment, however, leads to vanishing tunneling matrix elements ($\varphi_{\bm p^*}=\pi$ with $n=1$ in Eq.~\eqref{eq:matrix_elements}), causing the intensity of the hyperbola to diminish near the vertex. In Fig.~\ref{fig:chiral_sc}(b), the hyperbola connecting the two blue dots exhibits strong intensity near the vertex (at $k_F\sim 0.4k_M$), indicating that in Fig.~\ref{fig:chiral_sc}(c), the intersection on the right-hand side of $\bm K_{\theta_{1,2}}$ corresponds to the true point of tangency $\bm p^*$.

We reiterate that to extract the gap function along the sample’s Fermi line by the above method, we should use exclusively the type-(II) singularities in the $dI(k_F,\theta)/dV$ spectrum. However, the same spectrum may also exhibit type-(I) singularities - for example, the cone-shaped feature near $k_F\approx 0$ in Fig.~\ref{fig:chiral_sc}(b) - which contain additional information about the quasiparticle spectra. Moreover, the distinction between type-(I) and type-(II) singularities becomes subtle when the assumption of weak momentum dependence of the superconducting gap, Eq.~\eqref{eq:assumption}, is violated. These limitations of the ``triangulation" method motivate a more flexible framework for extracting band structure information from $dI/dV$, which will be presented in Appendix~\ref{sec:twist}.


\section{Band structure tomography}\label{sec:twist}
\begin{figure*}
    \centering
    \includegraphics[width=1\linewidth]{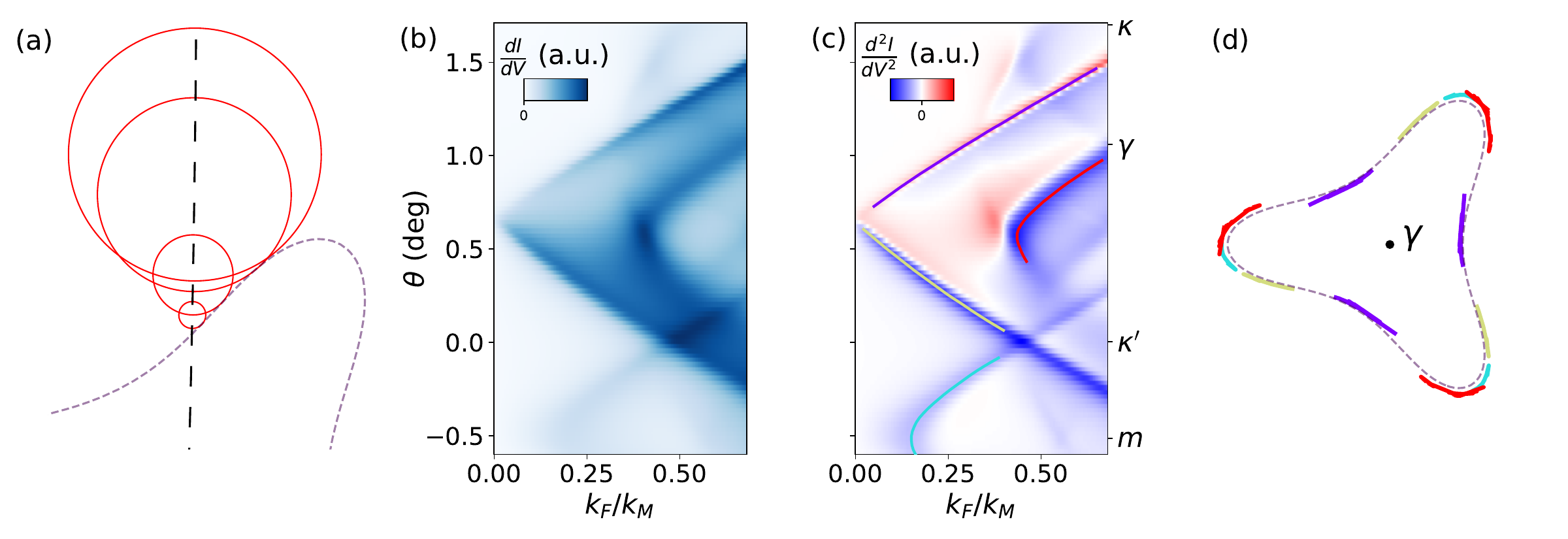}
    \caption{Reconstructing the Fermi line of the TBG normal state from the zero-bias tunneling conductance between the TBG and a MLG tip. (a) The solid red lines depict tip Fermi circles of varying radii and twist angles relative to the sample, each chosen to be tangent to the Fermi line of the sample (black dashed). The envelope of these Fermi circles matches the Fermi line of the sample. (b) Simulated $dI/dV$ map vs. the tip Fermi wave vector $k_F$ and the twist angle $\theta$ between the tip and the top layer of TBG. A $1.05^{\circ}-$TBG at filling $\nu=-3$ is considered. (c) $d^2I/dV^2$ map. Several local extrema $k_{F}(\theta)$ of $d^2I(k_F,\theta)/dV^2$ are marked by lines.  (d) Colored lines reconstructed from the lines of the same color in panel (c) using Eq.~\eqref{eq:reconstruction}. They match well with the Fermi line of the sample (black dashed). For these plots, we used $\gamma_{T}=1$~\si{meV} and $\eta=0.5$~\si{meV}.}
    \label{fig:fs_reconstruction}
\end{figure*}
%

In this section, we propose a method to reconstruct (segments of) the equal-energy contours $\epsilon_{\bm k}=-eV$ of a generic quasiparticle dispersion in the sample from the $dI(k_F, \theta)/dV$ spectrum at bias $-eV$, by generalizing the ``triangulation" method in Sec.~\ref{sec:nodal} of the main text and Appendix~\ref{sec:extract_gap}. 

Consider a map of $dI/dV$ collected at a fixed bias voltage and charge density of the sample while tuning the twist angle $\theta$ and the tip Fermi wave vector $k_F$ (at zero bias). 
Each line of $dI/dV$ singularity in the $(k_F, \theta)$ plane corresponds to a family of the tip's Fermi lines whose centers and sizes vary in a coordinated manner such that all Fermi lines stay tangent to the probed band at the energy level $\epsilon_{\bm p^*}^S=-eV$. Therefore, as shown in Fig.~\ref{fig:fs_reconstruction}, we can map out equal-energy contours of the sample by drawing the envelope of this family of Fermi lines.   

For a MLG tip, we can formulate the above idea as follows: Suppose that along a line of $dI/dV$ singularities in the $(k_F, \theta)$ plane, the Fermi wave vector of the tip at bias voltage $V$ is described by a function $k_F(\theta, V)$. The corresponding point of tangency $\bm p^*(\theta, V)$ moves along an equal-energy contour of the probed band, i.e., $\bm{v}_{\bm p^*}^{S}\cdot \partial_{\theta}\bm p^*(\theta, V)=0$. Therefore, 
\begin{equation}\label{eq:constraint}
    \bm v_{\bm p^*}^{T}\cdot \frac{\partial\bm p^*}{\partial\theta}=0.
\end{equation}
Due to the isotropic band structure of the tip, $\bm v_{\bm p^*}^{T} \propto \bm p^* - \bm K_{\theta}$ for $\bm p^*$ close to the $\bm K-$valley Dirac point $\bm K_{\theta}=|\bm K|(\cos\bar{\theta}, -\sin\bar{\theta})^{\text{t}}$. We define $\theta$ as the twist angle measured clockwise relative to the top-layer Dirac point of the TBG sample, and $\bar{\theta}=\theta+\theta_{\text{TBG}/2}$ as the angle relative to the $x-$axis, which lies midway between the top- and bottom-layer Dirac points of TBG. Solving the constraint Eq.~\eqref{eq:constraint}, we obtain the following expressions,
\begin{align}
    &\bm p^*(\theta, V) = \bm K_{\theta} - k_F(\theta,V)
    \begin{bmatrix}
    \sin (\bar{\theta}+\alpha(\theta,V)) \\
    \cos(\bar{\theta}+\alpha(\theta,V))
    \end{bmatrix}, \label{eq:reconstruction}\\
    &\cos\alpha(\theta,V) = -\frac{1}{|\bm K|}\frac{\partial k_F(\theta,V)}{\partial\theta}. \label{eq:cos_alpha}
\end{align}
Notice that Eq.~\eqref{eq:cos_alpha} admits two solutions for $\alpha\in [0,2 \pi)$, implying that $\bm p^*$ can only be determined up to a two-fold ambiguity by this method. 

Equations~\eqref{eq:reconstruction} and~\eqref{eq:cos_alpha} form the basis for reconstructing the equal-energy contours of the sample from the $dI/dV$ spectrum as a function of the twist angle and tip's Fermi wave vector $k_F$. One special application is to determine the position of the superconducting nodes in a nodal superconductor. In this scenario, Eqs.~\eqref{eq:reconstruction} and~\eqref{eq:cos_alpha} represent the limiting case of Eqs.~\eqref{eq:pn} and~\eqref{eq:alpha} as $\theta_{2}\rightarrow\theta_1$.

Another application is to measure the Fermi line of the normal state in the sample. Figure~\ref{fig:didv_tbg_s}(b) plots a simulated zero-temperature zero-bias $dI/dV$ spectrum for a normal phase of the $1.05^{\circ}-$TBG at filling factor $\nu=-3$.
The broadening of $dI/dV$ singularities due to finite quasiparticle relaxation introduces uncertainties in identifying their positions in the $(k_F,\theta)$ plane. To sharpen the singularities, we plot $d^2I/dV^2$ in panel (c). At fixed $\theta$, each peak in $dI/dV$ as a function of $k_F$ gives rise to a pair of local maximum and minimum in $d^2I/dV^2$. These two local extrema have different intensities, and the stronger one is taken as the position of the corresponding $dI/dV$ singularity.
The highlighted lines in (c) represent some of the most prominent $dI/dV$ singularities. 
By analyzing these lines with Eq.~\eqref{eq:alpha}, we obtain the trajectories of $\bm p^*$ and plot them (and their images under $C_{3z}$ rotations) in the mBZ in panel (d). We find that they match the dashed Fermi line of the tip reasonably well. The minor deviations can be attributed to the finite quasiparticle relaxation. Note that we have discarded the other half of the solutions to Eq.~\eqref{eq:reconstruction}, which would generate a mirror image of the correct Fermi line about $k_x=0$ (the vertical line across the $\bm\gamma$ point) in mBZ.

Similar to the ``triangluation" method, the band-structure tomography method described above is most effective if the system preserves $C_{3z}$ symmetry. Breaking $C_{3z}$ symmetry introduces additional three-fold ambiguity in identifying the point of tangency $\bm p^*$ from a $dI/dV$ singularity. This challenge is common in QTM-based band structure measurements, including the Dirac point spectroscopy of the QTM \cite{wei2025dirac}. Resolving these ambiguities (such as by applying in-plane magnetic field, strain, or current) could be an interesting topic for future investigation.



\section{Strong-coupling superconductors}\label{sec:strong-coupling}
In this section, we derive a zero-temperature tunneling conductance formula for a strong-coupling superconductor, where Cooper pairs and quasiparticles acquire finite lifetimes and the gap function becomes a frequency-dependent complex quantity. We assume that the superconducting gap is much smaller than the Fermi energy $\mu_S$, as in the derivation of Eq.~\eqref{eq:dIdV_ii_broadening2} in the main text.

We define the retarded Green's function in the Nambu space of a single-band superconductor, 
\begin{widetext}
\begin{equation}
    \begin{pmatrix}
    G_{\bm k\sigma; t}^R & F_{\bm k\sigma; t}^R\\
    F_{-\bm k\bar{\sigma}; t}^R & G_{-\bm k\bar{\sigma}; t}^R\\ 
    \end{pmatrix}
    \equiv 
    -i\Theta(t)
    \begin{pmatrix}
    \langle \{c_{\bm k \sigma}(t), c_{\bm k\sigma}^{\dagger}(0)\}\rangle &  \langle \{c_{\bm k\sigma}(t), c_{-\bm k\bar{\sigma}}(0)\}\rangle\\
    \langle \{c_{-\bm k\bar{\sigma}}^{\dagger}(t), c_{\bm k\sigma}^{\dagger}(0)\}\rangle & \langle \{c_{-\bm k\bar{\sigma}}^{\dagger}(t), c_{-\bm k\bar{\sigma}}(0)\}\rangle\\ 
    \end{pmatrix}.
\end{equation}
\end{widetext}
The equation of motion for the Green's function in the frequency domain reads 
\begin{equation}\label{eq:G_nambu}
    \begin{pmatrix}
    G_{\bm k\sigma;\omega}^R & F_{\bm k\sigma;\omega}^R\\
    F_{-\bm k\bar{\sigma};\omega}^R & G_{-\bm k\bar{\sigma};\omega}^R\\ 
    \end{pmatrix}=
    [\omega + i\eta- \xi_{\bm k}\tau_3 - \hat{\Sigma}^R(\bm k, \omega)]^{-1},
\end{equation}
where the self-energy can be formally written as follows,
\begin{equation}\label{eq:selfenergy_sc}
    \Sigma^R(\bm k,\omega) = \left(1- Z(\bm k,\omega)\right) \omega + \sum_{i=1,2}\Phi_i(\bm k, \omega)\tau_{i} + \chi(\bm k, \omega)\tau_3.
\end{equation}
Here, $\tau_{1,2,3}$ are Pauli matrices in the Nambu space and the identity matrix $\tau_0$ is omitted. $\chi$ represents the renormalization of the single-particle dispersion due to interactions with all other electrons including states far from the Fermi level. It does not depend on the superconducting order as long as the gap is much smaller than the Fermi energy. Thus, its energy independence is negligible for energy $\omega$ much lower than the Fermi energy, $|\omega|\ll |\mu_S|$. We define the renormalized dispersion $\epsilon_{\bm k} = \xi_{\bm k} + \chi(\bm k,0)$. The Fermi line of the sample is given by $\text{Re}(\epsilon_{\bm k}) = 0$.

Plugging Eq.~\eqref{eq:selfenergy_sc} into Eq.~\eqref{eq:G_nambu}, we find
\begin{equation}
    G_{\sigma/\bar{\sigma}}^{R}(\pm\bm k,\omega) = \frac{Z(\bm k,\omega)\omega \pm \epsilon_{\bm k}}{Z(\bm k,\omega)^2\omega^2-\epsilon_{\bm k}^2-\Phi(\bm k, \omega)^2},
\end{equation}
with $\Phi^2 \equiv \Phi_1^2+\Phi_2^2$. 
Taking the limit of $|\epsilon_{\bm k}|\gg |\sqrt{Z^2\omega^2-\Phi^2}|$, we find that
$\text{Im}(Z\omega\pm \epsilon)>0$ is a necessary condition for the positivity of the spectral function $A_{\sigma}(\bm k,\omega) = -\text{Im}G_{\sigma}^{R}(\bm k,\omega)/\pi$. This implies $\text{sgn}(\text{Im} Z(\bm k,\omega+i\eta))=\text{sgn}(\omega)$.

We assume that $Z,\Phi$ vary slowly over the scale of the Fermi wave vector. Thus, 
near a momentum $\bm k_0$ on the Fermi line of the sample, we can make the approximation $Z(\bm k,\omega)\approx Z(\bm k_0,\omega)$ and $\Phi(\bm k,\omega)\approx \Phi(\bm k_0,\omega)$. The momentum dependence of Green's function is then mainly attributed to $\epsilon_{\bm k}$, i.e., $G^{R}(\bm k,\omega)= \tilde{G}^R(\bm k_0, \epsilon_{\bm k}, \omega)$. Thus,
we can define an (angle-resolved) tunneling density of states,
\begin{align}
    N_{T}(\bm k_0,\omega) &= -\frac{N_{\bm k_0}}{\pi}\textrm{Im}\int d\epsilon_{\bm k}\ G^R(\bm k,\omega),
    \label{eq:N_T_def}\\
    \frac{N_{T}(\bm k_0,\omega)}{N_{\bm k_0}}& = \textrm{Re}\left\{\frac{|\omega|}{\sqrt{\omega^2-\Delta(\bm k_0,\omega)^2}}\right\} \label{eq:N_T},
\end{align}
where $N_{\bm k}\equiv N_f/\hbar v_{\bm k} (v_{\bm k}=|\partial_{\bm k}\epsilon_{\bm k}|)$ represents the tunneling density of states of the normal state at $\bm k$, and $\Delta(\bm k,\omega)\equiv \Phi(\bm k,\omega)/Z(\bm k,\omega)$ denotes the gap function. The branch cut of the square root functions is chosen as the positive real axis. To arrive at the second line, we used $\sqrt{Z(\bm k,\omega+i\eta)^2}=\text{sgn}(\omega)Z(\bm k,\omega+i\eta)$ and $\text{Im}\sqrt{Z^2\omega^2-\Phi^2}>0$. In the normal state $\Delta=0$, the tunneling density of states $N_{T}(\bm k, \omega)$ reduces to the single-particle density of state along the Fermi line. 
The tunneling density of states, Eq.~\eqref{eq:N_T}, exhibits a peak at energy $\omega = |\text{Re}\Delta(\bm k_0, \omega)|$, which can be defined as the superconducting gap at Fermi wave vector $\bm k_0$ of the sample. 

When the Fermi lines of the tip and sample intersect at the set of wave vectors $\mathcal{S}(k_F,\theta)$, the low-bias tunneling conductance in Eq.~\eqref{eq:dIdV_ii_broadening} simplifies to
\begin{equation}\label{eq:didv_strongcoupling}
    \frac{dI}{dV} = \sum_{\bm p^*\in \mathcal{S}(k_F, \theta)} G_{\bm p^*}\frac{N_T(\bm p^*, -eV)}{N_{\bm p^*}}.
\end{equation}
By substituting $\Delta(\bm k,\omega) = \omega|\Delta_{\bm k}|/(\omega + i\eta)$ and Eq.~\eqref{eq:N_T} into Eq.~\eqref{eq:didv_strongcoupling}, we can recover Eq.~\eqref{eq:dIdV_ii_broadening2} in the main text.

\bibliography{reference}

\begin{thebibliography}{74}%
\makeatletter
\providecommand \@ifxundefined [1]{%
 \@ifx{#1\undefined}
}%
\providecommand \@ifnum [1]{%
 \ifnum #1\expandafter \@firstoftwo
 \else \expandafter \@secondoftwo
 \fi
}%
\providecommand \@ifx [1]{%
 \ifx #1\expandafter \@firstoftwo
 \else \expandafter \@secondoftwo
 \fi
}%
\providecommand \natexlab [1]{#1}%
\providecommand \enquote  [1]{``#1''}%
\providecommand \bibnamefont  [1]{#1}%
\providecommand \bibfnamefont [1]{#1}%
\providecommand \citenamefont [1]{#1}%
\providecommand \href@noop [0]{\@secondoftwo}%
\providecommand \href [0]{\begingroup \@sanitize@url \@href}%
\providecommand \@href[1]{\@@startlink{#1}\@@href}%
\providecommand \@@href[1]{\endgroup#1\@@endlink}%
\providecommand \@sanitize@url [0]{\catcode `\\12\catcode `\$12\catcode `\&12\catcode `\#12\catcode `\^12\catcode `\_12\catcode `\%12\relax}%
\providecommand \@@startlink[1]{}%
\providecommand \@@endlink[0]{}%
\providecommand \url  [0]{\begingroup\@sanitize@url \@url }%
\providecommand \@url [1]{\endgroup\@href {#1}{\urlprefix }}%
\providecommand \urlprefix  [0]{URL }%
\providecommand \Eprint [0]{\href }%
\providecommand \doibase [0]{https://doi.org/}%
\providecommand \selectlanguage [0]{\@gobble}%
\providecommand \bibinfo  [0]{\@secondoftwo}%
\providecommand \bibfield  [0]{\@secondoftwo}%
\providecommand \translation [1]{[#1]}%
\providecommand \BibitemOpen [0]{}%
\providecommand \bibitemStop [0]{}%
\providecommand \bibitemNoStop [0]{.\EOS\space}%
\providecommand \EOS [0]{\spacefactor3000\relax}%
\providecommand \BibitemShut  [1]{\csname bibitem#1\endcsname}%
\let\auto@bib@innerbib\@empty
\bibitem [{\citenamefont {Wolf}(2011)}]{wolf2011principles}%
  \BibitemOpen
  \bibfield  {author} {\bibinfo {author} {\bibfnamefont {E.~L.}\ \bibnamefont {Wolf}},\ }\href@noop {} {\emph {\bibinfo {title} {Principles of electron tunneling spectroscopy}}},\ Vol.\ \bibinfo {volume} {152}\ (\bibinfo  {publisher} {OUP Oxford},\ \bibinfo {year} {2011})\BibitemShut {NoStop}%
\bibitem [{\citenamefont {Giaever}(1960{\natexlab{a}})}]{giaever1960energy}%
  \BibitemOpen
  \bibfield  {author} {\bibinfo {author} {\bibfnamefont {I.}~\bibnamefont {Giaever}},\ }\bibfield  {title} {\bibinfo {title} {Energy gap in superconductors measured by electron tunneling},\ }\href {https://doi.org/10.1103/PhysRevLett.5.147} {\bibfield  {journal} {\bibinfo  {journal} {Phys. Rev. Lett.}\ }\textbf {\bibinfo {volume} {5}},\ \bibinfo {pages} {147} (\bibinfo {year} {1960}{\natexlab{a}})}\BibitemShut {NoStop}%
\bibitem [{\citenamefont {Giaever}(1960{\natexlab{b}})}]{giaever1960electron}%
  \BibitemOpen
  \bibfield  {author} {\bibinfo {author} {\bibfnamefont {I.}~\bibnamefont {Giaever}},\ }\bibfield  {title} {\bibinfo {title} {Electron tunneling between two superconductors},\ }\href {https://doi.org/10.1103/PhysRevLett.5.464} {\bibfield  {journal} {\bibinfo  {journal} {Phys. Rev. Lett.}\ }\textbf {\bibinfo {volume} {5}},\ \bibinfo {pages} {464} (\bibinfo {year} {1960}{\natexlab{b}})}\BibitemShut {NoStop}%
\bibitem [{\citenamefont {Fischer}\ \emph {et~al.}(2007)\citenamefont {Fischer}, \citenamefont {Kugler}, \citenamefont {Maggio-Aprile}, \citenamefont {Berthod},\ and\ \citenamefont {Renner}}]{fischer2007scanning}%
  \BibitemOpen
  \bibfield  {author} {\bibinfo {author} {\bibfnamefont {{\O}.}~\bibnamefont {Fischer}}, \bibinfo {author} {\bibfnamefont {M.}~\bibnamefont {Kugler}}, \bibinfo {author} {\bibfnamefont {I.}~\bibnamefont {Maggio-Aprile}}, \bibinfo {author} {\bibfnamefont {C.}~\bibnamefont {Berthod}},\ and\ \bibinfo {author} {\bibfnamefont {C.}~\bibnamefont {Renner}},\ }\bibfield  {title} {\bibinfo {title} {Scanning tunneling spectroscopy of high-temperature superconductors},\ }\href {https://doi.org/10.1103/RevModPhys.79.353} {\bibfield  {journal} {\bibinfo  {journal} {Rev. Mod. Phys.}\ }\textbf {\bibinfo {volume} {79}},\ \bibinfo {pages} {353} (\bibinfo {year} {2007})}\BibitemShut {NoStop}%
\bibitem [{\citenamefont {Sukhachov}\ \emph {et~al.}(2023)\citenamefont {Sukhachov}, \citenamefont {von Oppen},\ and\ \citenamefont {Glazman}}]{sukhachov2023andreev}%
  \BibitemOpen
  \bibfield  {author} {\bibinfo {author} {\bibfnamefont {P.~O.}\ \bibnamefont {Sukhachov}}, \bibinfo {author} {\bibfnamefont {F.}~\bibnamefont {von Oppen}},\ and\ \bibinfo {author} {\bibfnamefont {L.~I.}\ \bibnamefont {Glazman}},\ }\bibfield  {title} {\bibinfo {title} {Andreev reflection in scanning tunneling spectroscopy of unconventional superconductors},\ }\href {https://doi.org/10.1103/PhysRevLett.130.216002} {\bibfield  {journal} {\bibinfo  {journal} {Phys. Rev. Lett.}\ }\textbf {\bibinfo {volume} {130}},\ \bibinfo {pages} {216002} (\bibinfo {year} {2023})}\BibitemShut {NoStop}%
\bibitem [{\citenamefont {Biswas}\ \emph {et~al.}(2025)\citenamefont {Biswas}, \citenamefont {Suman}, \citenamefont {Randeria},\ and\ \citenamefont {Sensarma}}]{biswas2025andreev}%
  \BibitemOpen
  \bibfield  {author} {\bibinfo {author} {\bibfnamefont {S.}~\bibnamefont {Biswas}}, \bibinfo {author} {\bibfnamefont {S.}~\bibnamefont {Suman}}, \bibinfo {author} {\bibfnamefont {M.}~\bibnamefont {Randeria}},\ and\ \bibinfo {author} {\bibfnamefont {R.}~\bibnamefont {Sensarma}},\ }\href {https://arxiv.org/abs/2503.07744} {\bibinfo {title} {Andreev versus tunneling spectroscopy of unconventional flat band superconductors}} (\bibinfo {year} {2025}),\ \Eprint {https://arxiv.org/abs/2503.07744} {arXiv:2503.07744 [cond-mat.supr-con]} \BibitemShut {NoStop}%
\bibitem [{\citenamefont {McMillan}\ and\ \citenamefont {Rowell}(1965)}]{mcmillan1965lead}%
  \BibitemOpen
  \bibfield  {author} {\bibinfo {author} {\bibfnamefont {W.~L.}\ \bibnamefont {McMillan}}\ and\ \bibinfo {author} {\bibfnamefont {J.~M.}\ \bibnamefont {Rowell}},\ }\bibfield  {title} {\bibinfo {title} {Lead phonon spectrum calculated from superconducting density of states},\ }\href {https://doi.org/10.1103/PhysRevLett.14.108} {\bibfield  {journal} {\bibinfo  {journal} {Phys. Rev. Lett.}\ }\textbf {\bibinfo {volume} {14}},\ \bibinfo {pages} {108} (\bibinfo {year} {1965})}\BibitemShut {NoStop}%
\bibitem [{\citenamefont {Parks}(2018)}]{parks2018superconductivity}%
  \BibitemOpen
  \bibfield  {author} {\bibinfo {author} {\bibfnamefont {R.~D.}\ \bibnamefont {Parks}},\ }\href@noop {} {\emph {\bibinfo {title} {Superconductivity: In Two Volumes: Volume 1}}}\ (\bibinfo  {publisher} {Routledge},\ \bibinfo {year} {2018})\BibitemShut {NoStop}%
\bibitem [{\citenamefont {Cao}\ \emph {et~al.}(2018)\citenamefont {Cao}, \citenamefont {Fatemi}, \citenamefont {Fang}, \citenamefont {Watanabe}, \citenamefont {Taniguchi}, \citenamefont {Kaxiras},\ and\ \citenamefont {Jarillo-Herrero}}]{cao2018unconventional}%
  \BibitemOpen
  \bibfield  {author} {\bibinfo {author} {\bibfnamefont {Y.}~\bibnamefont {Cao}}, \bibinfo {author} {\bibfnamefont {V.}~\bibnamefont {Fatemi}}, \bibinfo {author} {\bibfnamefont {S.}~\bibnamefont {Fang}}, \bibinfo {author} {\bibfnamefont {K.}~\bibnamefont {Watanabe}}, \bibinfo {author} {\bibfnamefont {T.}~\bibnamefont {Taniguchi}}, \bibinfo {author} {\bibfnamefont {E.}~\bibnamefont {Kaxiras}},\ and\ \bibinfo {author} {\bibfnamefont {P.}~\bibnamefont {Jarillo-Herrero}},\ }\bibfield  {title} {\bibinfo {title} {Unconventional superconductivity in magic-angle graphene superlattices},\ }\href {https://doi.org/10.1038/nature26160} {\bibfield  {journal} {\bibinfo  {journal} {Nature}\ }\textbf {\bibinfo {volume} {556}},\ \bibinfo {pages} {43} (\bibinfo {year} {2018})}\BibitemShut {NoStop}%
\bibitem [{\citenamefont {Park}\ \emph {et~al.}(2021)\citenamefont {Park}, \citenamefont {Cao}, \citenamefont {Watanabe}, \citenamefont {Taniguchi},\ and\ \citenamefont {Jarillo-Herrero}}]{park2021tunable}%
  \BibitemOpen
  \bibfield  {author} {\bibinfo {author} {\bibfnamefont {J.~M.}\ \bibnamefont {Park}}, \bibinfo {author} {\bibfnamefont {Y.}~\bibnamefont {Cao}}, \bibinfo {author} {\bibfnamefont {K.}~\bibnamefont {Watanabe}}, \bibinfo {author} {\bibfnamefont {T.}~\bibnamefont {Taniguchi}},\ and\ \bibinfo {author} {\bibfnamefont {P.}~\bibnamefont {Jarillo-Herrero}},\ }\bibfield  {title} {\bibinfo {title} {Tunable strongly coupled superconductivity in magic-angle twisted trilayer graphene},\ }\href {https://doi.org/10.1038/s41586-021-03192-0} {\bibfield  {journal} {\bibinfo  {journal} {Nature}\ }\textbf {\bibinfo {volume} {590}},\ \bibinfo {pages} {249} (\bibinfo {year} {2021})}\BibitemShut {NoStop}%
\bibitem [{\citenamefont {Hao}\ \emph {et~al.}(2021)\citenamefont {Hao}, \citenamefont {Zimmerman}, \citenamefont {Ledwith}, \citenamefont {Khalaf}, \citenamefont {Najafabadi}, \citenamefont {Watanabe}, \citenamefont {Taniguchi}, \citenamefont {Vishwanath},\ and\ \citenamefont {Kim}}]{hao2021electric}%
  \BibitemOpen
  \bibfield  {author} {\bibinfo {author} {\bibfnamefont {Z.}~\bibnamefont {Hao}}, \bibinfo {author} {\bibfnamefont {A.}~\bibnamefont {Zimmerman}}, \bibinfo {author} {\bibfnamefont {P.}~\bibnamefont {Ledwith}}, \bibinfo {author} {\bibfnamefont {E.}~\bibnamefont {Khalaf}}, \bibinfo {author} {\bibfnamefont {D.~H.}\ \bibnamefont {Najafabadi}}, \bibinfo {author} {\bibfnamefont {K.}~\bibnamefont {Watanabe}}, \bibinfo {author} {\bibfnamefont {T.}~\bibnamefont {Taniguchi}}, \bibinfo {author} {\bibfnamefont {A.}~\bibnamefont {Vishwanath}},\ and\ \bibinfo {author} {\bibfnamefont {P.}~\bibnamefont {Kim}},\ }\bibfield  {title} {\bibinfo {title} {Electric field–tunable superconductivity in alternating-twist magic-angle trilayer graphene},\ }\href {https://www.science.org/doi/10.1126/science.abg0399} {\bibfield  {journal} {\bibinfo  {journal} {Science}\ }\textbf {\bibinfo {volume} {371}},\ \bibinfo {pages} {1133} (\bibinfo {year} {2021})}\BibitemShut {NoStop}%
\bibitem [{\citenamefont {Zhang}\ \emph {et~al.}(2022)\citenamefont {Zhang}, \citenamefont {Polski}, \citenamefont {Lewandowski}, \citenamefont {Thomson}, \citenamefont {Peng}, \citenamefont {Choi}, \citenamefont {Kim}, \citenamefont {Watanabe}, \citenamefont {Taniguchi}, \citenamefont {Alicea} \emph {et~al.}}]{zhang2022promotion}%
  \BibitemOpen
  \bibfield  {author} {\bibinfo {author} {\bibfnamefont {Y.}~\bibnamefont {Zhang}}, \bibinfo {author} {\bibfnamefont {R.}~\bibnamefont {Polski}}, \bibinfo {author} {\bibfnamefont {C.}~\bibnamefont {Lewandowski}}, \bibinfo {author} {\bibfnamefont {A.}~\bibnamefont {Thomson}}, \bibinfo {author} {\bibfnamefont {Y.}~\bibnamefont {Peng}}, \bibinfo {author} {\bibfnamefont {Y.}~\bibnamefont {Choi}}, \bibinfo {author} {\bibfnamefont {H.}~\bibnamefont {Kim}}, \bibinfo {author} {\bibfnamefont {K.}~\bibnamefont {Watanabe}}, \bibinfo {author} {\bibfnamefont {T.}~\bibnamefont {Taniguchi}}, \bibinfo {author} {\bibfnamefont {J.}~\bibnamefont {Alicea}}, \emph {et~al.},\ }\bibfield  {title} {\bibinfo {title} {Promotion of superconductivity in magic-angle graphene multilayers},\ }\href {https://www.science.org/doi/10.1126/science.abn8585} {\bibfield  {journal} {\bibinfo  {journal} {Science}\ }\textbf {\bibinfo {volume} {377}},\ \bibinfo {pages} {1538} (\bibinfo {year} {2022})}\BibitemShut {NoStop}%
\bibitem [{\citenamefont {Park}\ \emph {et~al.}(2022)\citenamefont {Park}, \citenamefont {Cao}, \citenamefont {Xia}, \citenamefont {Sun}, \citenamefont {Watanabe}, \citenamefont {Taniguchi},\ and\ \citenamefont {Jarillo-Herrero}}]{park2022robust}%
  \BibitemOpen
  \bibfield  {author} {\bibinfo {author} {\bibfnamefont {J.~M.}\ \bibnamefont {Park}}, \bibinfo {author} {\bibfnamefont {Y.}~\bibnamefont {Cao}}, \bibinfo {author} {\bibfnamefont {L.-Q.}\ \bibnamefont {Xia}}, \bibinfo {author} {\bibfnamefont {S.}~\bibnamefont {Sun}}, \bibinfo {author} {\bibfnamefont {K.}~\bibnamefont {Watanabe}}, \bibinfo {author} {\bibfnamefont {T.}~\bibnamefont {Taniguchi}},\ and\ \bibinfo {author} {\bibfnamefont {P.}~\bibnamefont {Jarillo-Herrero}},\ }\bibfield  {title} {\bibinfo {title} {Robust superconductivity in magic-angle multilayer graphene family},\ }\href {https://www.nature.com/articles/s41563-022-01287-1} {\bibfield  {journal} {\bibinfo  {journal} {Nature Materials}\ }\textbf {\bibinfo {volume} {21}},\ \bibinfo {pages} {877} (\bibinfo {year} {2022})}\BibitemShut {NoStop}%
\bibitem [{\citenamefont {Burg}\ \emph {et~al.}(2022)\citenamefont {Burg}, \citenamefont {Khalaf}, \citenamefont {Wang}, \citenamefont {Watanabe}, \citenamefont {Taniguchi},\ and\ \citenamefont {Tutuc}}]{burg2022emergence}%
  \BibitemOpen
  \bibfield  {author} {\bibinfo {author} {\bibfnamefont {G.~W.}\ \bibnamefont {Burg}}, \bibinfo {author} {\bibfnamefont {E.}~\bibnamefont {Khalaf}}, \bibinfo {author} {\bibfnamefont {Y.}~\bibnamefont {Wang}}, \bibinfo {author} {\bibfnamefont {K.}~\bibnamefont {Watanabe}}, \bibinfo {author} {\bibfnamefont {T.}~\bibnamefont {Taniguchi}},\ and\ \bibinfo {author} {\bibfnamefont {E.}~\bibnamefont {Tutuc}},\ }\bibfield  {title} {\bibinfo {title} {Emergence of correlations in alternating twist quadrilayer graphene},\ }\href@noop {} {\bibfield  {journal} {\bibinfo  {journal} {Nature Materials}\ }\textbf {\bibinfo {volume} {21}},\ \bibinfo {pages} {884} (\bibinfo {year} {2022})}\BibitemShut {NoStop}%
\bibitem [{\citenamefont {Su}\ \emph {et~al.}(2023)\citenamefont {Su}, \citenamefont {Kuiri}, \citenamefont {Watanabe}, \citenamefont {Taniguchi},\ and\ \citenamefont {Folk}}]{su2023superconductivity}%
  \BibitemOpen
  \bibfield  {author} {\bibinfo {author} {\bibfnamefont {R.}~\bibnamefont {Su}}, \bibinfo {author} {\bibfnamefont {M.}~\bibnamefont {Kuiri}}, \bibinfo {author} {\bibfnamefont {K.}~\bibnamefont {Watanabe}}, \bibinfo {author} {\bibfnamefont {T.}~\bibnamefont {Taniguchi}},\ and\ \bibinfo {author} {\bibfnamefont {J.}~\bibnamefont {Folk}},\ }\bibfield  {title} {\bibinfo {title} {Superconductivity in twisted double bilayer graphene stabilized by wse2},\ }\href {https://www.nature.com/articles/s41563-023-01653-7} {\bibfield  {journal} {\bibinfo  {journal} {Nature Materials}\ }\textbf {\bibinfo {volume} {22}},\ \bibinfo {pages} {1332} (\bibinfo {year} {2023})}\BibitemShut {NoStop}%
\bibitem [{\citenamefont {Uri}\ \emph {et~al.}(2023)\citenamefont {Uri}, \citenamefont {de~la Barrera}, \citenamefont {Randeria}, \citenamefont {Rodan-Legrain}, \citenamefont {Devakul}, \citenamefont {Crowley}, \citenamefont {Paul}, \citenamefont {Watanabe}, \citenamefont {Taniguchi}, \citenamefont {Lifshitz} \emph {et~al.}}]{uri2023superconductivity}%
  \BibitemOpen
  \bibfield  {author} {\bibinfo {author} {\bibfnamefont {A.}~\bibnamefont {Uri}}, \bibinfo {author} {\bibfnamefont {S.~C.}\ \bibnamefont {de~la Barrera}}, \bibinfo {author} {\bibfnamefont {M.~T.}\ \bibnamefont {Randeria}}, \bibinfo {author} {\bibfnamefont {D.}~\bibnamefont {Rodan-Legrain}}, \bibinfo {author} {\bibfnamefont {T.}~\bibnamefont {Devakul}}, \bibinfo {author} {\bibfnamefont {P.~J.}\ \bibnamefont {Crowley}}, \bibinfo {author} {\bibfnamefont {N.}~\bibnamefont {Paul}}, \bibinfo {author} {\bibfnamefont {K.}~\bibnamefont {Watanabe}}, \bibinfo {author} {\bibfnamefont {T.}~\bibnamefont {Taniguchi}}, \bibinfo {author} {\bibfnamefont {R.}~\bibnamefont {Lifshitz}}, \emph {et~al.},\ }\bibfield  {title} {\bibinfo {title} {Superconductivity and strong interactions in a tunable moir{\'e} quasicrystal},\ }\href {https://doi.org/10.1038/s41586-023-06294-z} {\bibfield  {journal} {\bibinfo  {journal} {Nature}\ }\textbf {\bibinfo {volume} {620}},\ \bibinfo {pages} {762} (\bibinfo {year} {2023})}\BibitemShut {NoStop}%
\bibitem [{\citenamefont {Zhou}\ \emph {et~al.}(2021)\citenamefont {Zhou}, \citenamefont {Xie}, \citenamefont {Taniguchi}, \citenamefont {Watanabe},\ and\ \citenamefont {Young}}]{zhou2021superconductivity}%
  \BibitemOpen
  \bibfield  {author} {\bibinfo {author} {\bibfnamefont {H.}~\bibnamefont {Zhou}}, \bibinfo {author} {\bibfnamefont {T.}~\bibnamefont {Xie}}, \bibinfo {author} {\bibfnamefont {T.}~\bibnamefont {Taniguchi}}, \bibinfo {author} {\bibfnamefont {K.}~\bibnamefont {Watanabe}},\ and\ \bibinfo {author} {\bibfnamefont {A.~F.}\ \bibnamefont {Young}},\ }\bibfield  {title} {\bibinfo {title} {Superconductivity in rhombohedral trilayer graphene},\ }\href {https://www.nature.com/articles/s41586-021-03926-0} {\bibfield  {journal} {\bibinfo  {journal} {Nature}\ }\textbf {\bibinfo {volume} {598}},\ \bibinfo {pages} {434} (\bibinfo {year} {2021})}\BibitemShut {NoStop}%
\bibitem [{\citenamefont {Zhou}\ \emph {et~al.}(2022)\citenamefont {Zhou}, \citenamefont {Holleis}, \citenamefont {Saito}, \citenamefont {Cohen}, \citenamefont {Huynh}, \citenamefont {Patterson}, \citenamefont {Yang}, \citenamefont {Taniguchi}, \citenamefont {Watanabe},\ and\ \citenamefont {Young}}]{zhou2022isospin}%
  \BibitemOpen
  \bibfield  {author} {\bibinfo {author} {\bibfnamefont {H.}~\bibnamefont {Zhou}}, \bibinfo {author} {\bibfnamefont {L.}~\bibnamefont {Holleis}}, \bibinfo {author} {\bibfnamefont {Y.}~\bibnamefont {Saito}}, \bibinfo {author} {\bibfnamefont {L.}~\bibnamefont {Cohen}}, \bibinfo {author} {\bibfnamefont {W.}~\bibnamefont {Huynh}}, \bibinfo {author} {\bibfnamefont {C.~L.}\ \bibnamefont {Patterson}}, \bibinfo {author} {\bibfnamefont {F.}~\bibnamefont {Yang}}, \bibinfo {author} {\bibfnamefont {T.}~\bibnamefont {Taniguchi}}, \bibinfo {author} {\bibfnamefont {K.}~\bibnamefont {Watanabe}},\ and\ \bibinfo {author} {\bibfnamefont {A.~F.}\ \bibnamefont {Young}},\ }\bibfield  {title} {\bibinfo {title} {Isospin magnetism and spin-polarized superconductivity in bernal bilayer graphene},\ }\href {https://www.science.org/doi/10.1126/science.abm8386} {\bibfield  {journal} {\bibinfo  {journal} {Science}\ }\textbf {\bibinfo {volume} {375}},\ \bibinfo {pages} {774} (\bibinfo {year} {2022})}\BibitemShut {NoStop}%
\bibitem [{\citenamefont {Han}\ \emph {et~al.}(2025)\citenamefont {Han}, \citenamefont {Lu}, \citenamefont {Hadjri}, \citenamefont {Shi}, \citenamefont {Wu}, \citenamefont {Xu}, \citenamefont {Yao}, \citenamefont {Cotten}, \citenamefont {Sedeh}, \citenamefont {Weldeyesus}, \citenamefont {Yang}, \citenamefont {Seo}, \citenamefont {Ye}, \citenamefont {Zhou}, \citenamefont {Liu}, \citenamefont {Shi}, \citenamefont {Hua}, \citenamefont {Watanabe}, \citenamefont {Taniguchi}, \citenamefont {Xiong}, \citenamefont {Zumb\"uhl}, \citenamefont {Fu},\ and\ \citenamefont {Ju}}]{han2025chiral}%
  \BibitemOpen
  \bibfield  {author} {\bibinfo {author} {\bibfnamefont {T.}~\bibnamefont {Han}}, \bibinfo {author} {\bibfnamefont {Z.}~\bibnamefont {Lu}}, \bibinfo {author} {\bibfnamefont {Z.}~\bibnamefont {Hadjri}}, \bibinfo {author} {\bibfnamefont {L.}~\bibnamefont {Shi}}, \bibinfo {author} {\bibfnamefont {Z.}~\bibnamefont {Wu}}, \bibinfo {author} {\bibfnamefont {W.}~\bibnamefont {Xu}}, \bibinfo {author} {\bibfnamefont {Y.}~\bibnamefont {Yao}}, \bibinfo {author} {\bibfnamefont {A.~A.}\ \bibnamefont {Cotten}}, \bibinfo {author} {\bibfnamefont {O.~S.}\ \bibnamefont {Sedeh}}, \bibinfo {author} {\bibfnamefont {H.}~\bibnamefont {Weldeyesus}}, \bibinfo {author} {\bibfnamefont {J.}~\bibnamefont {Yang}}, \bibinfo {author} {\bibfnamefont {J.}~\bibnamefont {Seo}}, \bibinfo {author} {\bibfnamefont {S.}~\bibnamefont {Ye}}, \bibinfo {author} {\bibfnamefont {M.}~\bibnamefont {Zhou}}, \bibinfo {author} {\bibfnamefont {H.}~\bibnamefont {Liu}}, \bibinfo {author} {\bibfnamefont {G.}~\bibnamefont {Shi}}, \bibinfo {author} {\bibfnamefont
  {Z.}~\bibnamefont {Hua}}, \bibinfo {author} {\bibfnamefont {K.}~\bibnamefont {Watanabe}}, \bibinfo {author} {\bibfnamefont {T.}~\bibnamefont {Taniguchi}}, \bibinfo {author} {\bibfnamefont {P.}~\bibnamefont {Xiong}}, \bibinfo {author} {\bibfnamefont {D.~M.}\ \bibnamefont {Zumb\"uhl}}, \bibinfo {author} {\bibfnamefont {L.}~\bibnamefont {Fu}},\ and\ \bibinfo {author} {\bibfnamefont {L.}~\bibnamefont {Ju}},\ }\bibfield  {title} {\bibinfo {title} {Signatures of chiral superconductivity in rhombohedral graphene},\ }\href {https://www.nature.com/articles/s41586-025-09169-7} {\bibfield  {journal} {\bibinfo  {journal} {Nature}\ } (\bibinfo {year} {2025})}\BibitemShut {NoStop}%
\bibitem [{\citenamefont {Oh}\ \emph {et~al.}(2021)\citenamefont {Oh}, \citenamefont {Nuckolls}, \citenamefont {Wong}, \citenamefont {Lee}, \citenamefont {Liu}, \citenamefont {Watanabe}, \citenamefont {Taniguchi},\ and\ \citenamefont {Yazdani}}]{oh2021evidence}%
  \BibitemOpen
  \bibfield  {author} {\bibinfo {author} {\bibfnamefont {M.}~\bibnamefont {Oh}}, \bibinfo {author} {\bibfnamefont {K.~P.}\ \bibnamefont {Nuckolls}}, \bibinfo {author} {\bibfnamefont {D.}~\bibnamefont {Wong}}, \bibinfo {author} {\bibfnamefont {R.~L.}\ \bibnamefont {Lee}}, \bibinfo {author} {\bibfnamefont {X.}~\bibnamefont {Liu}}, \bibinfo {author} {\bibfnamefont {K.}~\bibnamefont {Watanabe}}, \bibinfo {author} {\bibfnamefont {T.}~\bibnamefont {Taniguchi}},\ and\ \bibinfo {author} {\bibfnamefont {A.}~\bibnamefont {Yazdani}},\ }\bibfield  {title} {\bibinfo {title} {Evidence for unconventional superconductivity in twisted bilayer graphene},\ }\href {https://www.nature.com/articles/s41586-021-04121-x} {\bibfield  {journal} {\bibinfo  {journal} {Nature}\ }\textbf {\bibinfo {volume} {600}},\ \bibinfo {pages} {240} (\bibinfo {year} {2021})}\BibitemShut {NoStop}%
\bibitem [{\citenamefont {Kim}\ \emph {et~al.}(2022)\citenamefont {Kim}, \citenamefont {Choi}, \citenamefont {Lewandowski}, \citenamefont {Thomson}, \citenamefont {Zhang}, \citenamefont {Polski}, \citenamefont {Watanabe}, \citenamefont {Taniguchi}, \citenamefont {Alicea},\ and\ \citenamefont {Nadj-Perge}}]{kim2022evidence}%
  \BibitemOpen
  \bibfield  {author} {\bibinfo {author} {\bibfnamefont {H.}~\bibnamefont {Kim}}, \bibinfo {author} {\bibfnamefont {Y.}~\bibnamefont {Choi}}, \bibinfo {author} {\bibfnamefont {C.}~\bibnamefont {Lewandowski}}, \bibinfo {author} {\bibfnamefont {A.}~\bibnamefont {Thomson}}, \bibinfo {author} {\bibfnamefont {Y.}~\bibnamefont {Zhang}}, \bibinfo {author} {\bibfnamefont {R.}~\bibnamefont {Polski}}, \bibinfo {author} {\bibfnamefont {K.}~\bibnamefont {Watanabe}}, \bibinfo {author} {\bibfnamefont {T.}~\bibnamefont {Taniguchi}}, \bibinfo {author} {\bibfnamefont {J.}~\bibnamefont {Alicea}},\ and\ \bibinfo {author} {\bibfnamefont {S.}~\bibnamefont {Nadj-Perge}},\ }\bibfield  {title} {\bibinfo {title} {Evidence for unconventional superconductivity in twisted trilayer graphene},\ }\href {https://www.nature.com/articles/s41586-022-04715-z} {\bibfield  {journal} {\bibinfo  {journal} {Nature}\ }\textbf {\bibinfo {volume} {606}},\ \bibinfo {pages} {494} (\bibinfo {year} {2022})}\BibitemShut {NoStop}%
\bibitem [{\citenamefont {Park}\ \emph {et~al.}(2025)\citenamefont {Park}, \citenamefont {Sun}, \citenamefont {Watanabe}, \citenamefont {Taniguchi},\ and\ \citenamefont {Jarillo-Herrero}}]{park2025simultaneous}%
  \BibitemOpen
  \bibfield  {author} {\bibinfo {author} {\bibfnamefont {J.~M.}\ \bibnamefont {Park}}, \bibinfo {author} {\bibfnamefont {S.}~\bibnamefont {Sun}}, \bibinfo {author} {\bibfnamefont {K.}~\bibnamefont {Watanabe}}, \bibinfo {author} {\bibfnamefont {T.}~\bibnamefont {Taniguchi}},\ and\ \bibinfo {author} {\bibfnamefont {P.}~\bibnamefont {Jarillo-Herrero}},\ }\href {https://arxiv.org/abs/2503.16410} {\bibinfo {title} {Simultaneous transport and tunneling spectroscopy of moir\'e graphene: Distinct observation of the superconducting gap and signatures of nodal superconductivity}} (\bibinfo {year} {2025}),\ \Eprint {https://arxiv.org/abs/2503.16410} {arXiv:2503.16410 [cond-mat.supr-con]} \BibitemShut {NoStop}%
\bibitem [{\citenamefont {Kim}\ \emph {et~al.}(2025)\citenamefont {Kim}, \citenamefont {Rai}, \citenamefont {Crippa}, \citenamefont {Călugăru}, \citenamefont {Hu}, \citenamefont {Choi}, \citenamefont {Kong}, \citenamefont {Baum}, \citenamefont {Zhang}, \citenamefont {Holleis}, \citenamefont {Watanabe}, \citenamefont {Taniguchi}, \citenamefont {Young}, \citenamefont {Bernevig}, \citenamefont {Valentí}, \citenamefont {Sangiovanni}, \citenamefont {Wehling},\ and\ \citenamefont {Nadj-Perge}}]{kim2025resolving}%
  \BibitemOpen
  \bibfield  {author} {\bibinfo {author} {\bibfnamefont {H.}~\bibnamefont {Kim}}, \bibinfo {author} {\bibfnamefont {G.}~\bibnamefont {Rai}}, \bibinfo {author} {\bibfnamefont {L.}~\bibnamefont {Crippa}}, \bibinfo {author} {\bibfnamefont {D.}~\bibnamefont {Călugăru}}, \bibinfo {author} {\bibfnamefont {H.}~\bibnamefont {Hu}}, \bibinfo {author} {\bibfnamefont {Y.}~\bibnamefont {Choi}}, \bibinfo {author} {\bibfnamefont {L.}~\bibnamefont {Kong}}, \bibinfo {author} {\bibfnamefont {E.}~\bibnamefont {Baum}}, \bibinfo {author} {\bibfnamefont {Y.}~\bibnamefont {Zhang}}, \bibinfo {author} {\bibfnamefont {L.}~\bibnamefont {Holleis}}, \bibinfo {author} {\bibfnamefont {K.}~\bibnamefont {Watanabe}}, \bibinfo {author} {\bibfnamefont {T.}~\bibnamefont {Taniguchi}}, \bibinfo {author} {\bibfnamefont {A.~F.}\ \bibnamefont {Young}}, \bibinfo {author} {\bibfnamefont {B.~A.}\ \bibnamefont {Bernevig}}, \bibinfo {author} {\bibfnamefont {R.}~\bibnamefont {Valentí}}, \bibinfo {author} {\bibfnamefont {G.}~\bibnamefont {Sangiovanni}},
  \bibinfo {author} {\bibfnamefont {T.}~\bibnamefont {Wehling}},\ and\ \bibinfo {author} {\bibfnamefont {S.}~\bibnamefont {Nadj-Perge}},\ }\href {https://arxiv.org/abs/2505.17200} {\bibinfo {title} {Resolving intervalley gaps and many-body resonances in moir\'e superconductor}} (\bibinfo {year} {2025}),\ \Eprint {https://arxiv.org/abs/2505.17200} {arXiv:2505.17200 [cond-mat.supr-con]} \BibitemShut {NoStop}%
\bibitem [{\citenamefont {Tanaka}\ \emph {et~al.}(2025)\citenamefont {Tanaka}, \citenamefont {Wang}, \citenamefont {Dinh}, \citenamefont {Rodan-Legrain}, \citenamefont {Zaman}, \citenamefont {Hays}, \citenamefont {Almanakly}, \citenamefont {Kannan}, \citenamefont {Kim}, \citenamefont {Niedzielski} \emph {et~al.}}]{tanaka2025superfluid}%
  \BibitemOpen
  \bibfield  {author} {\bibinfo {author} {\bibfnamefont {M.}~\bibnamefont {Tanaka}}, \bibinfo {author} {\bibfnamefont {J.~{\^I}.-j.}\ \bibnamefont {Wang}}, \bibinfo {author} {\bibfnamefont {T.~H.}\ \bibnamefont {Dinh}}, \bibinfo {author} {\bibfnamefont {D.}~\bibnamefont {Rodan-Legrain}}, \bibinfo {author} {\bibfnamefont {S.}~\bibnamefont {Zaman}}, \bibinfo {author} {\bibfnamefont {M.}~\bibnamefont {Hays}}, \bibinfo {author} {\bibfnamefont {A.}~\bibnamefont {Almanakly}}, \bibinfo {author} {\bibfnamefont {B.}~\bibnamefont {Kannan}}, \bibinfo {author} {\bibfnamefont {D.~K.}\ \bibnamefont {Kim}}, \bibinfo {author} {\bibfnamefont {B.~M.}\ \bibnamefont {Niedzielski}}, \emph {et~al.},\ }\bibfield  {title} {\bibinfo {title} {Superfluid stiffness of magic-angle twisted bilayer graphene},\ }\href {https://www.nature.com/articles/s41586-024-08494-7} {\bibfield  {journal} {\bibinfo  {journal} {Nature}\ }\textbf {\bibinfo {volume} {638}},\ \bibinfo {pages} {99} (\bibinfo {year} {2025})}\BibitemShut {NoStop}%
\bibitem [{\citenamefont {Banerjee}\ \emph {et~al.}(2025)\citenamefont {Banerjee}, \citenamefont {Hao}, \citenamefont {Kreidel}, \citenamefont {Ledwith}, \citenamefont {Phinney}, \citenamefont {Park}, \citenamefont {Zimmerman}, \citenamefont {Wesson}, \citenamefont {Watanabe}, \citenamefont {Taniguchi} \emph {et~al.}}]{banerjee2025superfluid}%
  \BibitemOpen
  \bibfield  {author} {\bibinfo {author} {\bibfnamefont {A.}~\bibnamefont {Banerjee}}, \bibinfo {author} {\bibfnamefont {Z.}~\bibnamefont {Hao}}, \bibinfo {author} {\bibfnamefont {M.}~\bibnamefont {Kreidel}}, \bibinfo {author} {\bibfnamefont {P.}~\bibnamefont {Ledwith}}, \bibinfo {author} {\bibfnamefont {I.}~\bibnamefont {Phinney}}, \bibinfo {author} {\bibfnamefont {J.~M.}\ \bibnamefont {Park}}, \bibinfo {author} {\bibfnamefont {A.}~\bibnamefont {Zimmerman}}, \bibinfo {author} {\bibfnamefont {M.~E.}\ \bibnamefont {Wesson}}, \bibinfo {author} {\bibfnamefont {K.}~\bibnamefont {Watanabe}}, \bibinfo {author} {\bibfnamefont {T.}~\bibnamefont {Taniguchi}}, \emph {et~al.},\ }\bibfield  {title} {\bibinfo {title} {Superfluid stiffness of twisted trilayer graphene superconductors},\ }\href {https://www.nature.com/articles/s41586-024-08444-3} {\bibfield  {journal} {\bibinfo  {journal} {Nature}\ }\textbf {\bibinfo {volume} {638}},\ \bibinfo {pages} {93} (\bibinfo {year} {2025})}\BibitemShut {NoStop}%
\bibitem [{\citenamefont {Cao}\ \emph {et~al.}(2021)\citenamefont {Cao}, \citenamefont {Rodan-Legrain}, \citenamefont {Park}, \citenamefont {Yuan}, \citenamefont {Watanabe}, \citenamefont {Taniguchi}, \citenamefont {Fernandes}, \citenamefont {Fu},\ and\ \citenamefont {Jarillo-Herrero}}]{cao2021nematicity}%
  \BibitemOpen
  \bibfield  {author} {\bibinfo {author} {\bibfnamefont {Y.}~\bibnamefont {Cao}}, \bibinfo {author} {\bibfnamefont {D.}~\bibnamefont {Rodan-Legrain}}, \bibinfo {author} {\bibfnamefont {J.~M.}\ \bibnamefont {Park}}, \bibinfo {author} {\bibfnamefont {N.~F.}\ \bibnamefont {Yuan}}, \bibinfo {author} {\bibfnamefont {K.}~\bibnamefont {Watanabe}}, \bibinfo {author} {\bibfnamefont {T.}~\bibnamefont {Taniguchi}}, \bibinfo {author} {\bibfnamefont {R.~M.}\ \bibnamefont {Fernandes}}, \bibinfo {author} {\bibfnamefont {L.}~\bibnamefont {Fu}},\ and\ \bibinfo {author} {\bibfnamefont {P.}~\bibnamefont {Jarillo-Herrero}},\ }\bibfield  {title} {\bibinfo {title} {Nematicity and competing orders in superconducting magic-angle graphene},\ }\href {https://doi.org/10.1126/science.abc2836} {\bibfield  {journal} {\bibinfo  {journal} {Science}\ }\textbf {\bibinfo {volume} {372}},\ \bibinfo {pages} {264} (\bibinfo {year} {2021})}\BibitemShut {NoStop}%
\bibitem [{\citenamefont {Inbar}\ \emph {et~al.}(2023)\citenamefont {Inbar}, \citenamefont {Birkbeck}, \citenamefont {Xiao}, \citenamefont {Taniguchi}, \citenamefont {Watanabe}, \citenamefont {Yan}, \citenamefont {Oreg}, \citenamefont {Stern}, \citenamefont {Berg},\ and\ \citenamefont {Ilani}}]{inbar2023quantum}%
  \BibitemOpen
  \bibfield  {author} {\bibinfo {author} {\bibfnamefont {A.}~\bibnamefont {Inbar}}, \bibinfo {author} {\bibfnamefont {J.}~\bibnamefont {Birkbeck}}, \bibinfo {author} {\bibfnamefont {J.}~\bibnamefont {Xiao}}, \bibinfo {author} {\bibfnamefont {T.}~\bibnamefont {Taniguchi}}, \bibinfo {author} {\bibfnamefont {K.}~\bibnamefont {Watanabe}}, \bibinfo {author} {\bibfnamefont {B.}~\bibnamefont {Yan}}, \bibinfo {author} {\bibfnamefont {Y.}~\bibnamefont {Oreg}}, \bibinfo {author} {\bibfnamefont {A.}~\bibnamefont {Stern}}, \bibinfo {author} {\bibfnamefont {E.}~\bibnamefont {Berg}},\ and\ \bibinfo {author} {\bibfnamefont {S.}~\bibnamefont {Ilani}},\ }\bibfield  {title} {\bibinfo {title} {The quantum twisting microscope},\ }\href {https://www.nature.com/articles/s41586-022-05685-y} {\bibfield  {journal} {\bibinfo  {journal} {Nature}\ }\textbf {\bibinfo {volume} {614}},\ \bibinfo {pages} {682} (\bibinfo {year} {2023})}\BibitemShut {NoStop}%
\bibitem [{\citenamefont {Birkbeck}\ \emph {et~al.}(2025)\citenamefont {Birkbeck}, \citenamefont {Xiao}, \citenamefont {Inbar}, \citenamefont {Taniguchi}, \citenamefont {Watanabe}, \citenamefont {Berg}, \citenamefont {Glazman}, \citenamefont {Guinea}, \citenamefont {von Oppen},\ and\ \citenamefont {Ilani}}]{birkbeck2024measuring}%
  \BibitemOpen
  \bibfield  {author} {\bibinfo {author} {\bibfnamefont {J.}~\bibnamefont {Birkbeck}}, \bibinfo {author} {\bibfnamefont {J.}~\bibnamefont {Xiao}}, \bibinfo {author} {\bibfnamefont {A.}~\bibnamefont {Inbar}}, \bibinfo {author} {\bibfnamefont {T.}~\bibnamefont {Taniguchi}}, \bibinfo {author} {\bibfnamefont {K.}~\bibnamefont {Watanabe}}, \bibinfo {author} {\bibfnamefont {E.}~\bibnamefont {Berg}}, \bibinfo {author} {\bibfnamefont {L.}~\bibnamefont {Glazman}}, \bibinfo {author} {\bibfnamefont {F.}~\bibnamefont {Guinea}}, \bibinfo {author} {\bibfnamefont {F.}~\bibnamefont {von Oppen}},\ and\ \bibinfo {author} {\bibfnamefont {S.}~\bibnamefont {Ilani}},\ }\bibfield  {title} {\bibinfo {title} {Quantum twisting microscopy of phonons in twisted bilayer graphene},\ }\href {https://doi.org/10.1038/s41586-025-08881-8} {\bibfield  {journal} {\bibinfo  {journal} {Nature}\ }\textbf {\bibinfo {volume} {641}},\ \bibinfo {pages} {345} (\bibinfo {year} {2025})}\BibitemShut {NoStop}%
\bibitem [{\citenamefont {Lee}\ \emph {et~al.}(2025)\citenamefont {Lee}, \citenamefont {Das}, \citenamefont {Herzog-Arbeitman}, \citenamefont {Papp}, \citenamefont {Li}, \citenamefont {Daschner}, \citenamefont {Zhou}, \citenamefont {Bhatt}, \citenamefont {Currle}, \citenamefont {Yu}, \citenamefont {Jiang}, \citenamefont {Becherer}, \citenamefont {Mittermeier}, \citenamefont {Altpeter}, \citenamefont {Obermayer}, \citenamefont {Lorenz}, \citenamefont {Chavez}, \citenamefont {Le}, \citenamefont {Williams}, \citenamefont {Watanabe}, \citenamefont {Taniguchi}, \citenamefont {Bernevig},\ and\ \citenamefont {Efetov}}]{lee2025revealing}%
  \BibitemOpen
  \bibfield  {author} {\bibinfo {author} {\bibfnamefont {M.}~\bibnamefont {Lee}}, \bibinfo {author} {\bibfnamefont {I.}~\bibnamefont {Das}}, \bibinfo {author} {\bibfnamefont {J.}~\bibnamefont {Herzog-Arbeitman}}, \bibinfo {author} {\bibfnamefont {J.}~\bibnamefont {Papp}}, \bibinfo {author} {\bibfnamefont {J.}~\bibnamefont {Li}}, \bibinfo {author} {\bibfnamefont {M.}~\bibnamefont {Daschner}}, \bibinfo {author} {\bibfnamefont {Z.}~\bibnamefont {Zhou}}, \bibinfo {author} {\bibfnamefont {M.}~\bibnamefont {Bhatt}}, \bibinfo {author} {\bibfnamefont {M.}~\bibnamefont {Currle}}, \bibinfo {author} {\bibfnamefont {J.}~\bibnamefont {Yu}}, \bibinfo {author} {\bibfnamefont {Y.}~\bibnamefont {Jiang}}, \bibinfo {author} {\bibfnamefont {M.}~\bibnamefont {Becherer}}, \bibinfo {author} {\bibfnamefont {R.}~\bibnamefont {Mittermeier}}, \bibinfo {author} {\bibfnamefont {P.}~\bibnamefont {Altpeter}}, \bibinfo {author} {\bibfnamefont {C.}~\bibnamefont {Obermayer}}, \bibinfo {author} {\bibfnamefont {H.}~\bibnamefont {Lorenz}},
  \bibinfo {author} {\bibfnamefont {G.}~\bibnamefont {Chavez}}, \bibinfo {author} {\bibfnamefont {B.~T.}\ \bibnamefont {Le}}, \bibinfo {author} {\bibfnamefont {J.}~\bibnamefont {Williams}}, \bibinfo {author} {\bibfnamefont {K.}~\bibnamefont {Watanabe}}, \bibinfo {author} {\bibfnamefont {T.}~\bibnamefont {Taniguchi}}, \bibinfo {author} {\bibfnamefont {B.~A.}\ \bibnamefont {Bernevig}},\ and\ \bibinfo {author} {\bibfnamefont {D.~K.}\ \bibnamefont {Efetov}},\ }\href {https://arxiv.org/abs/2507.03189} {\bibinfo {title} {Revealing electron-electron interactions in graphene at room temperature with the quantum twisting microscope}} (\bibinfo {year} {2025}),\ \Eprint {https://arxiv.org/abs/2507.03189} {arXiv:2507.03189 [cond-mat.mes-hall]} \BibitemShut {NoStop}%
\bibitem [{\citenamefont {Pichler}\ \emph {et~al.}(2024)\citenamefont {Pichler}, \citenamefont {Kadow}, \citenamefont {Kuhlenkamp},\ and\ \citenamefont {Knap}}]{pichler2024probing}%
  \BibitemOpen
  \bibfield  {author} {\bibinfo {author} {\bibfnamefont {F.}~\bibnamefont {Pichler}}, \bibinfo {author} {\bibfnamefont {W.}~\bibnamefont {Kadow}}, \bibinfo {author} {\bibfnamefont {C.}~\bibnamefont {Kuhlenkamp}},\ and\ \bibinfo {author} {\bibfnamefont {M.}~\bibnamefont {Knap}},\ }\bibfield  {title} {\bibinfo {title} {Probing magnetism in moir\'e heterostructures with quantum twisting microscopes},\ }\href {https://doi.org/10.1103/PhysRevB.110.045116} {\bibfield  {journal} {\bibinfo  {journal} {Phys. Rev. B}\ }\textbf {\bibinfo {volume} {110}},\ \bibinfo {pages} {045116} (\bibinfo {year} {2024})}\BibitemShut {NoStop}%
\bibitem [{\citenamefont {Xiao}\ \emph {et~al.}(2024)\citenamefont {Xiao}, \citenamefont {Berg}, \citenamefont {Glazman}, \citenamefont {Guinea}, \citenamefont {Ilani},\ and\ \citenamefont {von Oppen}}]{xiao2024theory}%
  \BibitemOpen
  \bibfield  {author} {\bibinfo {author} {\bibfnamefont {J.}~\bibnamefont {Xiao}}, \bibinfo {author} {\bibfnamefont {E.}~\bibnamefont {Berg}}, \bibinfo {author} {\bibfnamefont {L.~I.}\ \bibnamefont {Glazman}}, \bibinfo {author} {\bibfnamefont {F.}~\bibnamefont {Guinea}}, \bibinfo {author} {\bibfnamefont {S.}~\bibnamefont {Ilani}},\ and\ \bibinfo {author} {\bibfnamefont {F.}~\bibnamefont {von Oppen}},\ }\bibfield  {title} {\bibinfo {title} {Theory of phonon spectroscopy with the quantum twisting microscope},\ }\href {https://doi.org/10.1103/PhysRevB.110.205407} {\bibfield  {journal} {\bibinfo  {journal} {Phys. Rev. B}\ }\textbf {\bibinfo {volume} {110}},\ \bibinfo {pages} {205407} (\bibinfo {year} {2024})}\BibitemShut {NoStop}%
\bibitem [{\citenamefont {Wei}\ \emph {et~al.}(2025{\natexlab{a}})\citenamefont {Wei}, \citenamefont {von Oppen},\ and\ \citenamefont {Glazman}}]{wei2025dirac}%
  \BibitemOpen
  \bibfield  {author} {\bibinfo {author} {\bibfnamefont {N.}~\bibnamefont {Wei}}, \bibinfo {author} {\bibfnamefont {F.}~\bibnamefont {von Oppen}},\ and\ \bibinfo {author} {\bibfnamefont {L.~I.}\ \bibnamefont {Glazman}},\ }\bibfield  {title} {\bibinfo {title} {Dirac-point spectroscopy of flat-band systems with the quantum twisting microscope},\ }\href {https://doi.org/10.1103/PhysRevB.111.085128} {\bibfield  {journal} {\bibinfo  {journal} {Phys. Rev. B}\ }\textbf {\bibinfo {volume} {111}},\ \bibinfo {pages} {085128} (\bibinfo {year} {2025}{\natexlab{a}})}\BibitemShut {NoStop}%
\bibitem [{\citenamefont {Xiao}\ \emph {et~al.}(2025)\citenamefont {Xiao}, \citenamefont {Inbar}, \citenamefont {Birkbeck}, \citenamefont {Gershon}, \citenamefont {Zamir}, \citenamefont {Taniguchi}, \citenamefont {Watanabe}, \citenamefont {Berg},\ and\ \citenamefont {Ilani}}]{xiao2025interacting}%
  \BibitemOpen
  \bibfield  {author} {\bibinfo {author} {\bibfnamefont {J.}~\bibnamefont {Xiao}}, \bibinfo {author} {\bibfnamefont {A.}~\bibnamefont {Inbar}}, \bibinfo {author} {\bibfnamefont {J.}~\bibnamefont {Birkbeck}}, \bibinfo {author} {\bibfnamefont {N.}~\bibnamefont {Gershon}}, \bibinfo {author} {\bibfnamefont {Y.}~\bibnamefont {Zamir}}, \bibinfo {author} {\bibfnamefont {T.}~\bibnamefont {Taniguchi}}, \bibinfo {author} {\bibfnamefont {K.}~\bibnamefont {Watanabe}}, \bibinfo {author} {\bibfnamefont {E.}~\bibnamefont {Berg}},\ and\ \bibinfo {author} {\bibfnamefont {S.}~\bibnamefont {Ilani}},\ }\href {https://arxiv.org/abs/2506.20738} {\bibinfo {title} {The interacting energy bands of magic angle twisted bilayer graphene revealed by the quantum twisting microscope}} (\bibinfo {year} {2025}),\ \Eprint {https://arxiv.org/abs/2506.20738} {arXiv:2506.20738 [cond-mat.mes-hall]} \BibitemShut {NoStop}%
\bibitem [{\citenamefont {Wei}\ \emph {et~al.}(2025{\natexlab{b}})\citenamefont {Wei}, \citenamefont {Guinea}, \citenamefont {von Oppen},\ and\ \citenamefont {Glazman}}]{wei2025theory}%
  \BibitemOpen
  \bibfield  {author} {\bibinfo {author} {\bibfnamefont {N.}~\bibnamefont {Wei}}, \bibinfo {author} {\bibfnamefont {F.}~\bibnamefont {Guinea}}, \bibinfo {author} {\bibfnamefont {F.}~\bibnamefont {von Oppen}},\ and\ \bibinfo {author} {\bibfnamefont {L.~I.}\ \bibnamefont {Glazman}},\ }\href {https://arxiv.org/abs/2506.05485} {\bibinfo {title} {Theory of plasmon spectroscopy with the quantum twisting microscope}} (\bibinfo {year} {2025}{\natexlab{b}}),\ \Eprint {https://arxiv.org/abs/2506.05485} {arXiv:2506.05485 [cond-mat.str-el]} \BibitemShut {NoStop}%
\bibitem [{\citenamefont {Jang}\ \emph {et~al.}(2017)\citenamefont {Jang}, \citenamefont {Yoo}, \citenamefont {Pfeiffer}, \citenamefont {West}, \citenamefont {Baldwin},\ and\ \citenamefont {Ashoori}}]{jang2017full}%
  \BibitemOpen
  \bibfield  {author} {\bibinfo {author} {\bibfnamefont {J.}~\bibnamefont {Jang}}, \bibinfo {author} {\bibfnamefont {H.~M.}\ \bibnamefont {Yoo}}, \bibinfo {author} {\bibfnamefont {L.}~\bibnamefont {Pfeiffer}}, \bibinfo {author} {\bibfnamefont {K.}~\bibnamefont {West}}, \bibinfo {author} {\bibfnamefont {K.}~\bibnamefont {Baldwin}},\ and\ \bibinfo {author} {\bibfnamefont {R.~C.}\ \bibnamefont {Ashoori}},\ }\bibfield  {title} {\bibinfo {title} {Full momentum-and energy-resolved spectral function of a 2d electronic system},\ }\href {https://www.science.org/doi/full/10.1126/science.aam7073} {\bibfield  {journal} {\bibinfo  {journal} {Science}\ }\textbf {\bibinfo {volume} {358}},\ \bibinfo {pages} {901} (\bibinfo {year} {2017})}\BibitemShut {NoStop}%
\bibitem [{\citenamefont {Xiao}\ \emph {et~al.}(2023)\citenamefont {Xiao}, \citenamefont {Vituri},\ and\ \citenamefont {Berg}}]{xiao2023probing}%
  \BibitemOpen
  \bibfield  {author} {\bibinfo {author} {\bibfnamefont {J.}~\bibnamefont {Xiao}}, \bibinfo {author} {\bibfnamefont {Y.}~\bibnamefont {Vituri}},\ and\ \bibinfo {author} {\bibfnamefont {E.}~\bibnamefont {Berg}},\ }\bibfield  {title} {\bibinfo {title} {Probing the order parameter symmetry of two-dimensional superconductors by twisted josephson interferometry},\ }\href {https://doi.org/10.1103/PhysRevB.108.094520} {\bibfield  {journal} {\bibinfo  {journal} {Phys. Rev. B}\ }\textbf {\bibinfo {volume} {108}},\ \bibinfo {pages} {094520} (\bibinfo {year} {2023})}\BibitemShut {NoStop}%
\bibitem [{\citenamefont {Eisenstein}\ \emph {et~al.}(1991)\citenamefont {Eisenstein}, \citenamefont {Gramila}, \citenamefont {Pfeiffer},\ and\ \citenamefont {West}}]{eisenstein1991probing}%
  \BibitemOpen
  \bibfield  {author} {\bibinfo {author} {\bibfnamefont {J.~P.}\ \bibnamefont {Eisenstein}}, \bibinfo {author} {\bibfnamefont {T.~J.}\ \bibnamefont {Gramila}}, \bibinfo {author} {\bibfnamefont {L.~N.}\ \bibnamefont {Pfeiffer}},\ and\ \bibinfo {author} {\bibfnamefont {K.~W.}\ \bibnamefont {West}},\ }\bibfield  {title} {\bibinfo {title} {Probing a two-dimensional fermi surface by tunneling},\ }\href {https://doi.org/10.1103/PhysRevB.44.6511} {\bibfield  {journal} {\bibinfo  {journal} {Phys. Rev. B}\ }\textbf {\bibinfo {volume} {44}},\ \bibinfo {pages} {6511} (\bibinfo {year} {1991})}\BibitemShut {NoStop}%
\bibitem [{\citenamefont {Mahan}(2013)}]{mahan2013many}%
  \BibitemOpen
  \bibfield  {author} {\bibinfo {author} {\bibfnamefont {G.~D.}\ \bibnamefont {Mahan}},\ }\href@noop {} {\emph {\bibinfo {title} {Many-particle physics}}}\ (\bibinfo  {publisher} {Springer Science \& Business Media},\ \bibinfo {year} {2013})\BibitemShut {NoStop}%
\bibitem [{\citenamefont {Huang}\ \emph {et~al.}(2022)\citenamefont {Huang}, \citenamefont {Wei}, \citenamefont {Qin},\ and\ \citenamefont {MacDonald}}]{huang2022pseudospin}%
  \BibitemOpen
  \bibfield  {author} {\bibinfo {author} {\bibfnamefont {C.}~\bibnamefont {Huang}}, \bibinfo {author} {\bibfnamefont {N.}~\bibnamefont {Wei}}, \bibinfo {author} {\bibfnamefont {W.}~\bibnamefont {Qin}},\ and\ \bibinfo {author} {\bibfnamefont {A.~H.}\ \bibnamefont {MacDonald}},\ }\bibfield  {title} {\bibinfo {title} {Pseudospin paramagnons and the superconducting dome in magic angle twisted bilayer graphene},\ }\href {https://doi.org/10.1103/PhysRevLett.129.187001} {\bibfield  {journal} {\bibinfo  {journal} {Phys. Rev. Lett.}\ }\textbf {\bibinfo {volume} {129}},\ \bibinfo {pages} {187001} (\bibinfo {year} {2022})}\BibitemShut {NoStop}%
\bibitem [{\citenamefont {Qin}\ and\ \citenamefont {MacDonald}(2021)}]{qin2021inplane}%
  \BibitemOpen
  \bibfield  {author} {\bibinfo {author} {\bibfnamefont {W.}~\bibnamefont {Qin}}\ and\ \bibinfo {author} {\bibfnamefont {A.~H.}\ \bibnamefont {MacDonald}},\ }\bibfield  {title} {\bibinfo {title} {In-plane critical magnetic fields in magic-angle twisted trilayer graphene},\ }\href {https://doi.org/10.1103/PhysRevLett.127.097001} {\bibfield  {journal} {\bibinfo  {journal} {Phys. Rev. Lett.}\ }\textbf {\bibinfo {volume} {127}},\ \bibinfo {pages} {097001} (\bibinfo {year} {2021})}\BibitemShut {NoStop}%
\bibitem [{\citenamefont {Dong}\ \emph {et~al.}(2024)\citenamefont {Dong}, \citenamefont {Étienne Lantagne-Hurtubise},\ and\ \citenamefont {Alicea}}]{dong2024superconductivity}%
  \BibitemOpen
  \bibfield  {author} {\bibinfo {author} {\bibfnamefont {Z.}~\bibnamefont {Dong}}, \bibinfo {author} {\bibnamefont {Étienne Lantagne-Hurtubise}},\ and\ \bibinfo {author} {\bibfnamefont {J.}~\bibnamefont {Alicea}},\ }\href {https://arxiv.org/abs/2406.17036} {\bibinfo {title} {Superconductivity from spin-canting fluctuations in rhombohedral graphene}} (\bibinfo {year} {2024}),\ \Eprint {https://arxiv.org/abs/2406.17036} {arXiv:2406.17036 [cond-mat.supr-con]} \BibitemShut {NoStop}%
\bibitem [{\citenamefont {Dong}\ \emph {et~al.}(2023)\citenamefont {Dong}, \citenamefont {Chubukov},\ and\ \citenamefont {Levitov}}]{dong2023transformer}%
  \BibitemOpen
  \bibfield  {author} {\bibinfo {author} {\bibfnamefont {Z.}~\bibnamefont {Dong}}, \bibinfo {author} {\bibfnamefont {A.~V.}\ \bibnamefont {Chubukov}},\ and\ \bibinfo {author} {\bibfnamefont {L.}~\bibnamefont {Levitov}},\ }\bibfield  {title} {\bibinfo {title} {Transformer spin-triplet superconductivity at the onset of isospin order in bilayer graphene},\ }\href {https://doi.org/10.1103/PhysRevB.107.174512} {\bibfield  {journal} {\bibinfo  {journal} {Phys. Rev. B}\ }\textbf {\bibinfo {volume} {107}},\ \bibinfo {pages} {174512} (\bibinfo {year} {2023})}\BibitemShut {NoStop}%
\bibitem [{Note1()}]{Note1}%
  \BibitemOpen
  \bibinfo {note} {Figure~\ref {fig:schematic}(b) contains two intersections between the Fermi lines of tip and sample. We focus on one of them.}\BibitemShut {Stop}%
\bibitem [{\citenamefont {Zheng}\ and\ \citenamefont {MacDonald}(1993)}]{zheng1993tunneling}%
  \BibitemOpen
  \bibfield  {author} {\bibinfo {author} {\bibfnamefont {L.}~\bibnamefont {Zheng}}\ and\ \bibinfo {author} {\bibfnamefont {A.~H.}\ \bibnamefont {MacDonald}},\ }\bibfield  {title} {\bibinfo {title} {Tunneling conductance between parallel two-dimensional electron systems},\ }\href {https://doi.org/10.1103/PhysRevB.47.10619} {\bibfield  {journal} {\bibinfo  {journal} {Phys. Rev. B}\ }\textbf {\bibinfo {volume} {47}},\ \bibinfo {pages} {10619} (\bibinfo {year} {1993})}\BibitemShut {NoStop}%
\bibitem [{Note2()}]{Note2}%
  \BibitemOpen
  \bibinfo {note} {It does not matter whether the intersections between the Fermi lines of the tip and sample are chosen at zero bias, or at $-eV=|\Delta _{\protect \bm {p}^*}|$ or $-|\Delta _{\protect \bm {p}^*}|$. Using the other two definitions changes $\protect \bm {p}^*$ by $|\delta \protect \bm {p}^*|\sim |\Delta _{\protect \bm {p}^*}|/\hbar |\protect \bm {v}_{\protect \bm {p}}^T|$ and changes $|\Delta _{\protect \bm {p}^*}|$ by $\delta \Delta _{\protect \bm {p}^*}\sim |\Delta _{\protect \bm {p}^*}|\partial _{\protect \bm {p}^*}|\Delta _{\protect \bm {p}^*}|/\hbar v_{\protect \bm {p}^*}^{T}\ll |\Delta _{\protect \bm {p}^*}|$.}\BibitemShut {Stop}%
\bibitem [{\citenamefont {Dynes}\ \emph {et~al.}(1978)\citenamefont {Dynes}, \citenamefont {Narayanamurti},\ and\ \citenamefont {Garno}}]{dynes1978direct}%
  \BibitemOpen
  \bibfield  {author} {\bibinfo {author} {\bibfnamefont {R.~C.}\ \bibnamefont {Dynes}}, \bibinfo {author} {\bibfnamefont {V.}~\bibnamefont {Narayanamurti}},\ and\ \bibinfo {author} {\bibfnamefont {J.~P.}\ \bibnamefont {Garno}},\ }\bibfield  {title} {\bibinfo {title} {Direct measurement of quasiparticle-lifetime broadening in a strong-coupled superconductor},\ }\href {https://doi.org/10.1103/PhysRevLett.41.1509} {\bibfield  {journal} {\bibinfo  {journal} {Phys. Rev. Lett.}\ }\textbf {\bibinfo {volume} {41}},\ \bibinfo {pages} {1509} (\bibinfo {year} {1978})}\BibitemShut {NoStop}%
\bibitem [{\citenamefont {Wu}\ \emph {et~al.}(2018)\citenamefont {Wu}, \citenamefont {MacDonald},\ and\ \citenamefont {Martin}}]{wu2018theory}%
  \BibitemOpen
  \bibfield  {author} {\bibinfo {author} {\bibfnamefont {F.}~\bibnamefont {Wu}}, \bibinfo {author} {\bibfnamefont {A.~H.}\ \bibnamefont {MacDonald}},\ and\ \bibinfo {author} {\bibfnamefont {I.}~\bibnamefont {Martin}},\ }\bibfield  {title} {\bibinfo {title} {Theory of phonon-mediated superconductivity in twisted bilayer graphene},\ }\href {https://doi.org/10.1103/PhysRevLett.121.257001} {\bibfield  {journal} {\bibinfo  {journal} {Phys. Rev. Lett.}\ }\textbf {\bibinfo {volume} {121}},\ \bibinfo {pages} {257001} (\bibinfo {year} {2018})}\BibitemShut {NoStop}%
\bibitem [{\citenamefont {Wu}(2019)}]{wu2019topological}%
  \BibitemOpen
  \bibfield  {author} {\bibinfo {author} {\bibfnamefont {F.}~\bibnamefont {Wu}},\ }\bibfield  {title} {\bibinfo {title} {Topological chiral superconductivity with spontaneous vortices and supercurrent in twisted bilayer graphene},\ }\href {https://doi.org/10.1103/PhysRevB.99.195114} {\bibfield  {journal} {\bibinfo  {journal} {Phys. Rev. B}\ }\textbf {\bibinfo {volume} {99}},\ \bibinfo {pages} {195114} (\bibinfo {year} {2019})}\BibitemShut {NoStop}%
\bibitem [{\citenamefont {Kozii}\ \emph {et~al.}(2019)\citenamefont {Kozii}, \citenamefont {Isobe}, \citenamefont {Venderbos},\ and\ \citenamefont {Fu}}]{kozii2019nematic}%
  \BibitemOpen
  \bibfield  {author} {\bibinfo {author} {\bibfnamefont {V.}~\bibnamefont {Kozii}}, \bibinfo {author} {\bibfnamefont {H.}~\bibnamefont {Isobe}}, \bibinfo {author} {\bibfnamefont {J.~W.~F.}\ \bibnamefont {Venderbos}},\ and\ \bibinfo {author} {\bibfnamefont {L.}~\bibnamefont {Fu}},\ }\bibfield  {title} {\bibinfo {title} {Nematic superconductivity stabilized by density wave fluctuations: Possible application to twisted bilayer graphene},\ }\href {https://doi.org/10.1103/PhysRevB.99.144507} {\bibfield  {journal} {\bibinfo  {journal} {Phys. Rev. B}\ }\textbf {\bibinfo {volume} {99}},\ \bibinfo {pages} {144507} (\bibinfo {year} {2019})}\BibitemShut {NoStop}%
\bibitem [{\citenamefont {Lake}\ \emph {et~al.}(2022)\citenamefont {Lake}, \citenamefont {Patri},\ and\ \citenamefont {Senthil}}]{lake2022pairing}%
  \BibitemOpen
  \bibfield  {author} {\bibinfo {author} {\bibfnamefont {E.}~\bibnamefont {Lake}}, \bibinfo {author} {\bibfnamefont {A.~S.}\ \bibnamefont {Patri}},\ and\ \bibinfo {author} {\bibfnamefont {T.}~\bibnamefont {Senthil}},\ }\bibfield  {title} {\bibinfo {title} {Pairing symmetry of twisted bilayer graphene: A phenomenological synthesis},\ }\href {https://doi.org/10.1103/PhysRevB.106.104506} {\bibfield  {journal} {\bibinfo  {journal} {Phys. Rev. B}\ }\textbf {\bibinfo {volume} {106}},\ \bibinfo {pages} {104506} (\bibinfo {year} {2022})}\BibitemShut {NoStop}%
\bibitem [{\citenamefont {L{\"o}thman}\ \emph {et~al.}(2022)\citenamefont {L{\"o}thman}, \citenamefont {Schmidt}, \citenamefont {Parhizgar},\ and\ \citenamefont {Black-Schaffer}}]{lothman2022nematic}%
  \BibitemOpen
  \bibfield  {author} {\bibinfo {author} {\bibfnamefont {T.}~\bibnamefont {L{\"o}thman}}, \bibinfo {author} {\bibfnamefont {J.}~\bibnamefont {Schmidt}}, \bibinfo {author} {\bibfnamefont {F.}~\bibnamefont {Parhizgar}},\ and\ \bibinfo {author} {\bibfnamefont {A.~M.}\ \bibnamefont {Black-Schaffer}},\ }\bibfield  {title} {\bibinfo {title} {Nematic superconductivity in magic-angle twisted bilayer graphene from atomistic modeling},\ }\href {https://www.nature.com/articles/s42005-022-00860-z} {\bibfield  {journal} {\bibinfo  {journal} {Communications Physics}\ }\textbf {\bibinfo {volume} {5}},\ \bibinfo {pages} {92} (\bibinfo {year} {2022})}\BibitemShut {NoStop}%
\bibitem [{\citenamefont {Yu}\ \emph {et~al.}(2023)\citenamefont {Yu}, \citenamefont {Xie}, \citenamefont {Wu},\ and\ \citenamefont {Das~Sarma}}]{yu2023euler}%
  \BibitemOpen
  \bibfield  {author} {\bibinfo {author} {\bibfnamefont {J.}~\bibnamefont {Yu}}, \bibinfo {author} {\bibfnamefont {M.}~\bibnamefont {Xie}}, \bibinfo {author} {\bibfnamefont {F.}~\bibnamefont {Wu}},\ and\ \bibinfo {author} {\bibfnamefont {S.}~\bibnamefont {Das~Sarma}},\ }\bibfield  {title} {\bibinfo {title} {Euler-obstructed nematic nodal superconductivity in twisted bilayer graphene},\ }\href {https://doi.org/10.1103/PhysRevB.107.L201106} {\bibfield  {journal} {\bibinfo  {journal} {Phys. Rev. B}\ }\textbf {\bibinfo {volume} {107}},\ \bibinfo {pages} {L201106} (\bibinfo {year} {2023})}\BibitemShut {NoStop}%
\bibitem [{\citenamefont {Liu}\ \emph {et~al.}(2024)\citenamefont {Liu}, \citenamefont {Chen}, \citenamefont {Yazdani},\ and\ \citenamefont {Bernevig}}]{liu2024electron}%
  \BibitemOpen
  \bibfield  {author} {\bibinfo {author} {\bibfnamefont {C.-X.}\ \bibnamefont {Liu}}, \bibinfo {author} {\bibfnamefont {Y.}~\bibnamefont {Chen}}, \bibinfo {author} {\bibfnamefont {A.}~\bibnamefont {Yazdani}},\ and\ \bibinfo {author} {\bibfnamefont {B.~A.}\ \bibnamefont {Bernevig}},\ }\bibfield  {title} {\bibinfo {title} {Electron--$k$-phonon interaction in twisted bilayer graphene},\ }\href {https://doi.org/10.1103/PhysRevB.110.045133} {\bibfield  {journal} {\bibinfo  {journal} {Phys. Rev. B}\ }\textbf {\bibinfo {volume} {110}},\ \bibinfo {pages} {045133} (\bibinfo {year} {2024})}\BibitemShut {NoStop}%
\bibitem [{\citenamefont {Wang}\ \emph {et~al.}(2024)\citenamefont {Wang}, \citenamefont {Zhou}, \citenamefont {Peng}, \citenamefont {Lian},\ and\ \citenamefont {Song}}]{wang2024molecular}%
  \BibitemOpen
  \bibfield  {author} {\bibinfo {author} {\bibfnamefont {Y.-J.}\ \bibnamefont {Wang}}, \bibinfo {author} {\bibfnamefont {G.-D.}\ \bibnamefont {Zhou}}, \bibinfo {author} {\bibfnamefont {S.-Y.}\ \bibnamefont {Peng}}, \bibinfo {author} {\bibfnamefont {B.}~\bibnamefont {Lian}},\ and\ \bibinfo {author} {\bibfnamefont {Z.-D.}\ \bibnamefont {Song}},\ }\bibfield  {title} {\bibinfo {title} {Molecular pairing in twisted bilayer graphene superconductivity},\ }\href {https://doi.org/10.1103/PhysRevLett.133.146001} {\bibfield  {journal} {\bibinfo  {journal} {Phys. Rev. Lett.}\ }\textbf {\bibinfo {volume} {133}},\ \bibinfo {pages} {146001} (\bibinfo {year} {2024})}\BibitemShut {NoStop}%
\bibitem [{\citenamefont {Sengottaiyan}\ \emph {et~al.}(2024)\citenamefont {Sengottaiyan}, \citenamefont {Hara}, \citenamefont {Nagata}, \citenamefont {Mitsuboshi}, \citenamefont {Jeganathan},\ and\ \citenamefont {Yoshimura}}]{sengottaiyan2024large}%
  \BibitemOpen
  \bibfield  {author} {\bibinfo {author} {\bibfnamefont {C.}~\bibnamefont {Sengottaiyan}}, \bibinfo {author} {\bibfnamefont {M.}~\bibnamefont {Hara}}, \bibinfo {author} {\bibfnamefont {H.}~\bibnamefont {Nagata}}, \bibinfo {author} {\bibfnamefont {H.}~\bibnamefont {Mitsuboshi}}, \bibinfo {author} {\bibfnamefont {C.}~\bibnamefont {Jeganathan}},\ and\ \bibinfo {author} {\bibfnamefont {M.}~\bibnamefont {Yoshimura}},\ }\bibfield  {title} {\bibinfo {title} {Large-area synthesis and fabrication of few-layer hbn/monolayer rgo heterostructures for enhanced contact surface potential},\ }\href {https://pubs.acs.org/doi/full/10.1021/acsomega.4c02219} {\bibfield  {journal} {\bibinfo  {journal} {ACS omega}\ }\textbf {\bibinfo {volume} {9}},\ \bibinfo {pages} {26307} (\bibinfo {year} {2024})}\BibitemShut {NoStop}%
\bibitem [{\citenamefont {Yankowitz}\ \emph {et~al.}(2019)\citenamefont {Yankowitz}, \citenamefont {Chen}, \citenamefont {Polshyn}, \citenamefont {Zhang}, \citenamefont {Watanabe}, \citenamefont {Taniguchi}, \citenamefont {Graf}, \citenamefont {Young},\ and\ \citenamefont {Dean}}]{yankowitz2019tuning}%
  \BibitemOpen
  \bibfield  {author} {\bibinfo {author} {\bibfnamefont {M.}~\bibnamefont {Yankowitz}}, \bibinfo {author} {\bibfnamefont {S.}~\bibnamefont {Chen}}, \bibinfo {author} {\bibfnamefont {H.}~\bibnamefont {Polshyn}}, \bibinfo {author} {\bibfnamefont {Y.}~\bibnamefont {Zhang}}, \bibinfo {author} {\bibfnamefont {K.}~\bibnamefont {Watanabe}}, \bibinfo {author} {\bibfnamefont {T.}~\bibnamefont {Taniguchi}}, \bibinfo {author} {\bibfnamefont {D.}~\bibnamefont {Graf}}, \bibinfo {author} {\bibfnamefont {A.~F.}\ \bibnamefont {Young}},\ and\ \bibinfo {author} {\bibfnamefont {C.~R.}\ \bibnamefont {Dean}},\ }\bibfield  {title} {\bibinfo {title} {Tuning superconductivity in twisted bilayer graphene},\ }\href {https://www.science.org/doi/10.1126/science.aav1910} {\bibfield  {journal} {\bibinfo  {journal} {Science}\ }\textbf {\bibinfo {volume} {363}},\ \bibinfo {pages} {1059} (\bibinfo {year} {2019})}\BibitemShut {NoStop}%
\bibitem [{\citenamefont {Zhang}\ \emph {et~al.}(2023)\citenamefont {Zhang}, \citenamefont {Polski}, \citenamefont {Thomson}, \citenamefont {Lantagne-Hurtubise}, \citenamefont {Lewandowski}, \citenamefont {Zhou}, \citenamefont {Watanabe}, \citenamefont {Taniguchi}, \citenamefont {Alicea},\ and\ \citenamefont {Nadj-Perge}}]{zhang2023enhanced}%
  \BibitemOpen
  \bibfield  {author} {\bibinfo {author} {\bibfnamefont {Y.}~\bibnamefont {Zhang}}, \bibinfo {author} {\bibfnamefont {R.}~\bibnamefont {Polski}}, \bibinfo {author} {\bibfnamefont {A.}~\bibnamefont {Thomson}}, \bibinfo {author} {\bibfnamefont {{\'E}.}~\bibnamefont {Lantagne-Hurtubise}}, \bibinfo {author} {\bibfnamefont {C.}~\bibnamefont {Lewandowski}}, \bibinfo {author} {\bibfnamefont {H.}~\bibnamefont {Zhou}}, \bibinfo {author} {\bibfnamefont {K.}~\bibnamefont {Watanabe}}, \bibinfo {author} {\bibfnamefont {T.}~\bibnamefont {Taniguchi}}, \bibinfo {author} {\bibfnamefont {J.}~\bibnamefont {Alicea}},\ and\ \bibinfo {author} {\bibfnamefont {S.}~\bibnamefont {Nadj-Perge}},\ }\bibfield  {title} {\bibinfo {title} {Enhanced superconductivity in spin--orbit proximitized bilayer graphene},\ }\href {https://www.nature.com/articles/s41586-022-05446-x} {\bibfield  {journal} {\bibinfo  {journal} {Nature}\ }\textbf {\bibinfo {volume} {613}},\ \bibinfo {pages} {268} (\bibinfo {year} {2023})}\BibitemShut {NoStop}%
\bibitem [{\citenamefont {Zhang}\ \emph {et~al.}(2025)\citenamefont {Zhang}, \citenamefont {Shavit}, \citenamefont {Ma}, \citenamefont {Han}, \citenamefont {Siu}, \citenamefont {Mukherjee}, \citenamefont {Watanabe}, \citenamefont {Taniguchi}, \citenamefont {Hsieh}, \citenamefont {Lewandowski} \emph {et~al.}}]{zhang2025twist}%
  \BibitemOpen
  \bibfield  {author} {\bibinfo {author} {\bibfnamefont {Y.}~\bibnamefont {Zhang}}, \bibinfo {author} {\bibfnamefont {G.}~\bibnamefont {Shavit}}, \bibinfo {author} {\bibfnamefont {H.}~\bibnamefont {Ma}}, \bibinfo {author} {\bibfnamefont {Y.}~\bibnamefont {Han}}, \bibinfo {author} {\bibfnamefont {C.~W.}\ \bibnamefont {Siu}}, \bibinfo {author} {\bibfnamefont {A.}~\bibnamefont {Mukherjee}}, \bibinfo {author} {\bibfnamefont {K.}~\bibnamefont {Watanabe}}, \bibinfo {author} {\bibfnamefont {T.}~\bibnamefont {Taniguchi}}, \bibinfo {author} {\bibfnamefont {D.}~\bibnamefont {Hsieh}}, \bibinfo {author} {\bibfnamefont {C.}~\bibnamefont {Lewandowski}}, \emph {et~al.},\ }\bibfield  {title} {\bibinfo {title} {Twist-programmable superconductivity in spin--orbit-coupled bilayer graphene},\ }\href {https://www.nature.com/articles/s41586-025-08959-3} {\bibfield  {journal} {\bibinfo  {journal} {Nature}\ ,\ \bibinfo {pages} {1}} (\bibinfo {year} {2025})}\BibitemShut {NoStop}%
\bibitem [{\citenamefont {Yang}\ \emph {et~al.}(2025)\citenamefont {Yang}, \citenamefont {Shi}, \citenamefont {Ye}, \citenamefont {Yoon}, \citenamefont {Lu}, \citenamefont {Kakani}, \citenamefont {Han}, \citenamefont {Seo}, \citenamefont {Shi}, \citenamefont {Watanabe} \emph {et~al.}}]{yang2025impact}%
  \BibitemOpen
  \bibfield  {author} {\bibinfo {author} {\bibfnamefont {J.}~\bibnamefont {Yang}}, \bibinfo {author} {\bibfnamefont {X.}~\bibnamefont {Shi}}, \bibinfo {author} {\bibfnamefont {S.}~\bibnamefont {Ye}}, \bibinfo {author} {\bibfnamefont {C.}~\bibnamefont {Yoon}}, \bibinfo {author} {\bibfnamefont {Z.}~\bibnamefont {Lu}}, \bibinfo {author} {\bibfnamefont {V.}~\bibnamefont {Kakani}}, \bibinfo {author} {\bibfnamefont {T.}~\bibnamefont {Han}}, \bibinfo {author} {\bibfnamefont {J.}~\bibnamefont {Seo}}, \bibinfo {author} {\bibfnamefont {L.}~\bibnamefont {Shi}}, \bibinfo {author} {\bibfnamefont {K.}~\bibnamefont {Watanabe}}, \emph {et~al.},\ }\bibfield  {title} {\bibinfo {title} {Impact of spin--orbit coupling on superconductivity in rhombohedral graphene},\ }\href {https://www.nature.com/articles/s41563-025-02156-3} {\bibfield  {journal} {\bibinfo  {journal} {Nature Materials}\ ,\ \bibinfo {pages} {1}} (\bibinfo {year} {2025})}\BibitemShut {NoStop}%
\bibitem [{\citenamefont {Christos}\ \emph {et~al.}(2023)\citenamefont {Christos}, \citenamefont {Sachdev},\ and\ \citenamefont {Scheurer}}]{christos2023nodal}%
  \BibitemOpen
  \bibfield  {author} {\bibinfo {author} {\bibfnamefont {M.}~\bibnamefont {Christos}}, \bibinfo {author} {\bibfnamefont {S.}~\bibnamefont {Sachdev}},\ and\ \bibinfo {author} {\bibfnamefont {M.~S.}\ \bibnamefont {Scheurer}},\ }\bibfield  {title} {\bibinfo {title} {Nodal band-off-diagonal superconductivity in twisted graphene superlattices},\ }\href {https://www.nature.com/articles/s41467-023-42471-4} {\bibfield  {journal} {\bibinfo  {journal} {Nature Communications}\ }\textbf {\bibinfo {volume} {14}},\ \bibinfo {pages} {7134} (\bibinfo {year} {2023})}\BibitemShut {NoStop}%
\bibitem [{\citenamefont {Putzer}\ and\ \citenamefont {Scheurer}(2025)}]{Putzer2025eliashberg}%
  \BibitemOpen
  \bibfield  {author} {\bibinfo {author} {\bibfnamefont {B.}~\bibnamefont {Putzer}}\ and\ \bibinfo {author} {\bibfnamefont {M.~S.}\ \bibnamefont {Scheurer}},\ }\bibfield  {title} {\bibinfo {title} {Eliashberg theory and superfluid stiffness of band-off-diagonal pairing in twisted graphene},\ }\href {https://doi.org/10.1103/PhysRevB.111.144513} {\bibfield  {journal} {\bibinfo  {journal} {Phys. Rev. B}\ }\textbf {\bibinfo {volume} {111}},\ \bibinfo {pages} {144513} (\bibinfo {year} {2025})}\BibitemShut {NoStop}%
\bibitem [{\citenamefont {Chou}\ \emph {et~al.}(2024)\citenamefont {Chou}, \citenamefont {Tan}, \citenamefont {Wu},\ and\ \citenamefont {Das~Sarma}}]{chou2024topological}%
  \BibitemOpen
  \bibfield  {author} {\bibinfo {author} {\bibfnamefont {Y.-Z.}\ \bibnamefont {Chou}}, \bibinfo {author} {\bibfnamefont {Y.}~\bibnamefont {Tan}}, \bibinfo {author} {\bibfnamefont {F.}~\bibnamefont {Wu}},\ and\ \bibinfo {author} {\bibfnamefont {S.}~\bibnamefont {Das~Sarma}},\ }\bibfield  {title} {\bibinfo {title} {Topological flat bands, valley polarization, and interband superconductivity in magic-angle twisted bilayer graphene with proximitized spin-orbit couplings},\ }\href {https://doi.org/10.1103/PhysRevB.110.L041108} {\bibfield  {journal} {\bibinfo  {journal} {Phys. Rev. B}\ }\textbf {\bibinfo {volume} {110}},\ \bibinfo {pages} {L041108} (\bibinfo {year} {2024})}\BibitemShut {NoStop}%
\bibitem [{\citenamefont {Schrieffer}\ \emph {et~al.}(1963)\citenamefont {Schrieffer}, \citenamefont {Scalapino},\ and\ \citenamefont {Wilkins}}]{schrieffer1963effective}%
  \BibitemOpen
  \bibfield  {author} {\bibinfo {author} {\bibfnamefont {J.~R.}\ \bibnamefont {Schrieffer}}, \bibinfo {author} {\bibfnamefont {D.~J.}\ \bibnamefont {Scalapino}},\ and\ \bibinfo {author} {\bibfnamefont {J.~W.}\ \bibnamefont {Wilkins}},\ }\bibfield  {title} {\bibinfo {title} {Effective tunneling density of states in superconductors},\ }\href {https://doi.org/10.1103/PhysRevLett.10.336} {\bibfield  {journal} {\bibinfo  {journal} {Phys. Rev. Lett.}\ }\textbf {\bibinfo {volume} {10}},\ \bibinfo {pages} {336} (\bibinfo {year} {1963})}\BibitemShut {NoStop}%
\bibitem [{\citenamefont {Scalapino}\ \emph {et~al.}(1966)\citenamefont {Scalapino}, \citenamefont {Schrieffer},\ and\ \citenamefont {Wilkins}}]{scalapino1966strong}%
  \BibitemOpen
  \bibfield  {author} {\bibinfo {author} {\bibfnamefont {D.~J.}\ \bibnamefont {Scalapino}}, \bibinfo {author} {\bibfnamefont {J.~R.}\ \bibnamefont {Schrieffer}},\ and\ \bibinfo {author} {\bibfnamefont {J.~W.}\ \bibnamefont {Wilkins}},\ }\bibfield  {title} {\bibinfo {title} {Strong-coupling superconductivity. i},\ }\href {https://doi.org/10.1103/PhysRev.148.263} {\bibfield  {journal} {\bibinfo  {journal} {Phys. Rev.}\ }\textbf {\bibinfo {volume} {148}},\ \bibinfo {pages} {263} (\bibinfo {year} {1966})}\BibitemShut {NoStop}%
\bibitem [{\citenamefont {Xia}\ \emph {et~al.}(2025)\citenamefont {Xia}, \citenamefont {Han}, \citenamefont {Watanabe}, \citenamefont {Taniguchi}, \citenamefont {Shan},\ and\ \citenamefont {Mak}}]{xia2025superconductivity}%
  \BibitemOpen
  \bibfield  {author} {\bibinfo {author} {\bibfnamefont {Y.}~\bibnamefont {Xia}}, \bibinfo {author} {\bibfnamefont {Z.}~\bibnamefont {Han}}, \bibinfo {author} {\bibfnamefont {K.}~\bibnamefont {Watanabe}}, \bibinfo {author} {\bibfnamefont {T.}~\bibnamefont {Taniguchi}}, \bibinfo {author} {\bibfnamefont {J.}~\bibnamefont {Shan}},\ and\ \bibinfo {author} {\bibfnamefont {K.~F.}\ \bibnamefont {Mak}},\ }\bibfield  {title} {\bibinfo {title} {Superconductivity in twisted bilayer wse2},\ }\href {https://www.nature.com/articles/s41586-024-08116-2} {\bibfield  {journal} {\bibinfo  {journal} {Nature}\ }\textbf {\bibinfo {volume} {637}},\ \bibinfo {pages} {833} (\bibinfo {year} {2025})}\BibitemShut {NoStop}%
\bibitem [{\citenamefont {Guo}\ \emph {et~al.}(2025)\citenamefont {Guo}, \citenamefont {Pack}, \citenamefont {Swann}, \citenamefont {Holtzman}, \citenamefont {Cothrine}, \citenamefont {Watanabe}, \citenamefont {Taniguchi}, \citenamefont {Mandrus}, \citenamefont {Barmak}, \citenamefont {Hone} \emph {et~al.}}]{guo2025superconductivity}%
  \BibitemOpen
  \bibfield  {author} {\bibinfo {author} {\bibfnamefont {Y.}~\bibnamefont {Guo}}, \bibinfo {author} {\bibfnamefont {J.}~\bibnamefont {Pack}}, \bibinfo {author} {\bibfnamefont {J.}~\bibnamefont {Swann}}, \bibinfo {author} {\bibfnamefont {L.}~\bibnamefont {Holtzman}}, \bibinfo {author} {\bibfnamefont {M.}~\bibnamefont {Cothrine}}, \bibinfo {author} {\bibfnamefont {K.}~\bibnamefont {Watanabe}}, \bibinfo {author} {\bibfnamefont {T.}~\bibnamefont {Taniguchi}}, \bibinfo {author} {\bibfnamefont {D.~G.}\ \bibnamefont {Mandrus}}, \bibinfo {author} {\bibfnamefont {K.}~\bibnamefont {Barmak}}, \bibinfo {author} {\bibfnamefont {J.}~\bibnamefont {Hone}}, \emph {et~al.},\ }\bibfield  {title} {\bibinfo {title} {Superconductivity in 5.0° twisted bilayer wse2},\ }\href {https://www.nature.com/articles/s41586-024-08381-1} {\bibfield  {journal} {\bibinfo  {journal} {Nature}\ }\textbf {\bibinfo {volume} {637}},\ \bibinfo {pages} {839} (\bibinfo {year} {2025})}\BibitemShut {NoStop}%
\bibitem [{\citenamefont {Yang}\ and\ \citenamefont {Zhang}(2024)}]{yang2024topological}%
  \BibitemOpen
  \bibfield  {author} {\bibinfo {author} {\bibfnamefont {H.}~\bibnamefont {Yang}}\ and\ \bibinfo {author} {\bibfnamefont {Y.-H.}\ \bibnamefont {Zhang}},\ }\href {https://arxiv.org/abs/2411.02503} {\bibinfo {title} {Topological incommensurate fulde-ferrell-larkin-ovchinnikov superconductor and bogoliubov fermi surface in rhombohedral tetra-layer graphene}} (\bibinfo {year} {2024}),\ \Eprint {https://arxiv.org/abs/2411.02503} {arXiv:2411.02503 [cond-mat.supr-con]} \BibitemShut {NoStop}%
\bibitem [{\citenamefont {Qin}\ and\ \citenamefont {Wu}(2024)}]{qin2024chiral}%
  \BibitemOpen
  \bibfield  {author} {\bibinfo {author} {\bibfnamefont {Q.}~\bibnamefont {Qin}}\ and\ \bibinfo {author} {\bibfnamefont {C.}~\bibnamefont {Wu}},\ }\href {https://arxiv.org/abs/2412.07145} {\bibinfo {title} {Chiral finite-momentum superconductivity in the tetralayer graphene}} (\bibinfo {year} {2024}),\ \Eprint {https://arxiv.org/abs/2412.07145} {arXiv:2412.07145 [cond-mat.supr-con]} \BibitemShut {NoStop}%
\bibitem [{\citenamefont {Christos}\ \emph {et~al.}(2025)\citenamefont {Christos}, \citenamefont {Bonetti},\ and\ \citenamefont {Scheurer}}]{christos2025finite}%
  \BibitemOpen
  \bibfield  {author} {\bibinfo {author} {\bibfnamefont {M.}~\bibnamefont {Christos}}, \bibinfo {author} {\bibfnamefont {P.~M.}\ \bibnamefont {Bonetti}},\ and\ \bibinfo {author} {\bibfnamefont {M.~S.}\ \bibnamefont {Scheurer}},\ }\href {https://arxiv.org/abs/2503.15471} {\bibinfo {title} {Finite-momentum pairing and superlattice superconductivity in valley-imbalanced rhombohedral graphene}} (\bibinfo {year} {2025}),\ \Eprint {https://arxiv.org/abs/2503.15471} {arXiv:2503.15471 [cond-mat.str-el]} \BibitemShut {NoStop}%
\bibitem [{\citenamefont {Gaggioli}\ \emph {et~al.}(2025)\citenamefont {Gaggioli}, \citenamefont {Guerci},\ and\ \citenamefont {Fu}}]{gaggioli2025spontaneous}%
  \BibitemOpen
  \bibfield  {author} {\bibinfo {author} {\bibfnamefont {F.}~\bibnamefont {Gaggioli}}, \bibinfo {author} {\bibfnamefont {D.}~\bibnamefont {Guerci}},\ and\ \bibinfo {author} {\bibfnamefont {L.}~\bibnamefont {Fu}},\ }\href {https://arxiv.org/abs/2503.16384} {\bibinfo {title} {Spontaneous vortex-antivortex lattice and majorana fermions in rhombohedral graphene}} (\bibinfo {year} {2025}),\ \Eprint {https://arxiv.org/abs/2503.16384} {arXiv:2503.16384 [cond-mat.supr-con]} \BibitemShut {NoStop}%
\bibitem [{\citenamefont {Gil}\ and\ \citenamefont {Berg}(2025)}]{gil2025charge}%
  \BibitemOpen
  \bibfield  {author} {\bibinfo {author} {\bibfnamefont {A.}~\bibnamefont {Gil}}\ and\ \bibinfo {author} {\bibfnamefont {E.}~\bibnamefont {Berg}},\ }\href {https://arxiv.org/abs/2504.19321} {\bibinfo {title} {Charge and pair density waves in a spin and valley-polarized system at a van-hove singularity}} (\bibinfo {year} {2025}),\ \Eprint {https://arxiv.org/abs/2504.19321} {arXiv:2504.19321 [cond-mat.str-el]} \BibitemShut {NoStop}%
\bibitem [{\citenamefont {Sedov}\ and\ \citenamefont {Scheurer}(2025)}]{sedov2025probing}%
  \BibitemOpen
  \bibfield  {author} {\bibinfo {author} {\bibfnamefont {D.}~\bibnamefont {Sedov}}\ and\ \bibinfo {author} {\bibfnamefont {M.~S.}\ \bibnamefont {Scheurer}},\ }\href {https://arxiv.org/abs/2503.12650} {\bibinfo {title} {Probing superconductivity with tunneling spectroscopy in rhombohedral graphene}} (\bibinfo {year} {2025}),\ \Eprint {https://arxiv.org/abs/2503.12650} {arXiv:2503.12650 [cond-mat.supr-con]} \BibitemShut {NoStop}%
\bibitem [{\citenamefont {Chen}\ \emph {et~al.}(2025)\citenamefont {Chen}, \citenamefont {Xu}, \citenamefont {Zhang},\ and\ \citenamefont {Schrade}}]{chen2025finite}%
  \BibitemOpen
  \bibfield  {author} {\bibinfo {author} {\bibfnamefont {Y.}~\bibnamefont {Chen}}, \bibinfo {author} {\bibfnamefont {C.}~\bibnamefont {Xu}}, \bibinfo {author} {\bibfnamefont {Y.}~\bibnamefont {Zhang}},\ and\ \bibinfo {author} {\bibfnamefont {C.}~\bibnamefont {Schrade}},\ }\href {https://arxiv.org/abs/2506.18886} {\bibinfo {title} {Finite-momentum superconductivity from chiral bands in twisted mote$_2$}} (\bibinfo {year} {2025}),\ \Eprint {https://arxiv.org/abs/2506.18886} {arXiv:2506.18886 [cond-mat.supr-con]} \BibitemShut {NoStop}%
\bibitem [{Note3()}]{Note3}%
  \BibitemOpen
  \bibinfo {note} {Y. Waschitz, A. Stern, and Y. Oreg, in preparation.}\BibitemShut {Stop}%
\end{thebibliography}%

\end{document}